\RequirePackage{rotating}
\documentclass{jfm}
\usepackage{textcomp}
\usepackage{graphicx}
\usepackage{float}
\usepackage{natbib}
\usepackage{epsfig}
\usepackage{mathtools,esint,gensymb}
\usepackage[dvipsnames]{xcolor}
\usepackage{framed}
\usepackage{lipsum}
\graphicspath{{figures/}}
\usepackage{color}
\usepackage{bm}
\usepackage{soul}
\usepackage{caption}
\usepackage{rotating}
\usepackage[labelformat=simple]{subcaption}

\begin{document}
\newtheorem{lemma}{Lemma}
\newtheorem{corollary}{Corollary}
\newcommand{\red}[1]{{\color{red}{#1}}}
\newcommand{\blue}[1]{{\color{blue}{#1}}}


\title{Porous cylinder arrays for optimal wake and drag characteristics}
\author
 {
 Aishwarya Nair\aff{1},
 Amirkhosro Kazemi\aff{1},
 Oscar Curet\aff{1}
  \and 
  Siddhartha Verma\aff{1,2}
  \corresp{\email{vermas@fau.edu}}
  }

\affiliation
{
\aff{1}
Department of Ocean and Mechanical Engineering, Florida Atlantic University, Boca raton, Fl 33431, USA
\aff{2}
Harbor Branch Oceanographic Institute, Florida Atlantic University, Fort Pierce, FL 34946, USA
}

\maketitle
\section*{Abstract}
The root systems of mangroves, a tree species found in intertidal tropical and subtropical coastal zones, provide a natural barrier that dissipates wave energy effectively and reduces sediment erosion. Here, we use a combination of experiments and numerical simulations to examine the wake and drag characteristics of porous arrays of cylinders, which serve as simplified models of mangrove root networks. Optimal arrangements of the arrays are obtained by coupling Navier-Stokes simulations with a multi-objective optimization algorithm, which seeks configurations that minimize wake enstrophy and maximize drag on the porous structure. These optimal configurations are investigated using Particle Image Velocimetry, and the internal and external flow around the porous arrays are analyzed using a combination of Proper Orthogonal Decomposition and Lagrangian particle tracking. Large variations in drag and enstrophy are observed by varying the relative positions of the cylinders, which indicates that the geometrical arrangement of porous arrays plays a prominent role in determining wake and drag characteristics. A sensitivity analysis suggests that enstrophy is more sensitive than drag to specific cylinder placement, and depends on distinctive flow patterns that develop in the interior due to interactions among neighbouring cylinders. Arrays with higher drag involve a combination of larger projected frontal area and minimal flux through the interior, leading to increased wake enstrophy which is unfavourable for particle deposition and erosion. Based on the analysis of characteristics associated with the optimal arrays, several manually designed arrays are tested and they display the expected behaviour with regard to drag and enstrophy.

\section{Introduction}  
\label{sec:intro}
Coastal vegetation such as mangrove forests are among the most effective natural forms of coastal protection. Mangroves usually grow in tropical and subtropical regions, and provide many ecosystem services such as carbon dioxide sequestration~\citep{Sanderman2018}, nurturing juvenile fish~\citep{Mumby2006}, wave energy attenuation, and reducing coastal erosion~\citep{Furukawa1997,Narayan2016,Barbier2011,Mitsch2015}. The root systems of these mangrove forests have inspired novel designs for artificial seawalls~\citep{Takagi2017} with the aim of mitigating storm damage along coastlines. Compared to impervious breakwater structures, porous bioinspired designs require less material, and are less disruptive to natural habitats as they allow the natural ebb and flow of sediments. A recent experimental study demonstrated that mangrove-inspired roots models could also help reduce sediment erosion \citep{Kazemi2021}. While significant effort in the literature has been directed towards studying the influence of porous vegetation on the flow~\citep{Lynn2002,rominger2011,Kazemi2018}, the design of optimal porous breakwater structures requires further examination. Moreover, a detailed investigation of flow patterns that develop within such porous structures can help us understand their influence on sediment deposition as well as the intensity of vorticity and enstrophy generated in the wake.
 
\begin{figure}
\centering 
\includegraphics[width=15cm]{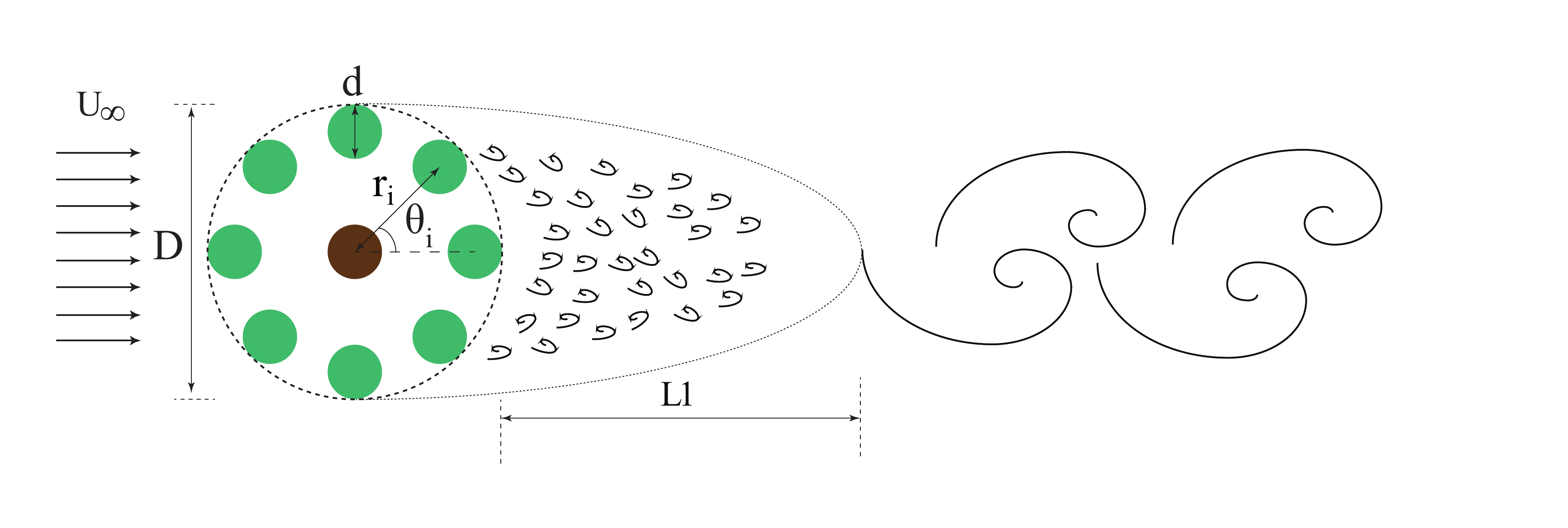}
	\caption{Schematic showing the horizontal cross section through a simplified model of a mangrove root system, with a uniform inflow imposed from left to right ($U_{\infty}$). The green cylinders represent auxiliary mangrove roots located around a main trunk indicated by the brown cylinder in the center. D represents the maximum nominal diameter of the array, whereas d represents the diameter of each individual cylinder in the array. $r_{i}$ and $\theta_{i}$ indicate the radial distance and azimuthal angle with respect to the central cylinder, and their values determine the overall layout of the array. L1 represents the steady wake region between the end of the patch and the onset of vortex shedding.}
\label{fig:patch}
\end{figure}

A number of studies have investigated various aspects of flow patterns generated by isolated finite-sized vegetation clusters. \citet{Anderson1982} found that while flexible plants reduce the turbulence  intensity in the oncoming flow, they also introduce significant unsteadiness due to vortex shedding by individual roots in the system. \citet{Zong2012} conducted experiments using cylinders to show that the flow velocity decreases immediately behind the vegetation cluster, due to increased drag exerted by the porous obstacle. \citet{Ricardo2016} showed that the incoming flow enters the porous patch and passes through to the wake, which can delay the onset of the von K\`arm\`an vortex street up to a certain distance behind the patch (e.g., L1 - Figure~\ref{fig:patch}). This length L1 is a function of the wake velocity, which depends on the freestream velocity and the extent of flow blockage (\citet{Chen2012}). Over this distance, the mean velocity is lower in magnitude compared to the freestream velocity, which leads to enhanced sediment deposition. Beyond this region, the von K\`arm\`an vortex street leads to a reduction in sediment deposition~\citep{Leonard1995}.

For seawalls, the drag force and extent of sediment deposition in the wake are two important functional characteristics. These two aspects have been studied primarily with regard to their dependence on root density, or the porosity of uniformly distributed arrays of cylinders, as in several of the studies mentioned previously. Root porosity is usually defined as the fraction of volume occupied by water in a given water-root space. Experiments conducted using model forests have demonstrated the existence of a high-drag region associated with the roots, resulting in reduced velocity within that zone~\citep{Maza2017}. Furthermore, \citet{Chang2015} used 3D simulations to demonstrate the dependence of drag and steady wake length on porosity. \citet{Shan2019} found that random physical arrangements of root models resulted in higher drag compared to uniform grid arrangements, and that the random arrangements resulted in higher variations in local velocity and drag on the individual cylinders. \citet{Norris2019} examined the role of turbulence within actual mangrove canopies, and found that high root densities enhance turbulent mixing which in turn is detrimental to sediment deposition. Several of these studies remark that for a group of cylinders, the drag force and the length of the sediment deposition region are two competing characteristics that depend on porosity and geometrical arrangement within the porous patch. Ideal designs should result in high drag and long sediment deposition regions. However, the drag force is known to decrease with increasing porosity \citep{Kazemi2021}, whereas the sediment deposition length increases at lower porosity \citep{Kazemi2017}. The trade-off between drag force and length of sediment deposition underscores the need for an effective method for optimizing the arrangement of the individual cylinders that constitute the porous patch.

A few studies have investigated the compromise between drag force and sediment deposition length. For instance, \citet{Ricardo2016} examined the wakes for cylinders arranged in random arrays using Particle Imaging Velocimetry (PIV), and found that vorticity-cancellation caused by neighbouring cylinders within the array was an important factor in suppressing the wake. However, there are few studies that aim to leverage this finding to optimize the drag or turbulence intensity generated by a porous array of cylinders. Most studies that examine the wake characteristics' dependence on the relative locations of cylinders consist of cylinders arranged in pre-determined geometrical configurations. Moreover, most such arrays are either spread uniformly across the area under consideration in a staggered or in-line arrangement, or arranged completely randomly \citep{Chen2012, Gijon2021}. \cite{Maza2015} determined using three-dimensional simulations of cylinder array distributions that wave-induced forces and wave attenuation were affected notably by the geometrical arrangement, and that using uniform models as a basis for studying natural vegetation could lead to an incorrect estimation of certain characteristics. This can be attributed to differences in local velocity that are dependent on an individual cylinder's placement within the array, resulting in large variations in drag force experienced by individual cylinders~\citep{Tinoco2020}.

Other studies of porous arrays have adopted simplified mathematical models to focus on rigid cylinder arrays with various arrangements in uniform flow ~\citep{Nepf1999,Nicolle2011,VanRooijen2018}. However, these models can involve certain limitations and simplifications with regard to important physical phenomena such as viscous effects and vortex shedding, and often do not provide detailed flow field information. Most of the experimental studies mentioned previously also entail certain limitations as described by \citet{Tinoco2020}, such as their time-consuming and labour-intensive nature, which in turn necessitates the use of simplifying parameters such as uniform arrangements. This makes it difficult to examine a large and varied assortment of cylinder arrangements when studying optimal characteristics experimentally. Numerical simulations present a useful complement to experiments, at least at low to moderate Reynolds numbers, since they enable the evaluation of a large number of samples which is a crucial requirement for conducting optimization studies. In the present work, we couple two-dimensional Direct Numerical Simulations (DNSs) of flow around porous cylinder arrays with a multi-objective optimization algorithm, in order to discover geometrical arrangements that optimize drag force and wake enstrophy simultaneously. We note that while the present study is limited to two-dimensional flows owing to the high computational cost involved with optimization, three-dimensional flow structure is important when considering sediment transport and deposition which is modulated primarily by the near-bed flow. This near-bed flow depends heavily on the formation and interaction of horseshoe vortices near the base of individual cylinders.

The main aim of the present work is to discover optimal array configurations that maximize drag and minimize wake enstrophy, and to examine the underlying mechanisms that influence these characteristics. Higher drag leads to a higher velocity deficit in the wake \citep{Maza2017,Gijon2021}, which is preferable for coastal-defense structures inspired from mangrove vegetation since it is more effective in mitigating the impact of the incoming flow. However, while high drag leads to increased energy dissipation in the wake, the turbulent kinetic energy produced in this process generates stronger vortices (i.e., higher wake enstrophy) and hinders sediment deposition and promotes resuspension \citep{Chen2012,Tinoco2018}. Thus, a multi-objective optimization approach is adopted in the present work in an attempt to simultaneously attain the conflicting metrics of high drag and low enstrophy. The optimal cylinder arrangements obtained in this manner are also examined experimentally using PIV to support the numerical findings.

Details of the optimization algorithm, the simulation methodology, and the PIV setup are provided in \S\ref{sec:methods}. The optimal cylinder arrangements discovered by the optimization procedure are discussed in \S\ref{sec:results}, and the flow field both within and outside the porous arrays is examined in detail. A short discussion is presented in \S\ref{sec:discussion}, followed by concluding remarks in \S\ref{sec:conclusion}.

\section{Methods}
\label{sec:methods}
In the present work, arrays of circular cylinders are used as simplified models for mangrove roots. This approach is consistent with several of the studies discussed in \S\ref{sec:intro}, which have shown that three-dimensional effects are of second order with regard to the wake structure \citep{Zong2012, Verschoren2016}. The individual mangrove roots are represented as two-dimensional rigid cylinders with diameter $d$, which are grouped together in various arrangements to form a porous patch with maximum nominal diameter $D$, as shown in Figure~\ref{fig:patch}. The Reynolds number for all the configurations examined in this work is $Re_d = U_{\infty}d/\nu = 500$, based on the diameter of an individual cylinder $d$ and the uniform inflow velocity $U_{\infty}$. The porosity of the 9-cylinder arrays, defined as the solid volume fraction within a given reference volume \citep{Chang2015, Nicolle2011} is $\phi= 9(d/D)^{2} = 0.316$, and is constant for all the 2D arrays discussed here since they contain the same number of identical cylinders per unit reference area.

\subsection{Numerical Simulations}
The simulations used in this work solve the two-dimensional incompressible Navier-Stokes equations. Brinkman penalization is used to enforce the no-slip boundary condition at the fluid-solid interfaces with the help of a penalty forcing term:
\begin{equation} \label{eq:momentum}
    \frac{\partial \bm{u}}{\partial t} + \bm{u}\cdot\nabla\bm{u}= \frac{-\nabla P}{\rho} + \nu \nabla ^2 \bm{u} + \lambda \chi (\bm{u_s - u})
\end{equation}
\newline
Here, $\lambda=1/dt$ is the penalization parameter (with $dt$ being the time-step size) and $\chi$ is the characteristic function that represents the solid in discretized form on a Cartesian grid. Grid cells with $\chi = 0$ are occupied entirely by the fluid, and those with $\chi = 1$ are occupied entirely by the solid. $\chi$ transitions smoothly from $1$ to $0$ within 2 grid points at the interface, using a discrete representation of the Heaviside function. $\bm{u_s}$ is the pointwise velocity of the discretized solid and accounts for the motion of the solid object, if any. In the present work, $\bm{u_s} = 0$ for the rigid stationary cylinders.
 
To solve the Navier-Stokes equations, we use an open-source solver based on the remeshed vortex method \citep{ROSSINELLI20151, Verma2017}. This method uses the vorticity form of the momentum equation, which is obtained by taking the curl of Eq.~\ref{eq:momentum}.
\begin{equation} \label{eq:momentum_vort}
    \frac{\partial \omega}{\partial t} + \bm{u}\cdot\nabla\omega=  \nu \nabla ^2 \omega + \lambda \nabla \times \big(\chi (\bm{u_s - u}))
\end{equation}
In obtaining this equation, we have used $\nabla\cdot\bm{u} = 0$, and the fact that the vortex-stretching term $\bm{\omega}\cdot\nabla \bm{u}$ is absent in two-dimensions. Godunov splitting is used to perform time-splitting of advection, diffusion, and the penalty forcing term. The sequential steps used for solving Eq.~\ref{eq:momentum_vort} can be found in \citet{Verma2017}. We note that the simulations use the free-space boundary condition, which allows the use of Green’s function for solving the Poisson’s equation in the vortex method. A validation of the numerical simulations for flow around 2-cylinder configurations is shown in Appendix~\ref{app:validation}.

\subsection{Optimization Algorithm}
\label{sec:opt_algorithm}
\begin{figure}
\centering
\begin{subfigure}{0.24\textwidth}
  \centering
  \includegraphics[scale=0.16]{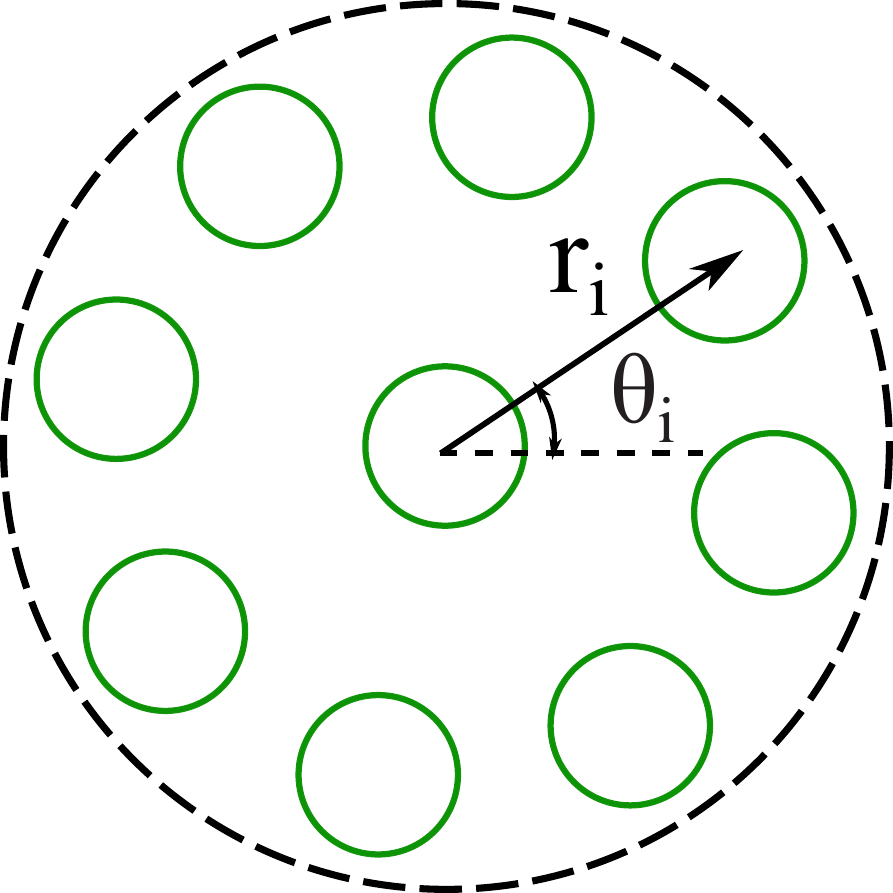}
  \caption{}
\end{subfigure}
\begin{subfigure}{0.24\textwidth}
  \centering
  \includegraphics[scale=0.16]{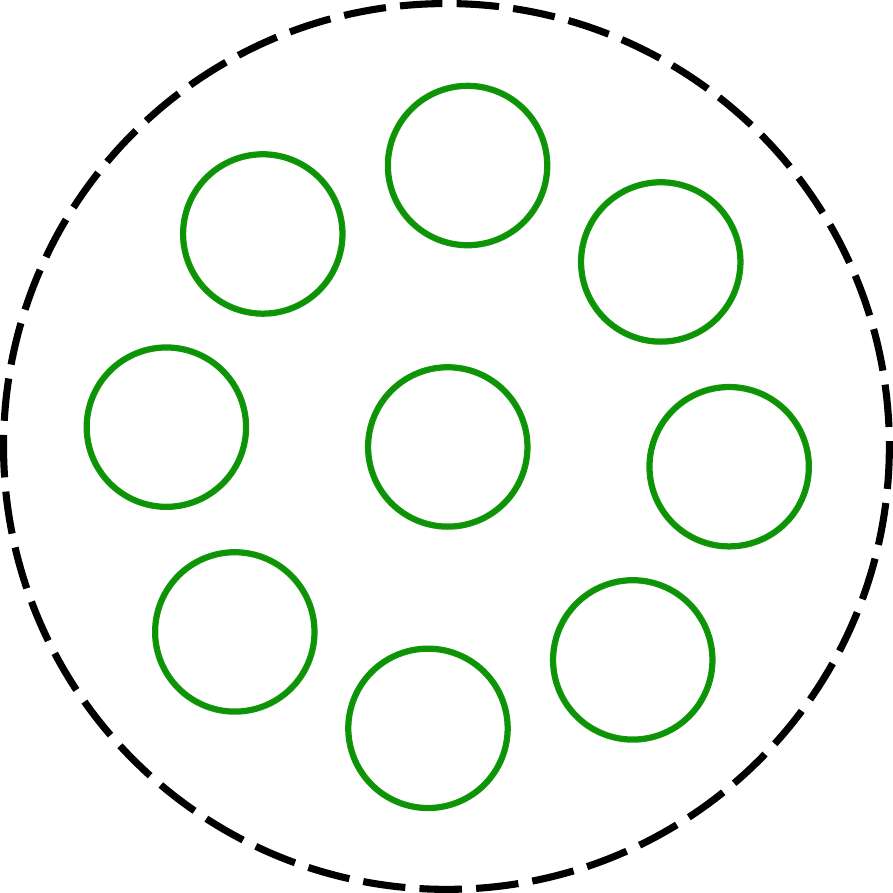}
  \caption{}
\end{subfigure}
\begin{subfigure}{0.24\textwidth}
  \centering
  \includegraphics[scale=0.16]{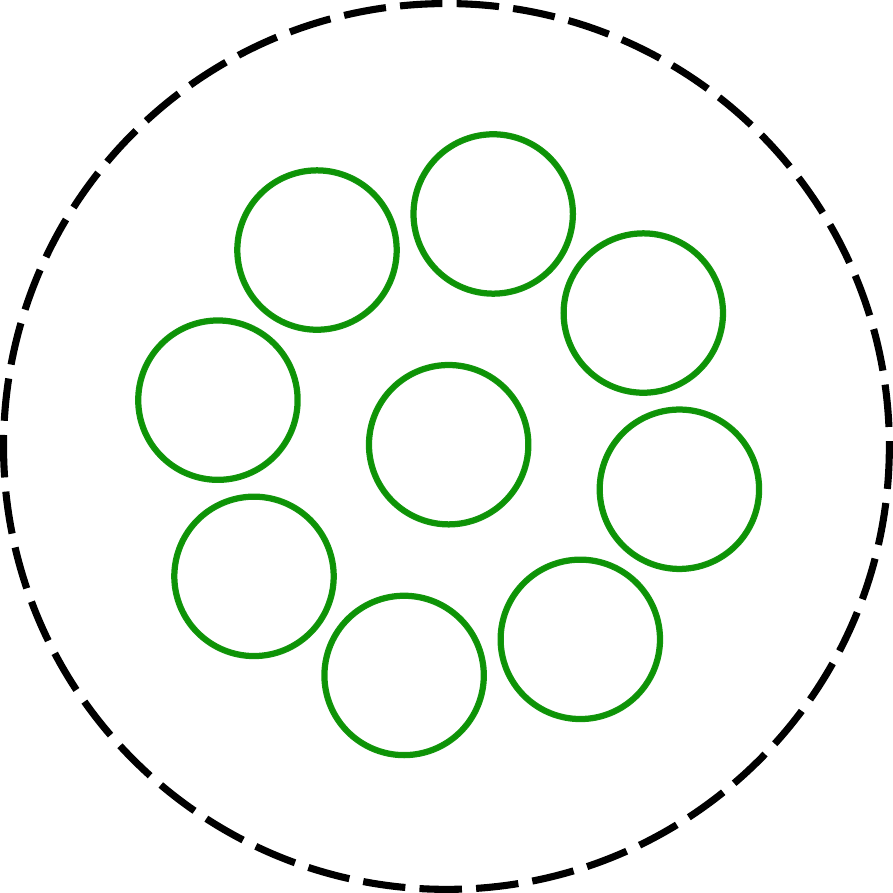}
  \caption{}
\end{subfigure}
\begin{subfigure}{0.24\textwidth}
  \centering
  \includegraphics[scale=0.16]{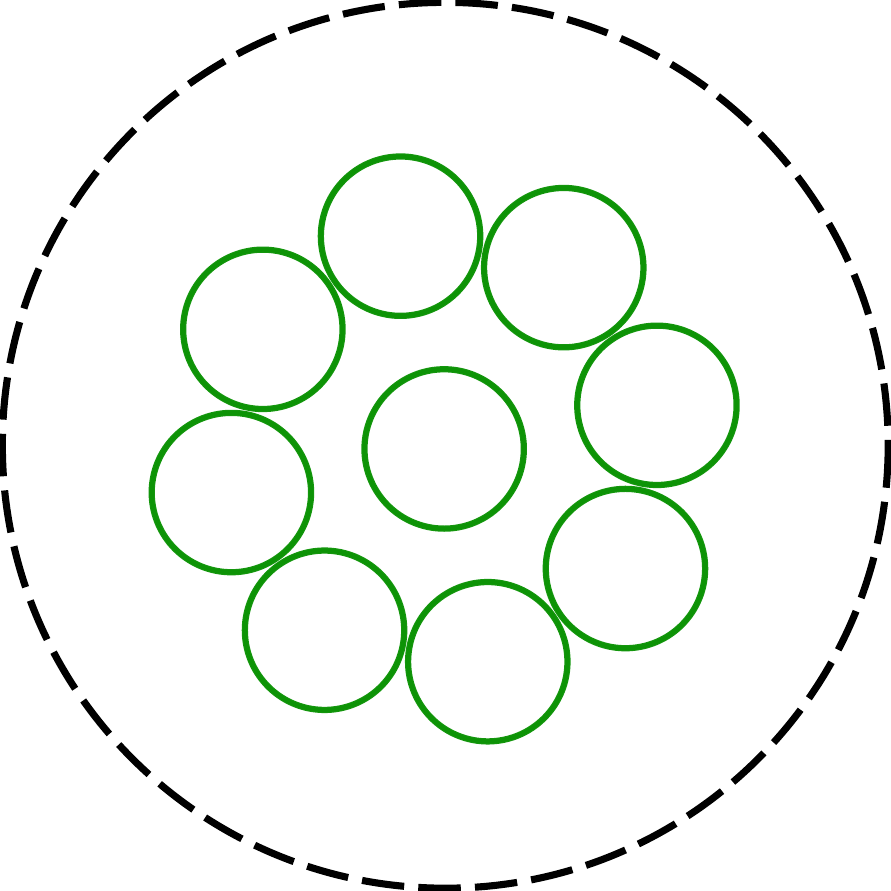}
  \caption{}
\end{subfigure}
\newline
\\
\begin{subfigure}{0.24\textwidth}
  \centering
  \includegraphics[scale=0.16]{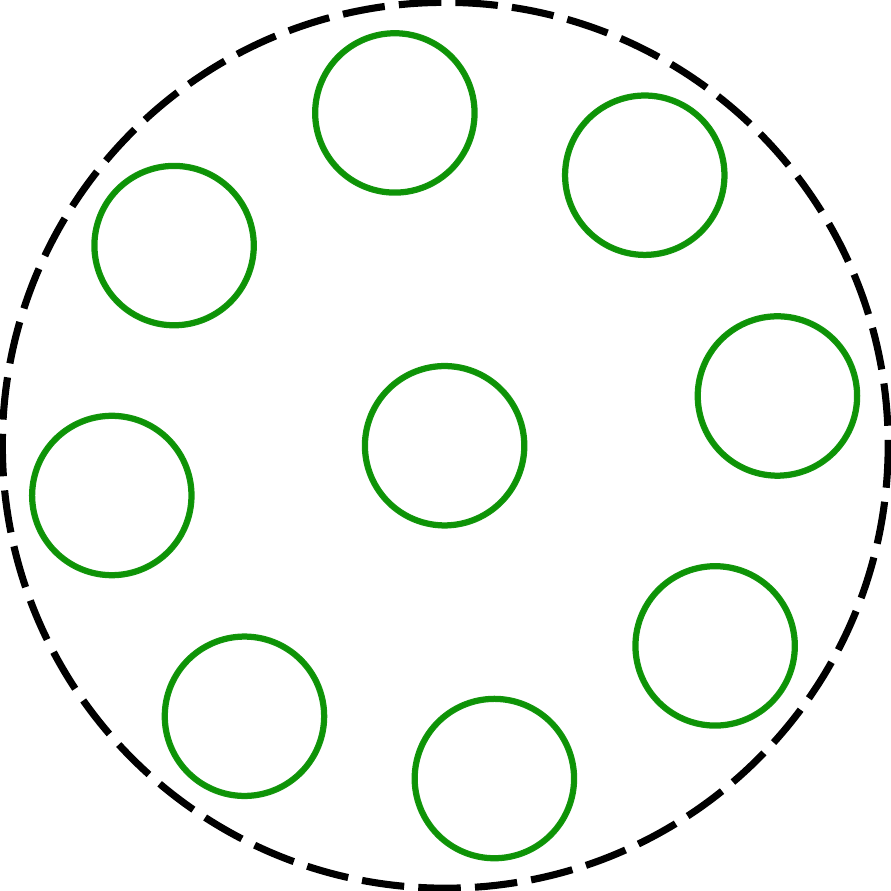}
  \caption{}
\end{subfigure}
\begin{subfigure}{0.24\textwidth}
  \centering
  \includegraphics[scale=0.16]{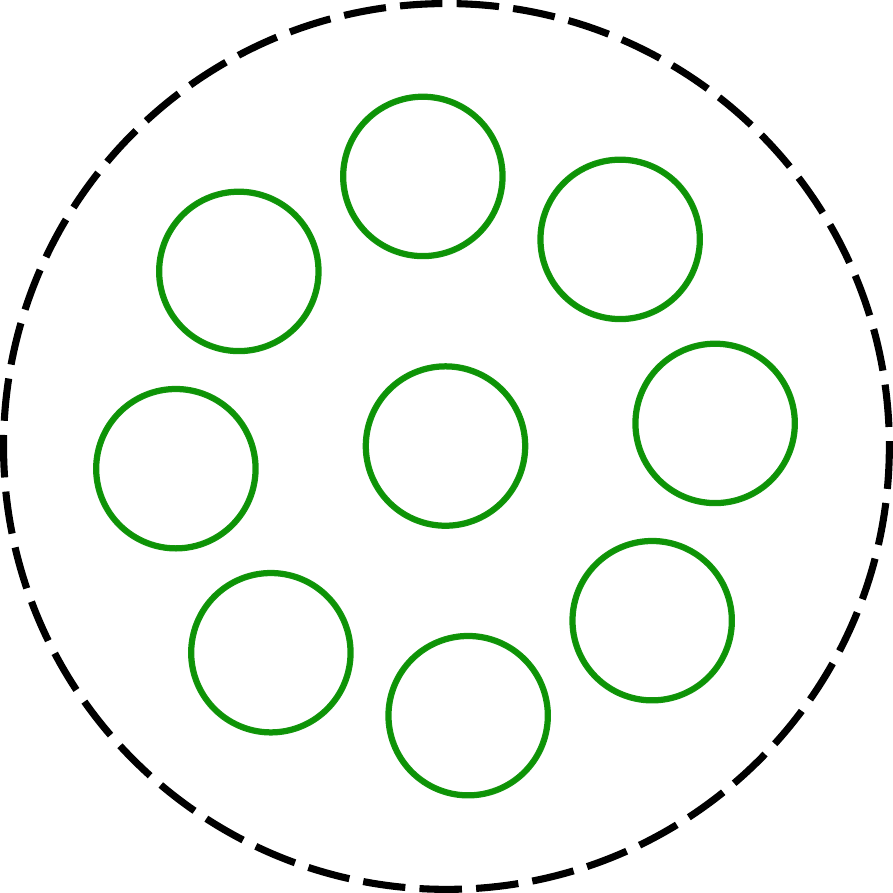}
    \caption{}
\end{subfigure}
\begin{subfigure}{0.24\textwidth}
  \centering
  \includegraphics[scale=0.16]{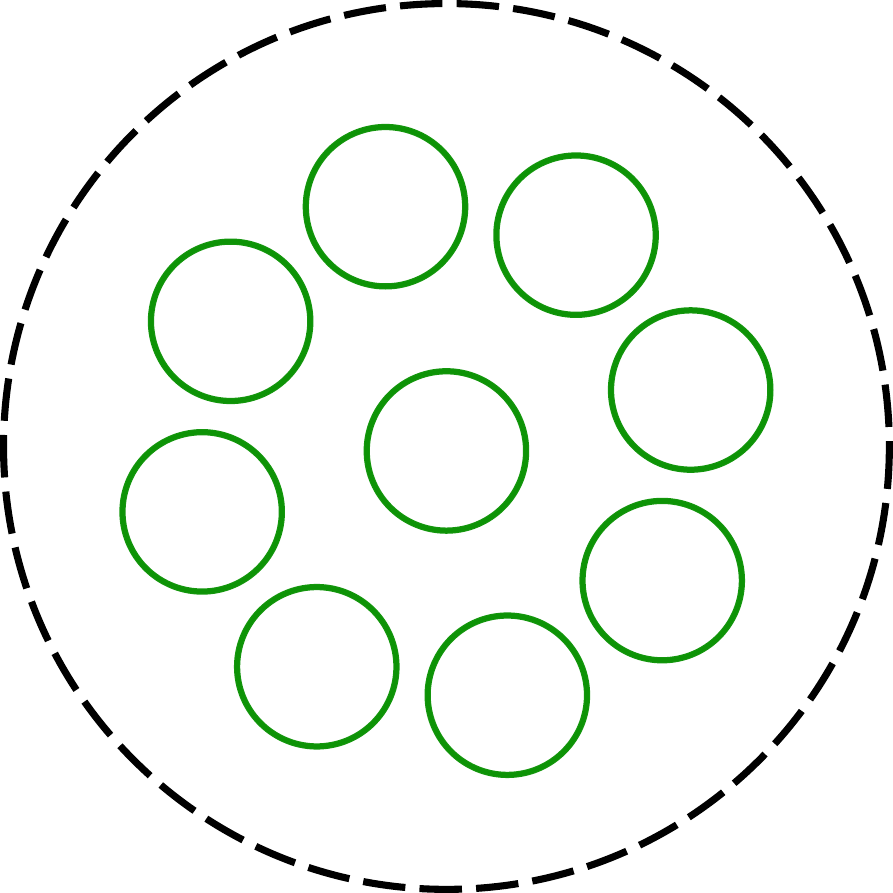}
    \caption{}
\end{subfigure}
\begin{subfigure}{0.24\textwidth}
  \centering
  \includegraphics[scale=0.16]{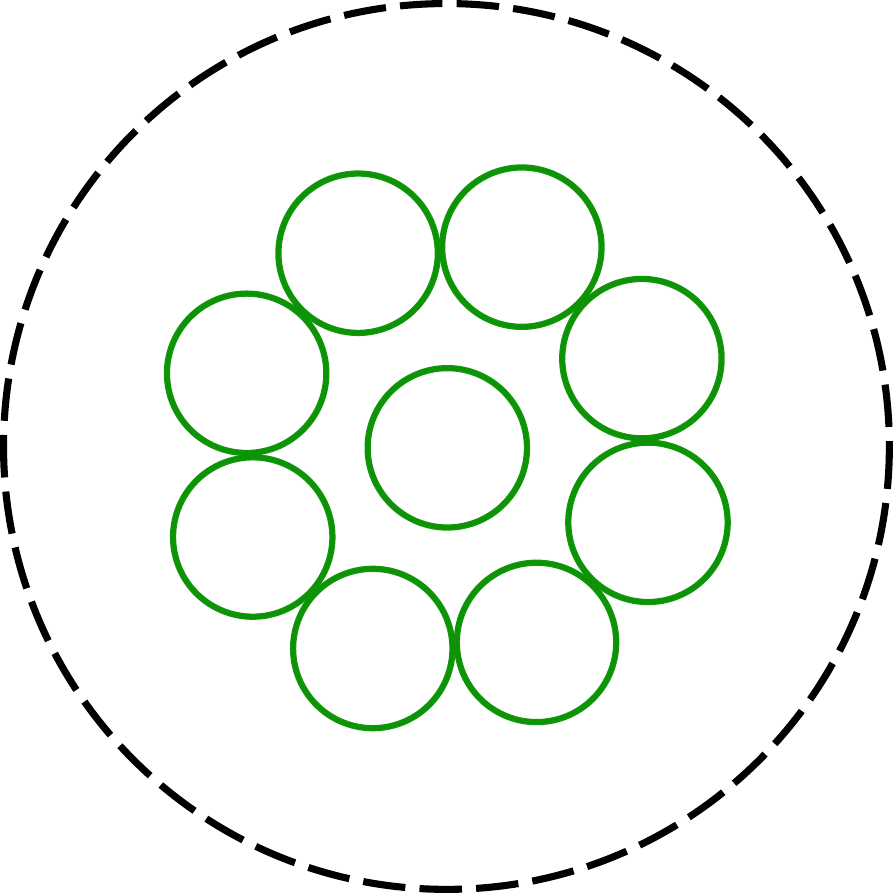}
    \caption{}
\end{subfigure}
\caption{Randomly generated individuals from the first generation used to initialize the optimization algorithm. Only 8 out of the 128 individuals generated for the first generation are shown here. Each of these randomized arrangements consists of one central cylinder, and 8 equally spaced cylinders that are also equidistant from the center. Subsequent generations created by the optimization procedure are not constrained to using equal radii and azimuthal angles for the 8 surrounding cylinders.}
\label{fig:gen1_arrangement}
\end{figure}

To determine the best possible configuration of cylinder arrangements, we employ a multi-objective genetic optimization algorithm called Non-dominated Sorting Genetic Algorithm II (NSGA II) \citep{Debsander2002}. The optimizer attempts to simultaneously maximize the drag acting on the structure and minimize the intensity of vortices in the wake, which is represented by the enstrophy in this case. In general, genetic algorithms operate on a collection of `individuals' and iteratively improve a population's characteristics by assigning each individual a certain `fitness' value or rank. The fitness value represents the eligibility of a particular individual to pass on its characteristics to future generations. In the present work, each `individual' corresponds to one distinct arrangement of 8 cylinders around a central fixed cylinder, and is characterized by 8 pairs of $r_{i}$ and $\theta_{i}$ values as shown in Figure~\ref{fig:gen1_arrangement}. A large number of distinct individuals are created and evaluated throughout the optimization process.
Fitness values are assigned to each individual based on the average enstrophy observed in the wake, and the net drag experienced by the cylinder array. These quantities are determined via two-dimensional Direct Numerical Simulations, where the particular individual (i.e., array arrangement) being evaluated is placed in a uniform inflow, as depicted in Figure \ref{fig:patch}. The average wake enstrophy is computed within a predefined window $\Omega \in [3.21D$ x $2.58D]$ downstream of the arrays, which is sufficiently large to accommodate the spanwise extent of the vortices generated in the wake. Each individual is evaluated for a total of 120 units of nondimensionalized time $(t^{*} = tU_{\infty}/d)$, during which the time-averaged enstrophy and net drag are computed. The enstrophy $E$ is computed as the double integral of the square of the vorticity magnitude $\omega$ within the predefined region $\Omega$ as follows:

\begin{equation}
    E = \iint\limits_\Omega\mid \bm{\omega}\mid ^2 dA
\end{equation}

\begin{figure}
    \centering
    \includegraphics[width=0.9\textwidth]{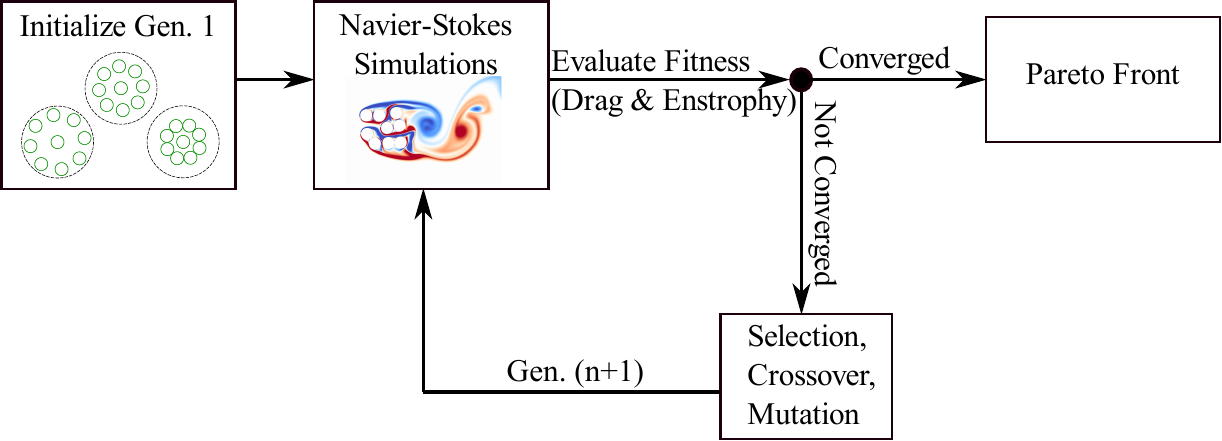}
    \caption{Flowchart depicting the multi-objective optimization of the porous cylinder arrays.}
    \label{fig:flowchart}
\end{figure}

The drag is computed using the streamwise component of the penalty-force determined from the 2D simulations as follows:
\begin{subequations}
\begin{align}
	\bm{F} &= \iint \rho \lambda \chi \bm{u} \ dA \\
	F_{Drag} &= \dfrac{\bm{F} \cdot \bm{u}}{\lVert \bm{u} \rVert}
\end{align}
\label{eq:penaltyDrag}
\end{subequations}
Once the two fitness values, i.e., drag and enstrophy, are determined for each individual using the numerical simulations, the individuals can be ordered using the concept of Pareto dominance. A member $\alpha_1$ of a population dominates another member $\alpha_2$ if: 1) $\alpha_1$ is no worse than $\alpha_2$ for all the objectives under consideration; and 2) $\alpha_1$ is strictly better than $\alpha_2$ in at least one objective. Each individual of the population is assigned a rank according to its level of non-dominance. Individuals with the same non-dominance ranks form a front, and are ranked amongst themselves using the crowding distance parameter, as described by \citet{Debsander2002}. The crowding distance parameter is an estimate of the average side-length of the cuboid formed by an individual's two closest neighbours. Individuals with a higher crowding distance are preferred because they increase diversity in the solution. Once the individuals are ranked according to their dominance and crowding distance parameter, the best ones are selected to act as parents for creating the next generation of individuals using tournament selection, crossover, and mutation \citep{Debsander2002}. In the present work, the probability of crossover was set to 0.9, with a crossover distribution index of 5, whereas the probability of mutation was set to 0.5 with a mutation distribution index of 10. Furthermore, an additional constraint was applied so that members of the new generation which resulted in overlapping cylinders were discarded, and a new replacement individual was sampled by the optimizer. The initial condition for the optimization algorithm was a set of 128 configurations that consisted of a central cylinder and 8 surrounding cylinders in a regular circular arrangement at a randomized but equal azimuthal angle and radial distance from the center (Figure~\ref{fig:gen1_arrangement}). The optimization algorithm was run for a total of 33 generations, until notable changes in the Pareto front were no longer observed. A flowchart depicting the optimization process is shown in Figure \ref{fig:flowchart}.

\subsection{Particle Image Velocimetry (PIV)}
\begin{figure}
\centering
\begin{subfigure}{\textwidth}
	\includegraphics[width=\textwidth]{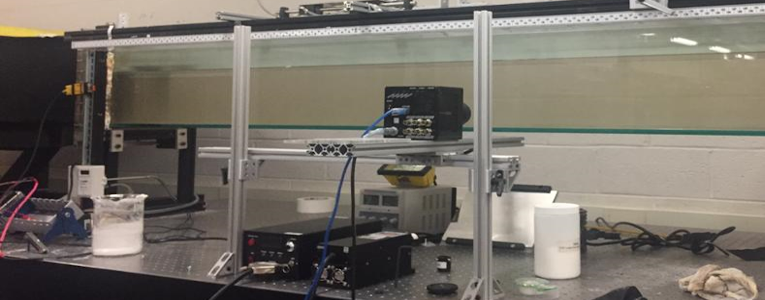}
	\caption{}
	\label{fig:sub1}
\end{subfigure}
\begin{subfigure}{.45\textwidth}
	\centering
	\includegraphics[width=\textwidth]{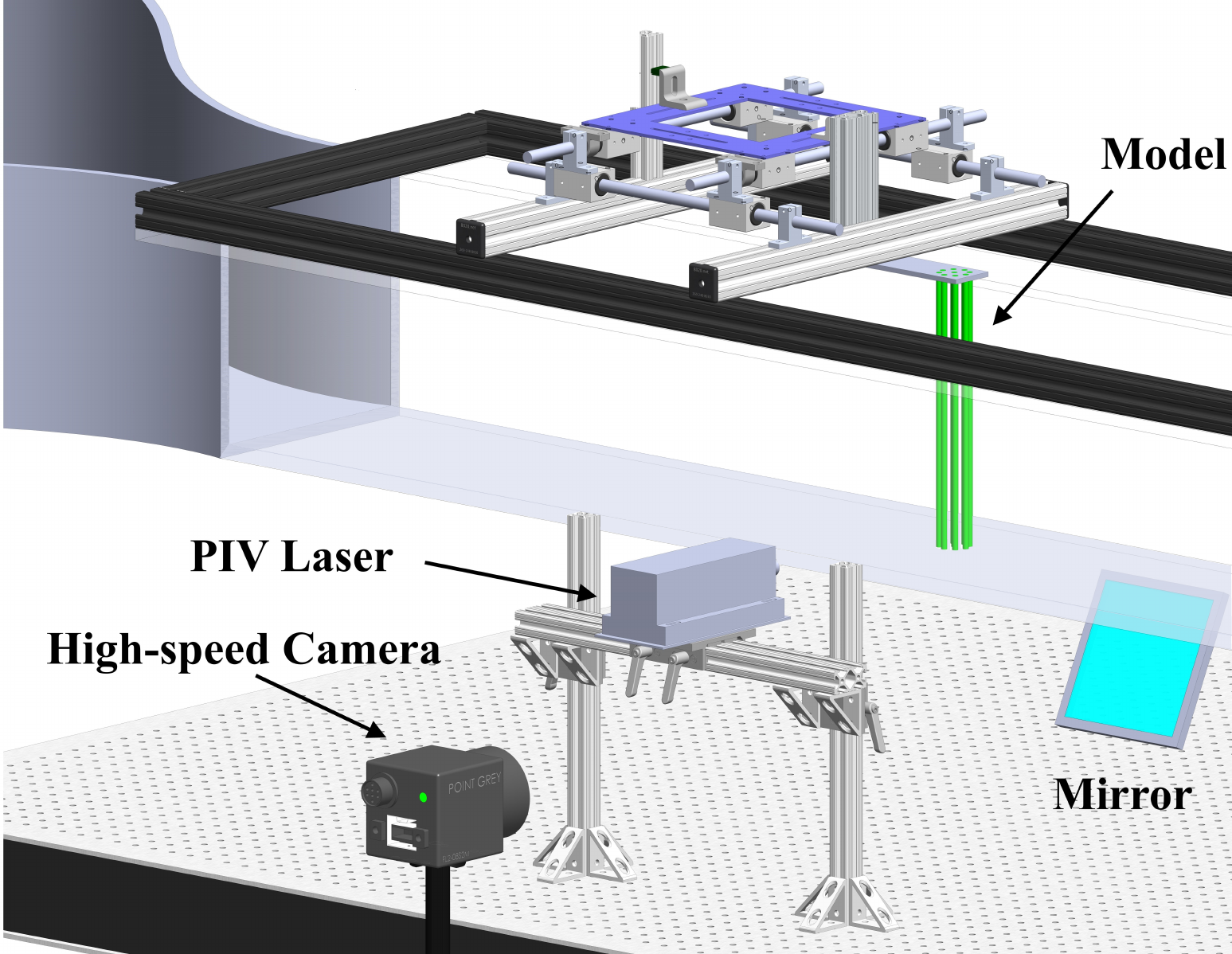}
	\caption{}
	\label{fig:sub2}
\end{subfigure} \qquad
\begin{subfigure}{.49\textwidth}
	\centering
	\includegraphics[width=\textwidth]{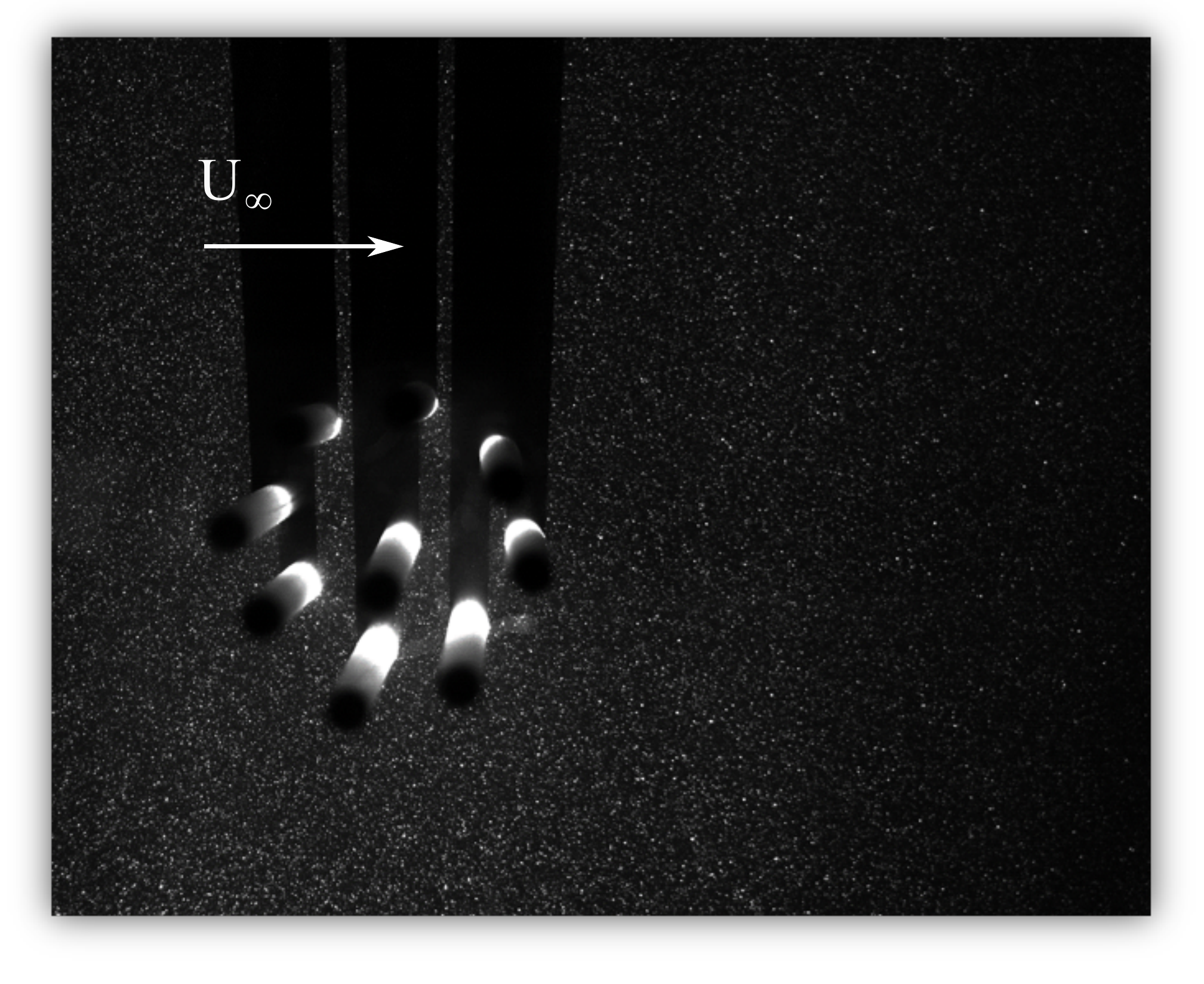}
	\caption{}
	\label{fig:sub3}
\end{subfigure}
\caption{\subref{fig:sub1} Particle Image Velocimetry setup used for analyzing the wakes of the cylinder arrays. \subref{fig:sub2} Schematic of the basic components used in the setup. \subref{fig:sub3} An image-capture showing the seeded particles used for velocimetry.}
\label{fig:piv}
\end{figure}

The behaviour of the optimal cylinder arrangements discovered via numerical optimization was also examined using a series of PIV experiments carried out in a closed-loop water flume with dimensions $2m\times0.25m\times0.25m$ (Figure~\ref{fig:piv}). 
The optimized cylinder arrays were  mounted at a distance of $0.7m$ ($27.96$ D) from the inlet. The blockage ratio of the test section, defined as the patch diameter divided by the channel width, was approximately $10\%$. The inflow velocity was set to $0.1m/s$ which corresponds to a Reynolds number of $Re_d = U_\infty d/\nu$ = 500 based on individual cylinder diameter $d$. The diameter of each individual cylinder in the experiments was $d=5mm$, and the maximum allowable nominal patch diameter was set to $D = 26.7mm$. The PIV images were recorded after reaching a quasi-steady state for the periodic vortex shedding, i.e., approximately three minutes after starting the inflow. The flow was seeded with hollow glass spherical particles of diameter $10\mu m$, which were illuminated with a continuous $5W$ laser with a $532 nm$ wavelength. The particles were recorded using a high-speed camera (Photron Fastcam Mini UX50) with a resolution of $1280\times1024$ pixels, at 125 frames per second using a shutter speed of $1/250s$. A total of 1000 images were recorded for each experimental run, which is equivalent to approximately 8 vortex-shedding cycles. The background was subtracted and a fast Fourier transform correlator was used for the PIV analysis. The velocity vector field was computed from particle displacement within a fixed elapsed time ($8ms$) from two sequential images. Measurements of the free stream velocity indicated that the turbulence intensity in the free stream was less than $0.5 \%$. 

\section{Results}
\label{sec:results}

\subsection{Optimization results}
\label{sec:opt_results}
\begin{figure}
\begin{subfigure}{0.5\textwidth}
  \centering
  \includegraphics{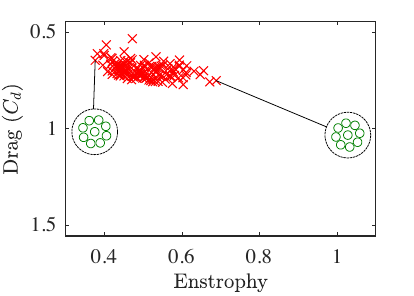}
  \caption{}
  \label{fig:gen1}
\end{subfigure}
\begin{subfigure}{0.5\textwidth}
  \centering
  \includegraphics{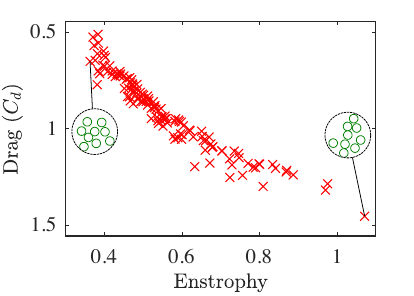}
  \caption{}
  \label{fig:gen11}
\end{subfigure}
\newline
\begin{subfigure}{0.5\textwidth}
  \centering
  \includegraphics{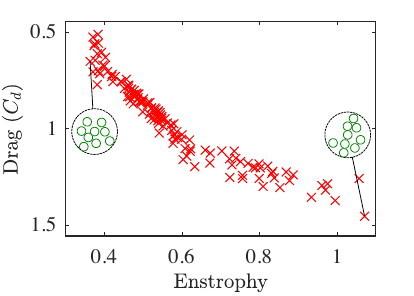}
  \caption{}
  \label{fig:gen22}
\end{subfigure}
\begin{subfigure}{0.5\textwidth}
  \centering
  \includegraphics{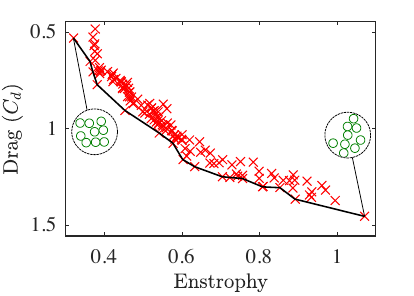}
  \caption{}
  \label{fig:gen33}
\end{subfigure}
	\caption {\label{fig:nsga} The evolution of a population of 128 distinct cylinder array arrangements with respect to the chosen fitness values, i.e., drag and average enstrophy, across successive generations: \subref{fig:gen1} generation 1; \subref{fig:gen11} generation 11; \subref{fig:gen22} generation 22; and \subref{fig:gen33} generation 33. The drag values shown represent the average drag coefficient for the arrays, $C_d = F_{Drag}/(0.5\rho U_\infty^2 \cdot 9d)$. Each `$\times$' symbol represents a unique arrangement of 8 cylinders around a central cylinder. The two extreme-most Pareto-optimal individuals for each generation are shown as inset.} 
\end{figure}

The results obtained from the optimization procedure are shown in Figure \ref{fig:nsga}, and indicate that the algorithm improves the desired wake characteristics of the porous cylinder arrangements with successive generations.
Each symbol on the plots represents a unique arrangement of cylinders within the confined patch diameter $D$ as described in \S\ref{sec:opt_algorithm}, with the axes depicting the corresponding drag and enstrophy values. The drag values shown represent the average drag coefficient $C_d = F_{Drag}/(0.5\rho U_\infty^2\cdot 9d)$ for each 9-cylinder array. We also note that the drag axis is inverted, i.e., the individuals are arranged such that the drag decreases along the positive vertical axis on the plots. This is necessary since the optimizer attempts to minimize wake enstrophy, while at the same time trying to maximize the drag acting on the structure, which are two mutually conflicting objectives as can be observed from Figure~\ref{fig:nsga}; in general, the wake enstrophy increases with increasing drag and vice versa.

In Figure~\ref{fig:gen1}, we observe a dense clustering of drag and enstrophy values for the initial 128 individuals, a subset of which are shown in Figure~\ref{fig:gen1_arrangement}. Although there is some variation in enstrophy, the drag coefficient is limited to approximately 0.7. Within the next 10 generations, the optimizer is able to discover cylinder arrangements that significantly increase drag, as can be seen in Figure~\ref{fig:gen11}. This indicates that the arrangement of cylinders within a porous patch must be considered, in addition to patch porosity, for obtaining the desired wake characteristics. The influence of projected surface area and velocity flux through the interior of the array are explored at a later point.

\begin{figure}
    \centering
    \begin{subfigure}{\textwidth}
    \centering
    \includegraphics{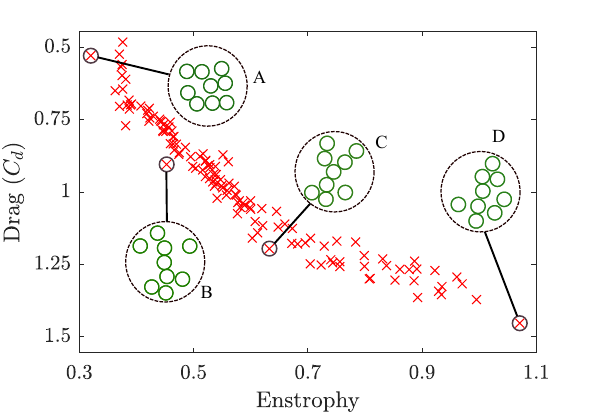}
    \subcaption{}
    \label{fig:gen33results}
    \end{subfigure}
    \begin{subfigure}{0.24\textwidth}
  \centering
  \includegraphics{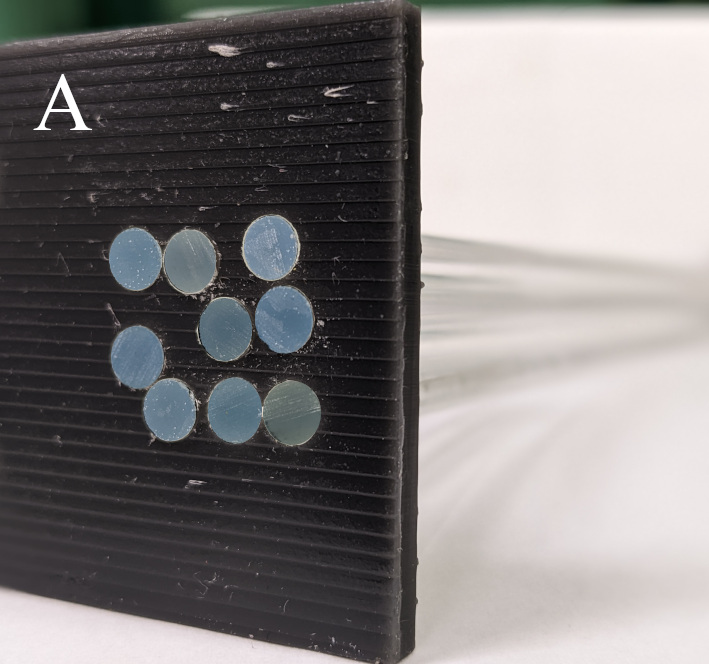}
  \subcaption{}
  \label{fig:piv1}
\end{subfigure}
\begin{subfigure}{0.24\textwidth}
  \centering
  \includegraphics{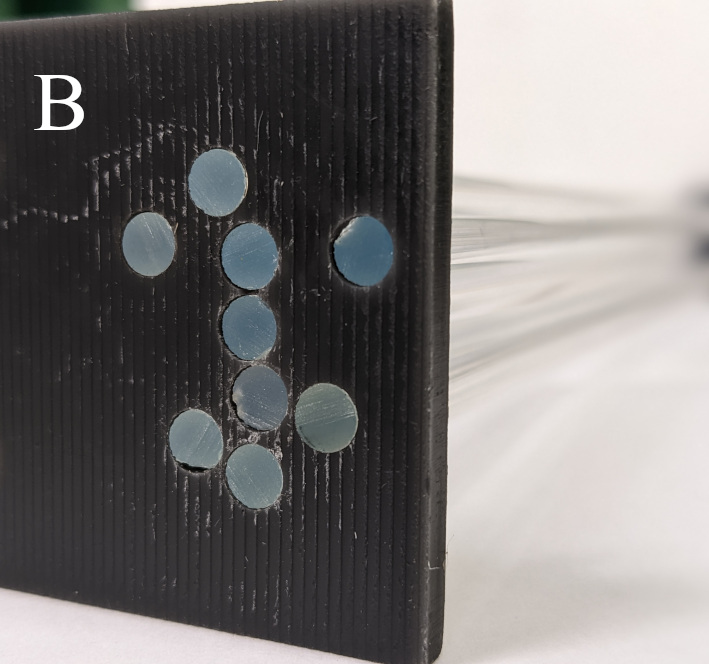}
  \subcaption{}
  \label{fig:piv2}
\end{subfigure}
\begin{subfigure}{0.24\textwidth}
  \centering
  \includegraphics{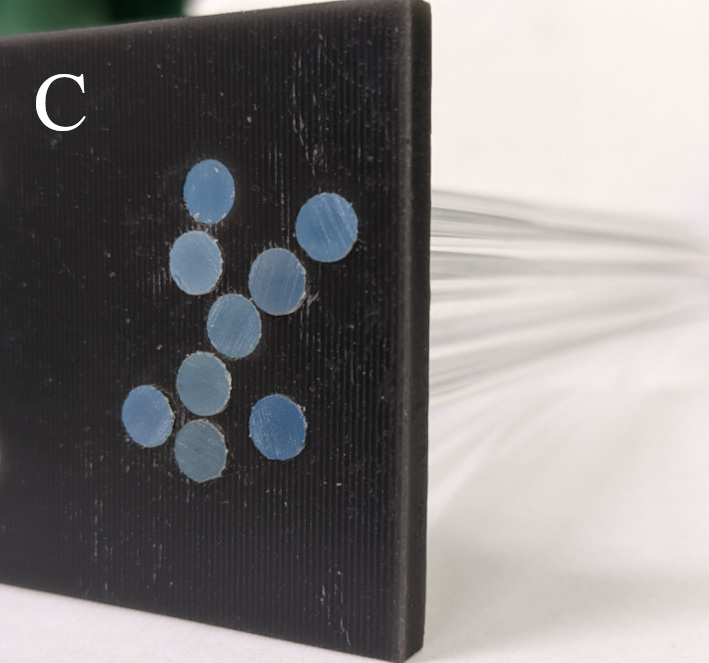}
  \subcaption{}
  \label{fig:piv3}
\end{subfigure}
\begin{subfigure}{0.24\textwidth}
  \centering
  \includegraphics{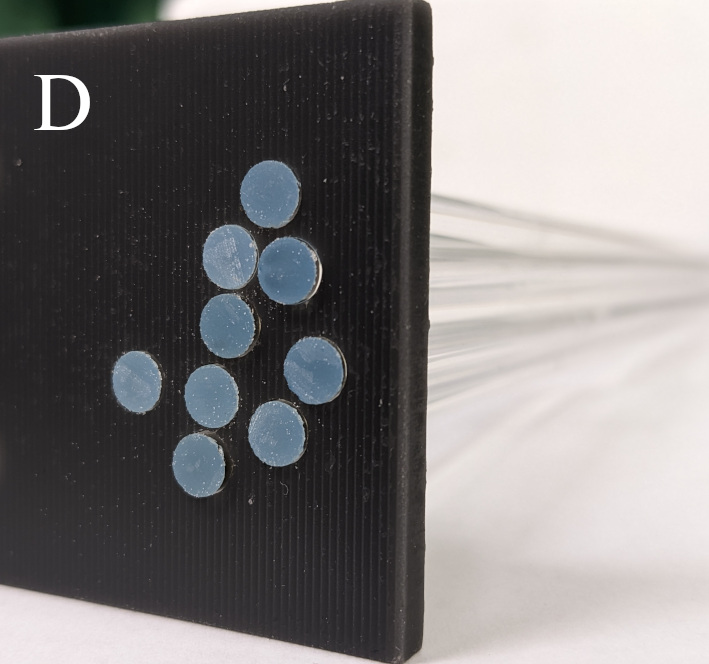}
  \subcaption{}
  \label{fig:piv4}
\end{subfigure}
	\caption{\subref{fig:gen33results} Results from generation 33 of the optimization process. The insets show the Pareto-optimal arrangements that were selected for further analysis. \subref{fig:piv1} - \subref{fig:piv4} Images of the corresponding physical models used in the PIV experiments ($5mm$ diameter borosilicate glass rods).}
    \label{fig:chosenones}
\end{figure}

Comparing the distribution of individuals from generations 11, 22 and 33 in Figure~\ref{fig:nsga}, we do not observe a significant change in the overall distribution of the results, although there is an increase in the number of high-drag individuals in later generations. Given this convergence, the optimization procedure was stopped after 33 generations. Individuals that comprise the final Pareto front are identified in Figure~\ref{fig:gen33}. This front consists of various distinct cylinder arrangements, all of which are optimal in their own right, i.e., no other solution is better than these Pareto-optimal individuals in \emph{both} fitness metrics (i.e., lower enstrophy and higher drag simultaneously). Four of these Pareto-optimal individuals were selected for further analysis, namely, the two individuals at the extreme ends of the Pareto front, as well as two intermediate individuals. These individuals are depicted in Figure \ref{fig:chosenones}, along with images of the physical models used for investigating the wake flow using the PIV experiments.

\subsection{Examining the wake flow}
\label{subsec:wakeflow}

\begin{figure}
\centering
\begin{subfigure}{0.45\textwidth}
  \centering
  \includegraphics[width=\textwidth]{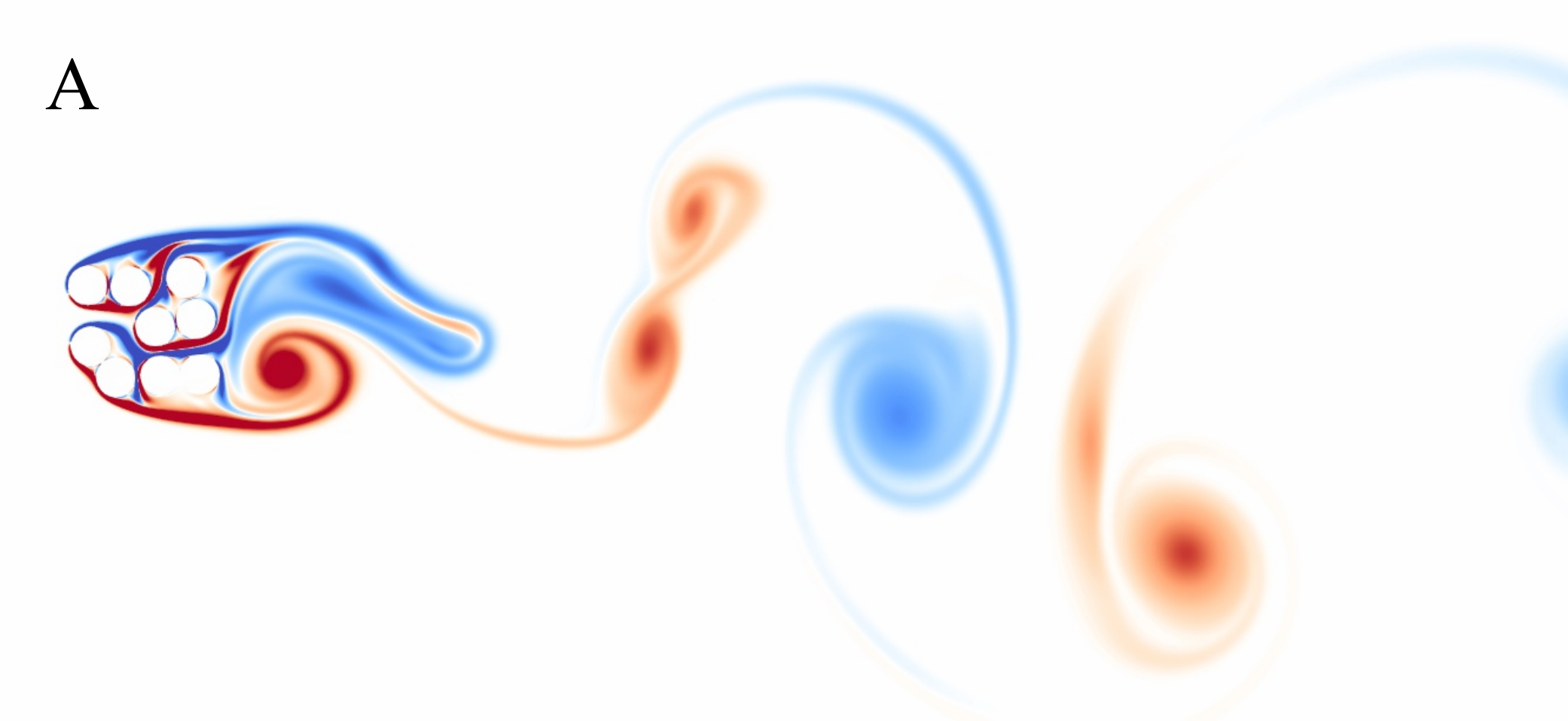}
  \subcaption{}
  \label{fig:wake1}
\end{subfigure}
\begin{subfigure}{0.45\textwidth}
  \centering
  \includegraphics[width=\textwidth]{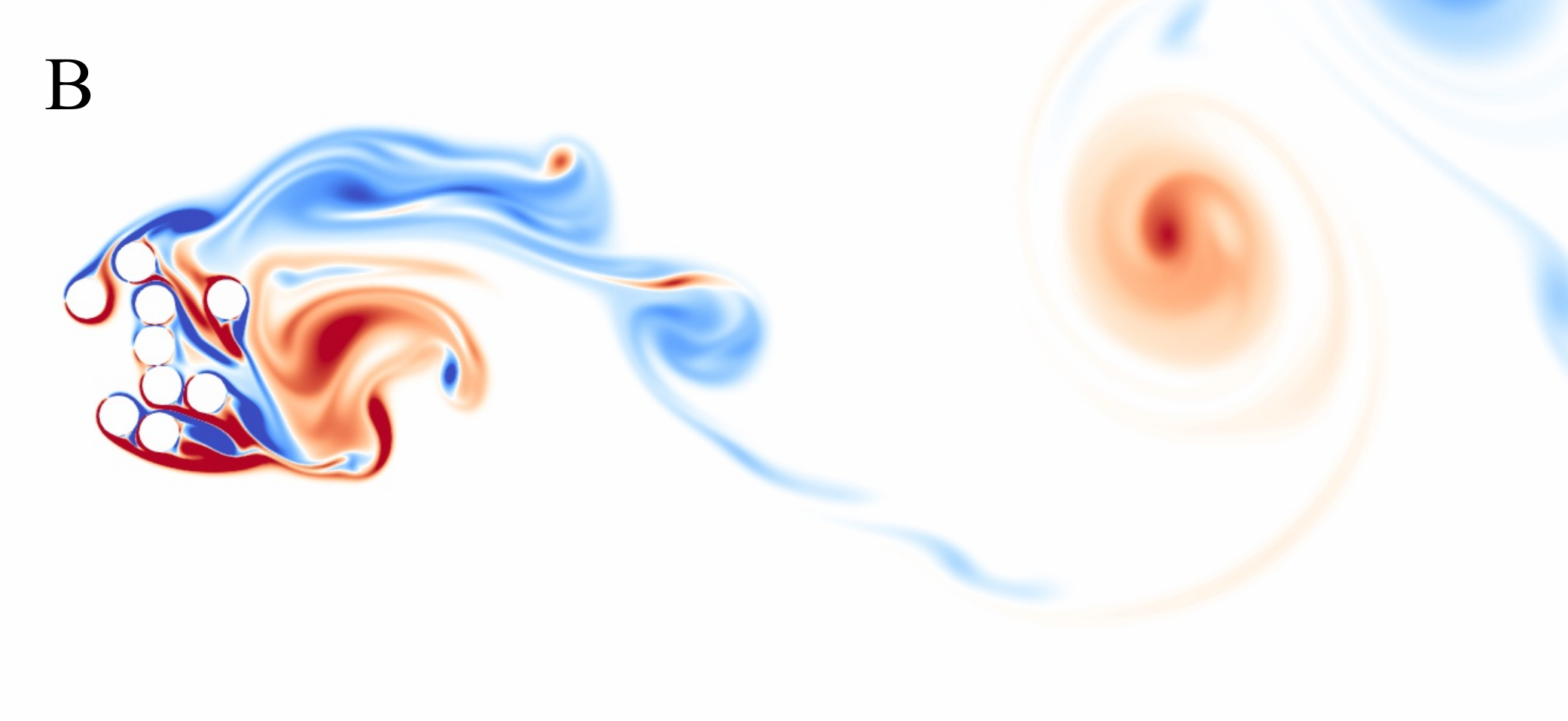}
  \subcaption{}
  \label{fig:wake2}
\end{subfigure}
\begin{subfigure}{0.45\textwidth}
  \centering
  \includegraphics[width=\textwidth]{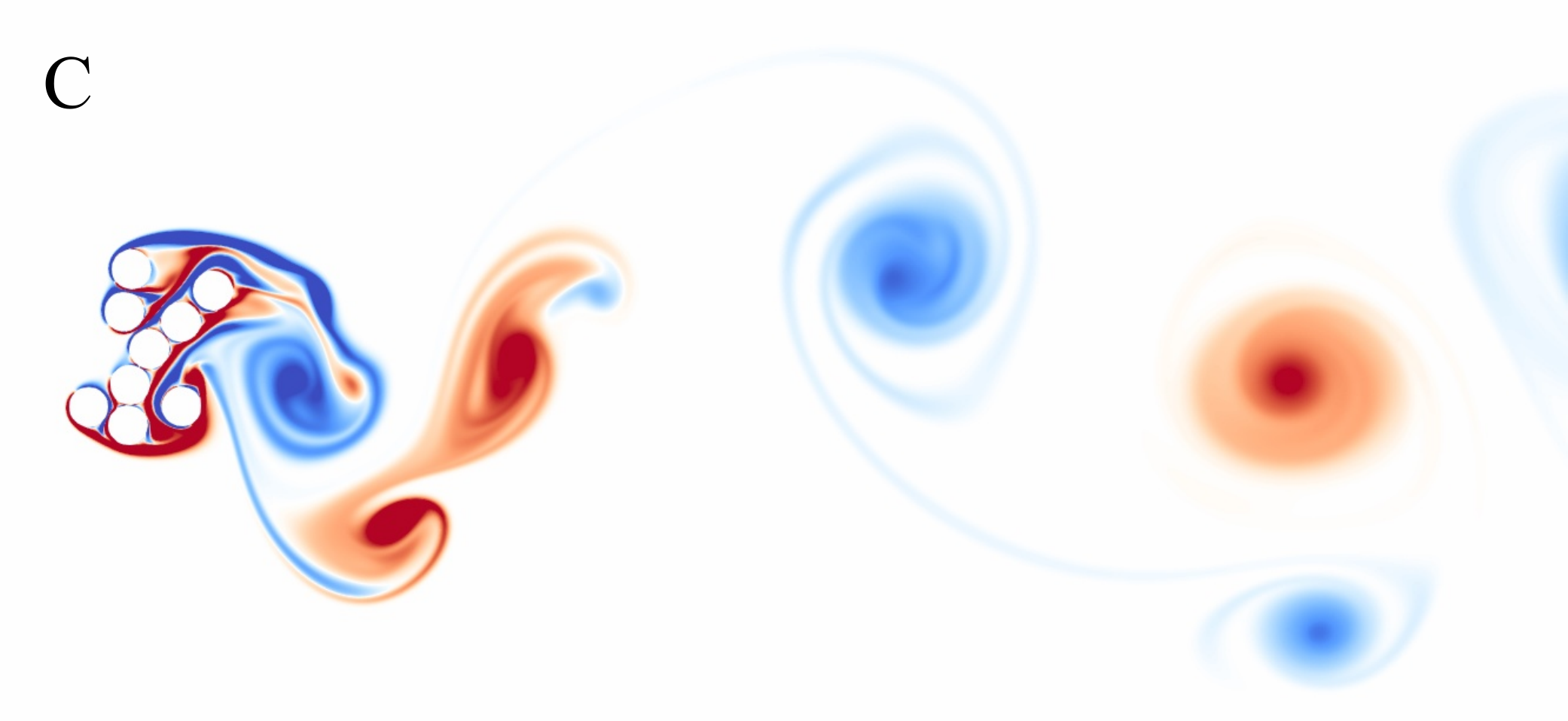}
  \subcaption{}
  \label{fig:wake3}
\end{subfigure}
\begin{subfigure}{0.45\textwidth}
  \centering
  \includegraphics[width=\textwidth]{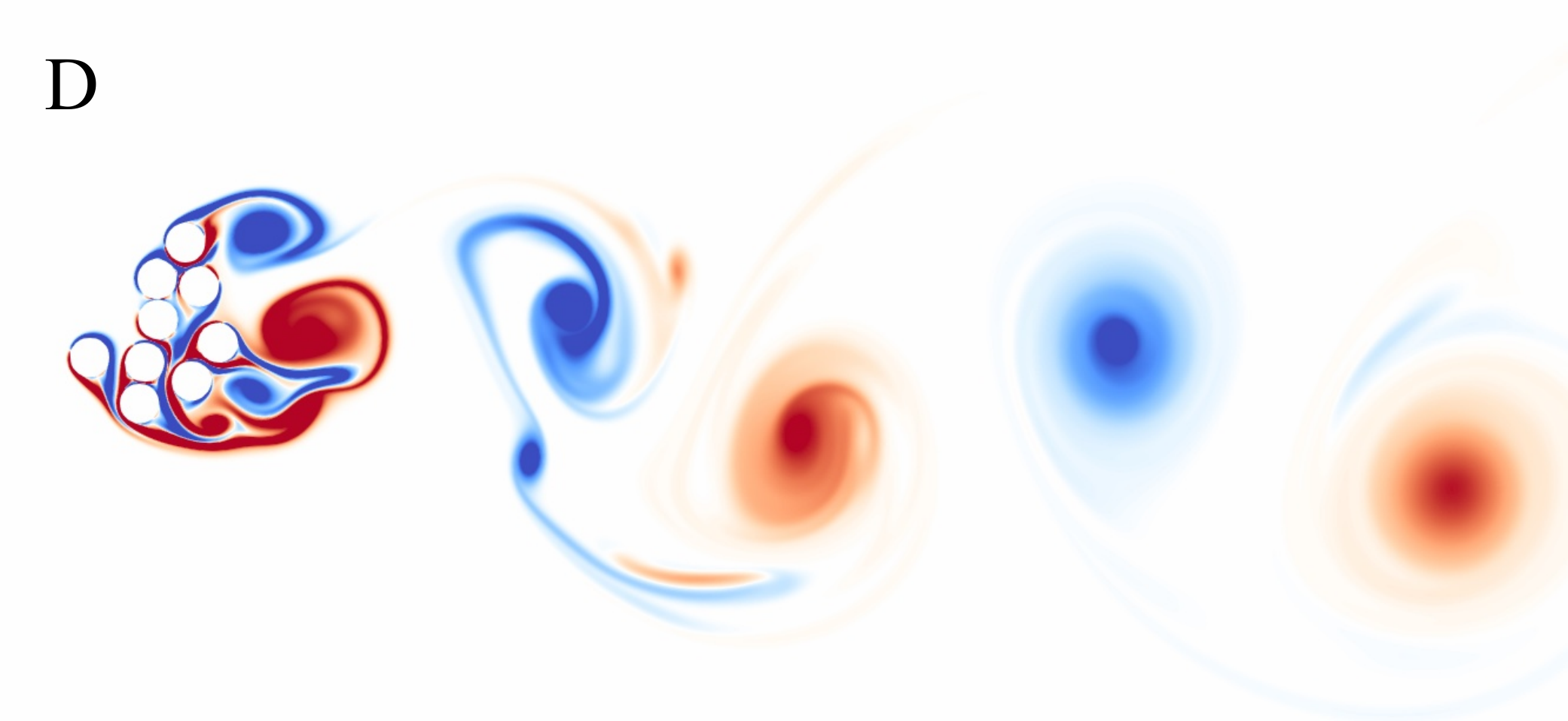}
  \subcaption{}
  \label{fig:wake4}
\end{subfigure}
\begin{subfigure}{\textwidth}
\centering
\includegraphics[width=0.75\textwidth]{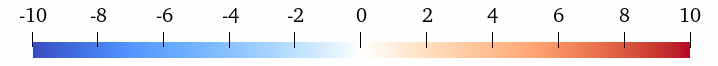}
\end{subfigure}
	\caption{The wake generated by each of the four Pareto-optimal arrangements shown in Figure \ref{fig:chosenones} when a uniform inflow is imposed from left to right. The data was obtained using DNS, and the colours indicate vorticity. Corresponding animations are provided in Supplementary Movie 1.}
\label{fig:4wakes}
\end{figure}

We now consider wake characteristics obtained using 2D Navier-Stokes simulations for the four different arrangements shown in Figure \ref{fig:chosenones}. Figure \ref{fig:wake1} depicts the array with the lowest average wake enstrophy (individual A), whereas Figure \ref{fig:wake4} shows the array with the highest wake enstrophy (individual D). Qualitatively, we observe that there is an increasing tendency to form distinct high-intensity vortices as we go from \ref{fig:wake1} to \ref{fig:wake4}. This trend is expected, given that higher enstrophy corresponds to higher vorticity magnitude. In addition to enstrophy, the configurations shown in Figure \ref{fig:4wakes} are also arranged in order of increasing drag, which was the second objective considered by the optimization procedure.

\begin{figure}
\centering
\begin{subfigure}{0.24\textwidth}
  \centering
  \includegraphics[width=\textwidth]{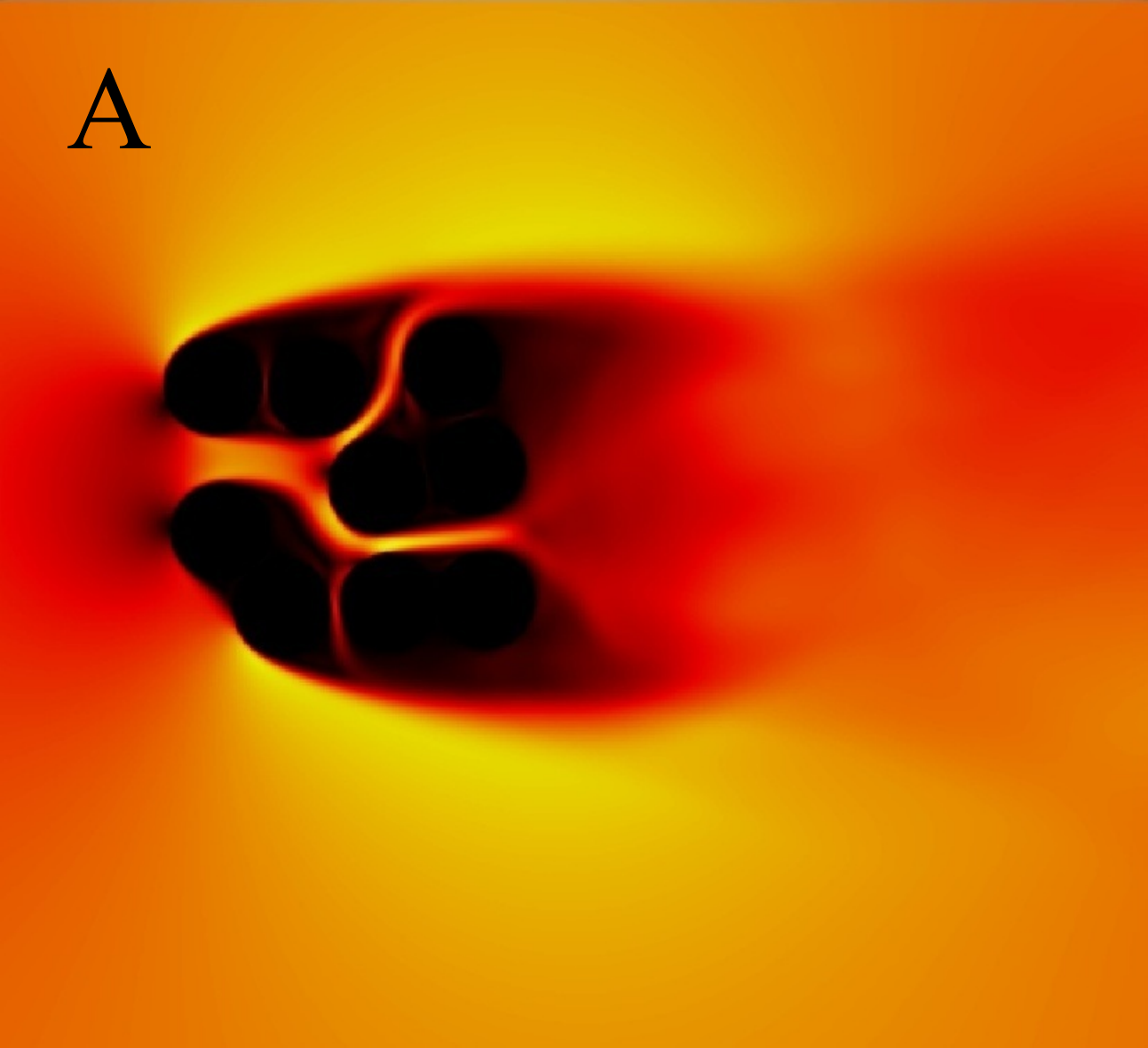}
  \subcaption{}
  \label{fig:avg1}
\end{subfigure}
\begin{subfigure}{0.24\textwidth}
  \centering
  \includegraphics[width=\textwidth]{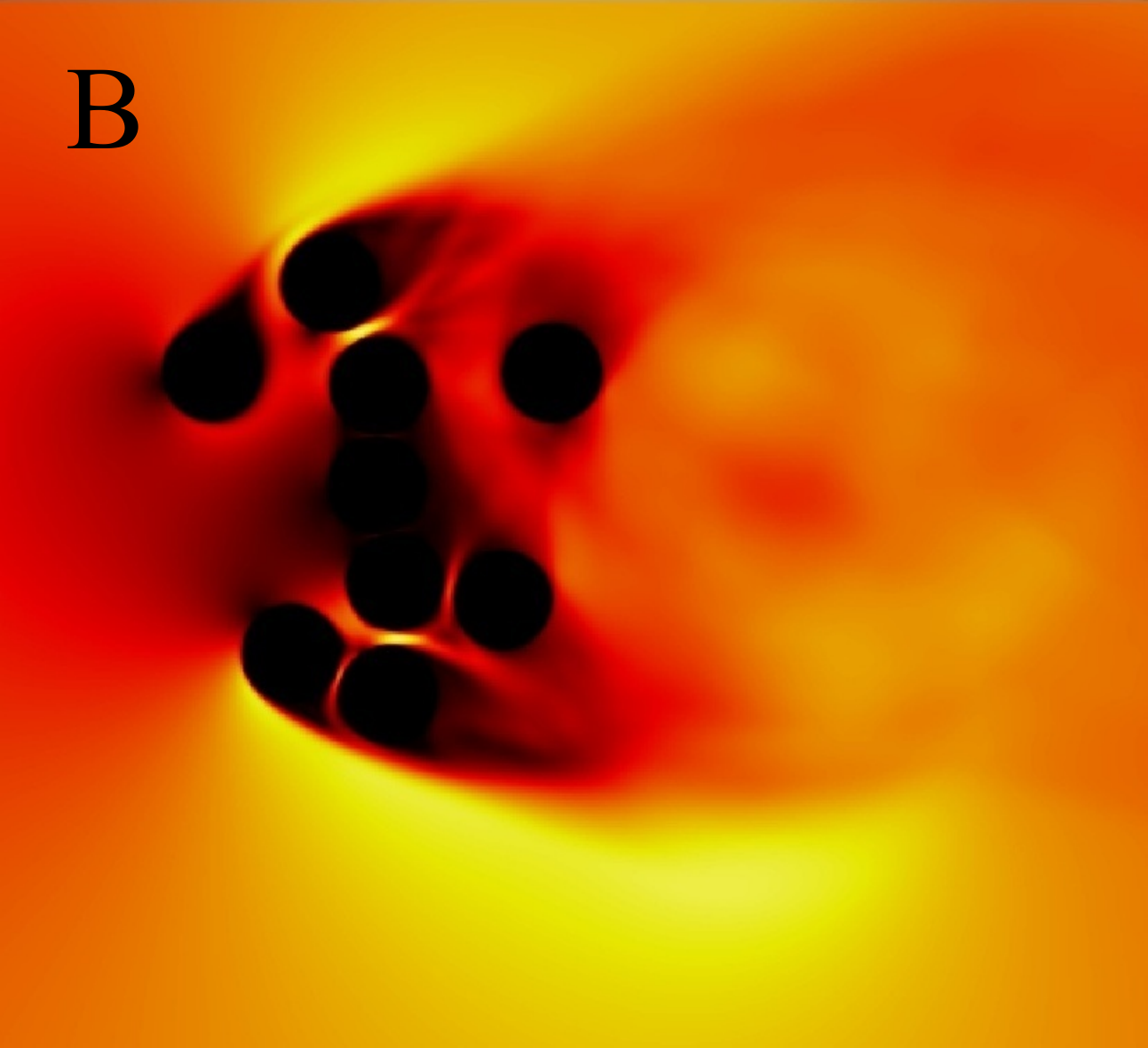}
  \subcaption{}
  \label{fig:avg2}
\end{subfigure}
\begin{subfigure}{0.24\textwidth}
  \centering
  \includegraphics[width=\textwidth]{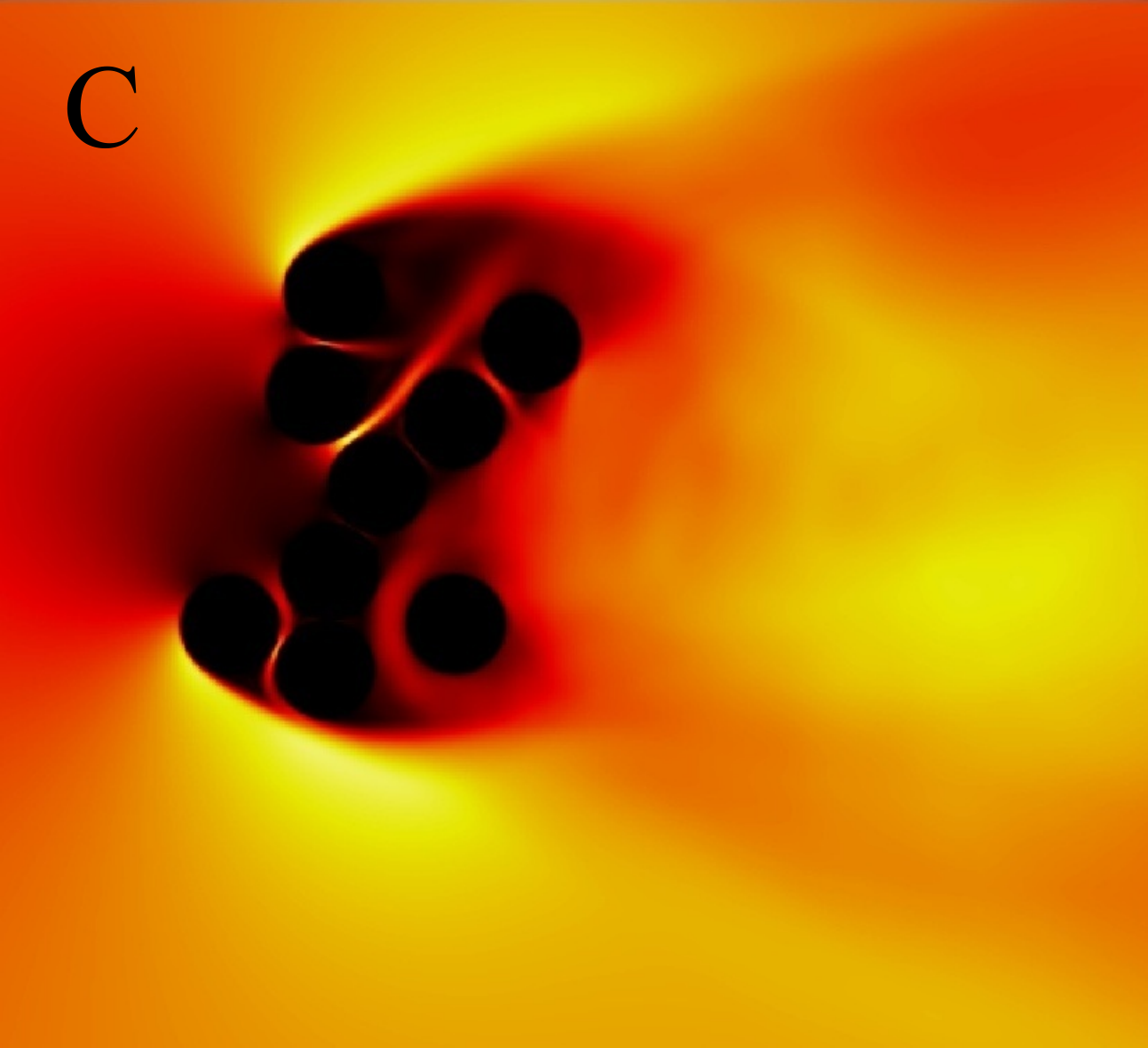}
  \subcaption{}
  \label{fig:avg3}
\end{subfigure}
\begin{subfigure}{0.24\textwidth}
  \centering
  \includegraphics[width=\textwidth]{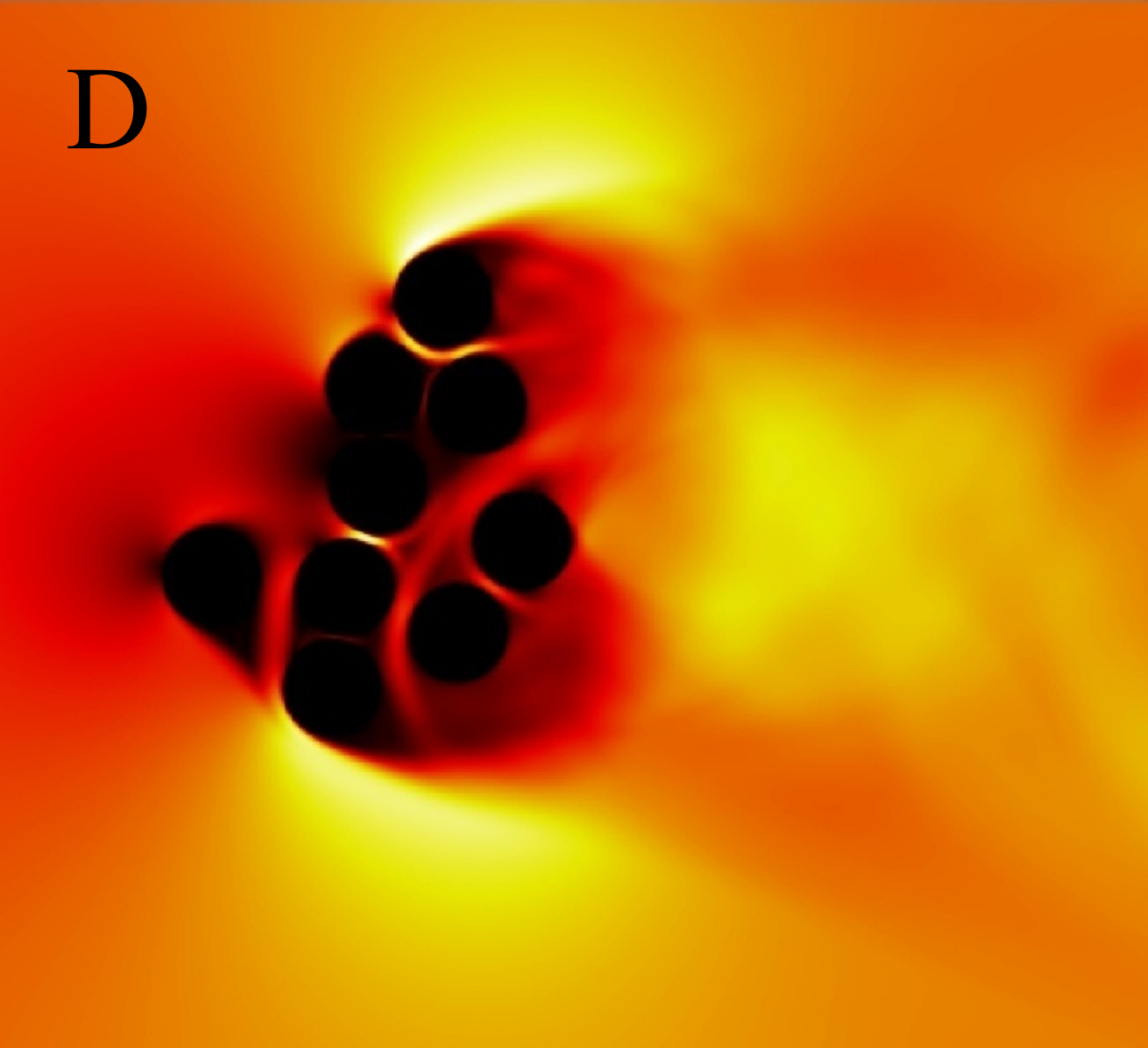}
  \subcaption{}
  \label{fig:avg4}
\end{subfigure}
\begin{subfigure}{\textwidth}
\centering
\includegraphics[width=0.75\textwidth]{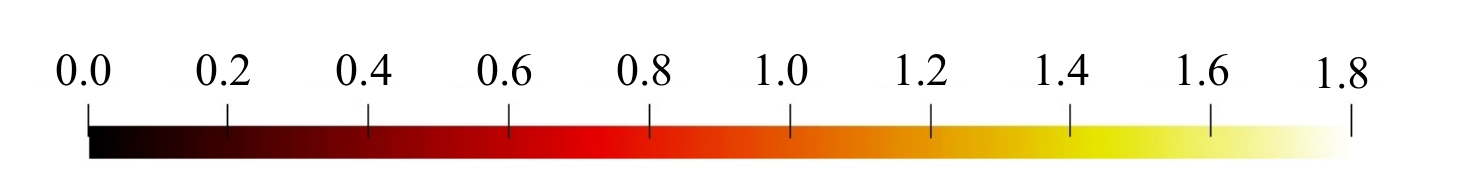}
\end{subfigure}
	\caption{Flow speed normalized with respect to the freestream velocity ($\lVert \bm{u}\rVert/U_\infty$). The data was obtained using DNS and time-averaged over one shedding time period for each of the four selected individuals A to D (\subref{fig:avg1}-\subref{fig:avg4}). Localized high-speed regions are visible as bright spots in the arrays' interior.}
\label{fig:avg_vel}
\end{figure}

Considering the internal porous structure of these arrays, we note that the low-enstrophy low-drag case (individual A) has noticeable gaps within the array that resemble channels, which provide a path for the flow to pass through the interior of the array, as observed in the visualization of speed shown in Figure \ref{fig:avg1}. This has the dual impact of reducing drag, as well as disrupting the vortices shed in the wake, which in turn reduces the average wake enstrophy. Individual B in Figure \ref{fig:avg2} also has small gaps which allow flow to pass through the porous structure, but to a lesser extent. Additionally, we observe a low-speed region in front of the array, which indicates that a majority of the freestream flow gets redirected around the array instead of passing through it. We observe that the high-drag high-enstrophy cases (Figure \ref{fig:avg3} and \ref{fig:avg4}) have cylinders that display comparatively tighter clustering, with minimal streamwise flow allowed through the arrays. Quantitative details regarding the impact of the arrays' internal porous structure are discussed at a later point.

\begin{figure}
\centering
\includegraphics{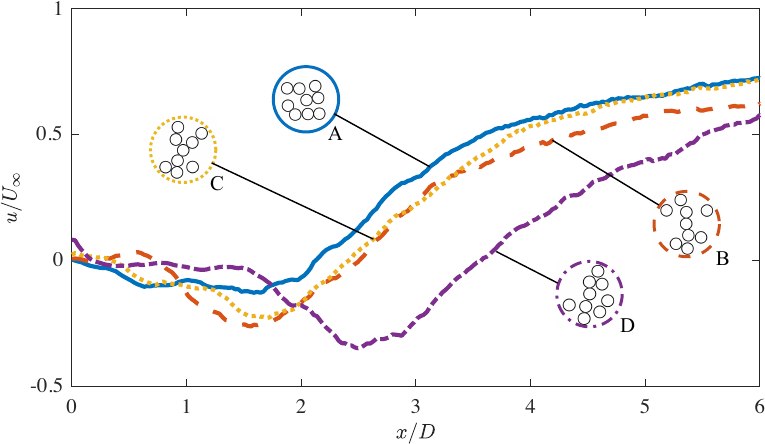}
        \caption{Normalized streamwise velocity along the midline for the four selected Pareto-optimal arrangements. The velocity measurements shown here were obtained experimentally using Particle Image Velocimetry (PIV), and were time-averaged over 8 vortex-shedding cycles.}
\label{fig:midline}
\end{figure}

Figure \ref{fig:midline} shows velocity data obtained using PIV experiments in the wakes of the four Pareto-optimal arrangements. The variation of the normalized streamwise velocity ($u/U_{\infty}$) along a line cut is shown in the figure, plotted against the normalized downstream distance from the rear edge of the clusters (i.e., $x/D$). This line cut (also referred to as the midline here) was selected to be the straight line in the streamwise direction that passes through the array's center. The general trend in Figure~\ref{fig:midline} indicates a reduction in the streamwise velocity a short distance behind the cluster, after which there is a gradual recovery up to a maximum value which is lower than the freesetream value. Furthermore, we observe that the most noticeable drop in $u/U_{\infty}$ occurs for individual D, which experiences the highest flow-induced drag as well as the highest wake enstrophy. At the same time, individual A with the minimum drag and lowest wake enstrophy displays the least severe drop. Individuals B and C are also shown for comparison, but neither are extreme examples with regard to drag or enstrophy. We observe that the recovered streamline velocity is highest for individuals A and C at 0.72$U_{\infty}$, and lower for B at $0.62U_\infty$ and D at $0.58U_\infty$. We also observe that the minimum for individual D occurs farther downstream compared to the other arrays at $x=2.5D$, with a velocity drop of $-0.35U_\infty$. The minima for B and C occur at approximately $x=1.6D$ and display similar velocity drops of $-0.25U_\infty$. These negative values are indicative of the formation of strong recirculation regions in the wakes of these arrays. The velocity profile for individual A does not display a distinctive minimum, but instead we observe a comparatively long region with a moderate velocity drop of $-0.12U_\infty$. Overall, the most prominent difference among the velocity profiles is between that of individuals A and D, with the profiles for individuals B and C not differing significantly from that of A. These observations indicate the need for utilizing other aspects besides a 1-dimensional line cut in order to fully characterize the arrays' performance characteristics.

\begin{figure}
\centering
\includegraphics{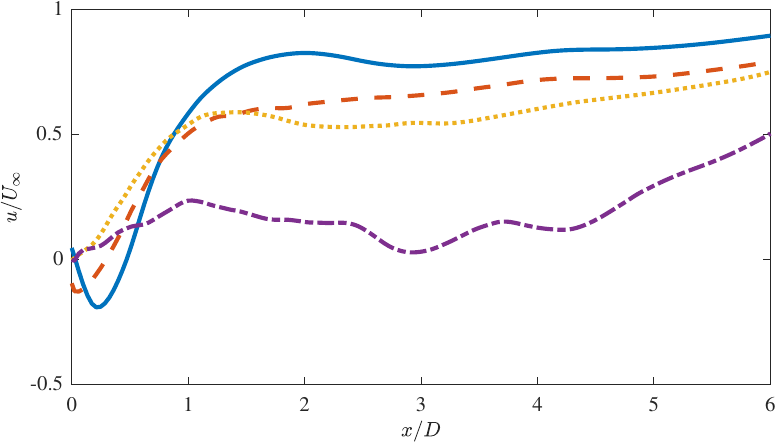}
	\caption{Normalized streamwise velocity along the midline from DNS data, time-averaged over 8 vortex-shedding cycles. The line-types correspond to those shown in Figure~\ref{fig:midline}.}
\label{fig:midline_sim}
\end{figure}

Figure~\ref{fig:midline_sim} shows the time-averaged streamwise velocity along the midline from DNS data for the four individuals. We observe a decreasing trend in far-wake velocity going from individual A to D, with A being the highest at 0.9$U_{\infty}$, followed by individual B at 0.8$U_{\infty}$, C at 0.75$U_{\infty}$, and finally D at 0.5$U_{\infty}$. Individual A has the largest drop in velocity in the region behind the cluster ($-0.2U_{\infty}$), however, velocity recovery occurs rapidly. The recovery in velocity slows down noticeably as we go from configuration A to D, i.e., with increasing drag and enstrophy. Comparing Figures~\ref{fig:midline} and~\ref{fig:midline_sim}, we observe that the largest drop in velocity occurs closer to the arrays in the simulations compared to the experiments, in addition to faster recovery observed for the simulations. Several potential reasons may explain the differences observed between velocity profiles obtained from experiments and simulations, and these are discussed at a later point. However, the near-wake velocity field from the DNS was confirmed to match 2D simulation results from a separate open-source solver (OpenFOAM), which uses finite volume methods and body-fitted meshes instead of the vortex methods and Brinkmann penalization approach adopted in the present work.

It is expected that lower flow speeds will favour the sedimentation of suspended particles by increasing the time available for gravitational settling. However, examining only the streamwise component along a 1-dimensional line cut may not provide a complete picture of the sedimentation process. For instance, in experiments conducted by \cite{Chen2012}, porous cylinder arrangements with higher overall midline streamwise velocity were found to be more conducive to soil deposition, which seems contrary to the expected behaviour.  As discussed earlier, increased drag leads to higher velocity deficit and energy dissipation in the wake \citep{Maza2017,Gijon2021}, which can promote gravitational settling. However, this process is also accompanied by increased turbulent kinetic energy production \citep{Chen2012,Tinoco2018}, which in turn may hinder sediment deposition and promote resuspension. Thus, it is the combined influence of drag and wake enstrophy over the entire wake volume that determines particle sedimentation levels. It is likely that for the cases explored here with higher drag, high wake enstrophy might make the configurations unfavourable for sediment deposition (\cite{Chen2012}, \cite{Norris2019}) even though the streamwise wake velocity is notably lower than $U_\infty$. On the other hand, for the cases with low enstrophy, the overall speed in the wake may be higher due to low drag, which is also unfavourable for gravitational settling (\cite{Yamasaki2019}). This implies that a balance must be struck between drag and wake enstrophy to promote sediment deposition. 

\begin{figure}
\centering
\begin{subfigure}{0.495\textwidth}
  \centering
  \includegraphics{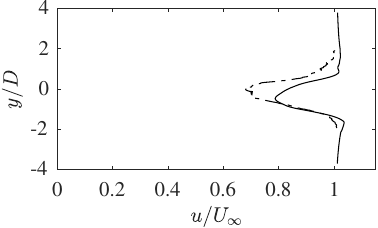}
  \subcaption{}
  \label{fig:lcut1}
\end{subfigure}
\begin{subfigure}{0.495\textwidth}
  \centering
  \includegraphics{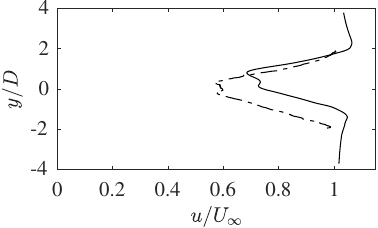}
  \subcaption{}
  \label{fig:lcut2}
\end{subfigure}
\begin{subfigure}{0.495\textwidth}
  \centering
  \includegraphics{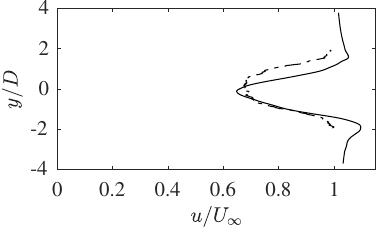}
  \subcaption{}
  \label{fig:lcut3}
\end{subfigure}
\begin{subfigure}{0.495\textwidth}
  \centering
  \includegraphics{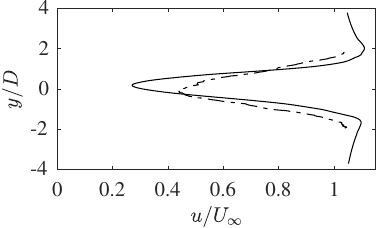}
  \subcaption{}
  \label{fig:lcut4}
\end{subfigure}
	\caption{Cross-stream velocity profiles for the selected Pareto-optimal individuals A through D (\subref{fig:lcut1}-\subref{fig:lcut4}). Profiles obtained using DNS are shown as the solid lines, whereas those obtained using PIV are shown as the dash-dotted lines. The profiles were computed using time-averaged streamwise velocity over at least 8 shedding cycles, at a distance $5D$ downstream from the arrays' rear edges.}
\label{fig:linecut}
\end{figure}

\begin{figure}
\centering
\includegraphics{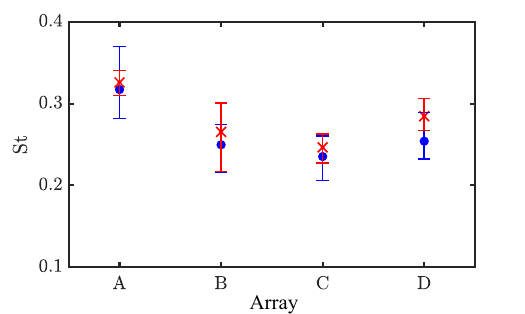}
	\caption{Comparison of the Strouhal number ($St = fD/U_\infty$, where $f$ is the frequency of vortex shedding, and $U_\infty$ is the freestream velocity) between the PIV experiments ($\times$) and DNS ($\bullet$). The symbols represent the mean over 8 shedding cycles and the error bars represent the standard deviation.}
\label{fig:stpivsim}
\end{figure}

In addition to the streamwise line cuts shown in Figures~\ref{fig:midline} and~\ref{fig:midline_sim}, cross-stream line cuts for the time-averaged streamwise velocity $u$ were examined using both PIV and DNS data. The resulting comparison is shown in Figure~\ref{fig:linecut}. The plots show reasonably good agreement between data from the experiments and 2D simulations, and the differences observed may be due to several potential reasons such as sensitivity to array orientation and relative positioning of the cylinders in the experiments, flow blockage within the experimental channel, and three-dimensional effects that are absent from the 2D simulations, to name a few. 
The time-averaged cross-stream velocity profiles shown in Figure~\ref{fig:linecut} indicate an increase in velocity deficit going from A to D, and comparable deficits for individuals B and C. We observe that individual D displays the largest reduction in velocity, although this deficit is confined to a relatively narrow cross-stream region. In comparison, individuals B and C experience a less severe reduction in streamwise velocity, albeit a larger cross-stream area is affected as indicated by the broader profiles. In addition to the velocity profiles, the Strouhal number ($St=fD/U_\infty$) for all four arrays was also computed using both PIV and DNS data, and a comparison is shown in Figure~\ref{fig:stpivsim}. We observe good agreement between the values of $St$ obtained from the experiments and simulations. We note that it is difficult to compute enstrophy from PIV data since velocity gradients cannot be computed with high accuracy due to spatial averaging involved in the PIV analysis. 

\begin{figure}
\centering
\begin{subfigure}{\textwidth}
    \centering
	\includegraphics[scale=0.9]{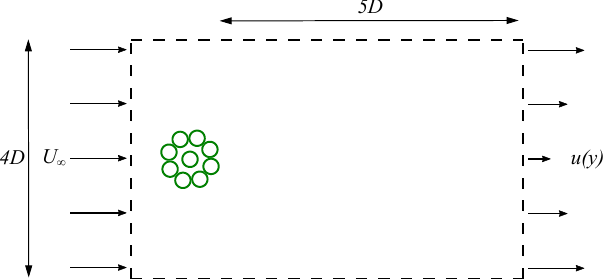}
    \caption{}
    \label{fig:controlVolume}
\end{subfigure}
\begin{subfigure}{\textwidth}
    \centering
    \includegraphics{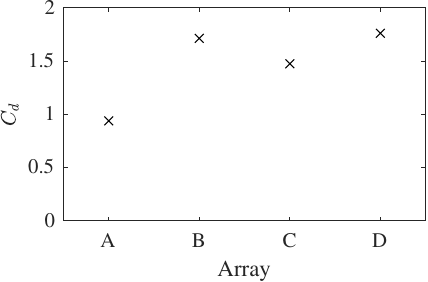}
    \caption{}
    \label{fig:dragPIV}
\end{subfigure}
        \caption{\label{fig:PIVdrag} \subref{fig:controlVolume} Schematic of the control-volume used for estimating the drag force on the porous arrays using PIV data. The outflow velocity profile $u(y)$ was measured at a distance $5D$ downstream of the array edges, and the corresponding profiles are shown in Figure~\ref{fig:linecut}. \subref{fig:dragPIV} Drag coefficient computed using control-volume analysis, with the drag force determined using equation~\ref{eq:controlVolumeDrag}.}
\end{figure}

To estimate the drag acting on the arrays using the PIV data, a control-volume analysis was performed using the velocity profiles shown in Figure~\ref{fig:linecut}. A simple schematic of the control-volume setup is shown in Figure~\ref{fig:controlVolume}. The average drag force on the arrays was determined as follows (assuming unit span):
\begin{subequations}
\begin{align}
	F_{Drag} &= \left(-\varoiint (\rho \bm{u}) \bm{u} \cdot \hat{\bm{n}} dA \right) \cdot \hat{\bm{i}} \\
	 &= 4D\rho U_\infty^2 - \int\limits_{-2D}^{2D}\rho u^2(y) dy
\end{align}
\label{eq:controlVolumeDrag}
\end{subequations}
The corresponding drag coefficient for each 9-cylinder array was then calculated as $C_d = F_{Drag}/(0.5\rho U_\infty^2 9d)$, and the resulting values are shown in Figure~\ref{fig:dragPIV}.
We observe an increasing trend in $C_d$ going from configuration A to B to D, however there is a slight decrease for configuration C. The value of $C_d$ for configuration D is considerably higher than that for A, and slightly higher than (but close to) that for B. We note that the trend for $C_d$ computed from DNS using equation~\ref{eq:controlVolumeDrag} displayed the same trend as that observed in Figure~\ref{fig:dragPIV}, but the magnitude tends to differ between the 2D DNS and PIV. This can be explained by the differences observed between the cross-stream velocity profiles in Figure~\ref{fig:linecut}, since these profiles form the basis of the control-volume based $C_d$ calculations. We remark that the lower value of $C_d$ for array C compared to array B in Figure~\ref{fig:dragPIV} is due to the control-volume analysis being done in the steady shedding state, which is necessary due to the use of time-averaged velocity profiles. This is also the reason why the far-wake velocity for configuration C recovers to a higher value than for B in Figure~\ref{fig:midline}. The $C_d$ computed during optimization (Figure~\ref{fig:gen33results}) included the unsteady startup phase (i.e., the initial transient), and the subsequent time-averaging resulted in lower drag for array B than for C (Appendix Figure~\ref{fig:timeVaryingCd}). During optimization it is not feasible to predict beforehand when a specific array configuration might reach a steady shedding state, or what the resulting shedding frequency might be, making it necessary to use time-averaging over a long duration. We note that the very early stages of the startup can be dependent on numerical ramp-up characteristics. To prevent any non-physical effects that might arise from this ramp-up, the initial non-dimensionalized time from 0 to $4t^*$ was excluded from all DNS runs for calculating the time-averaged $C_d$.

\subsection{Flow Within the Porous Arrays}
\label{subsec:flowwithin}

\begin{figure}
\begin{subfigure}{\textwidth}
\fbox{
\begin{minipage}{0.96\textwidth}
  \centering
  \includegraphics[width=0.24\textwidth]{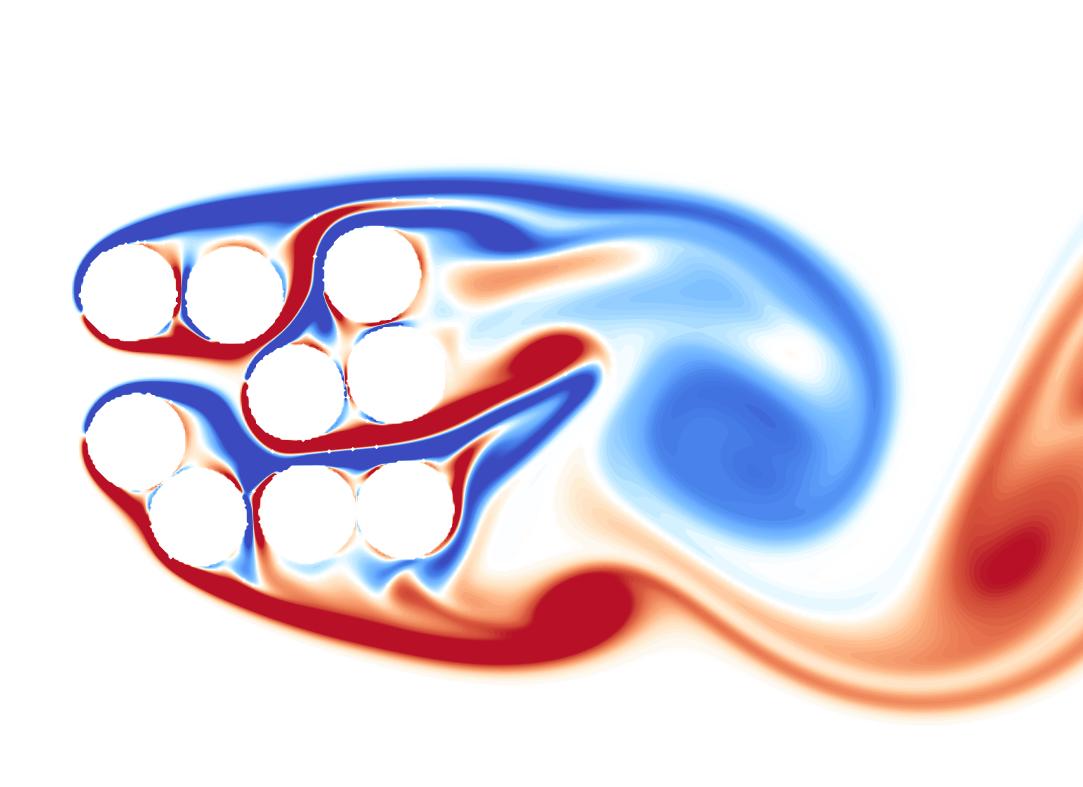}
  \hfill
  \includegraphics[width=0.24\textwidth]{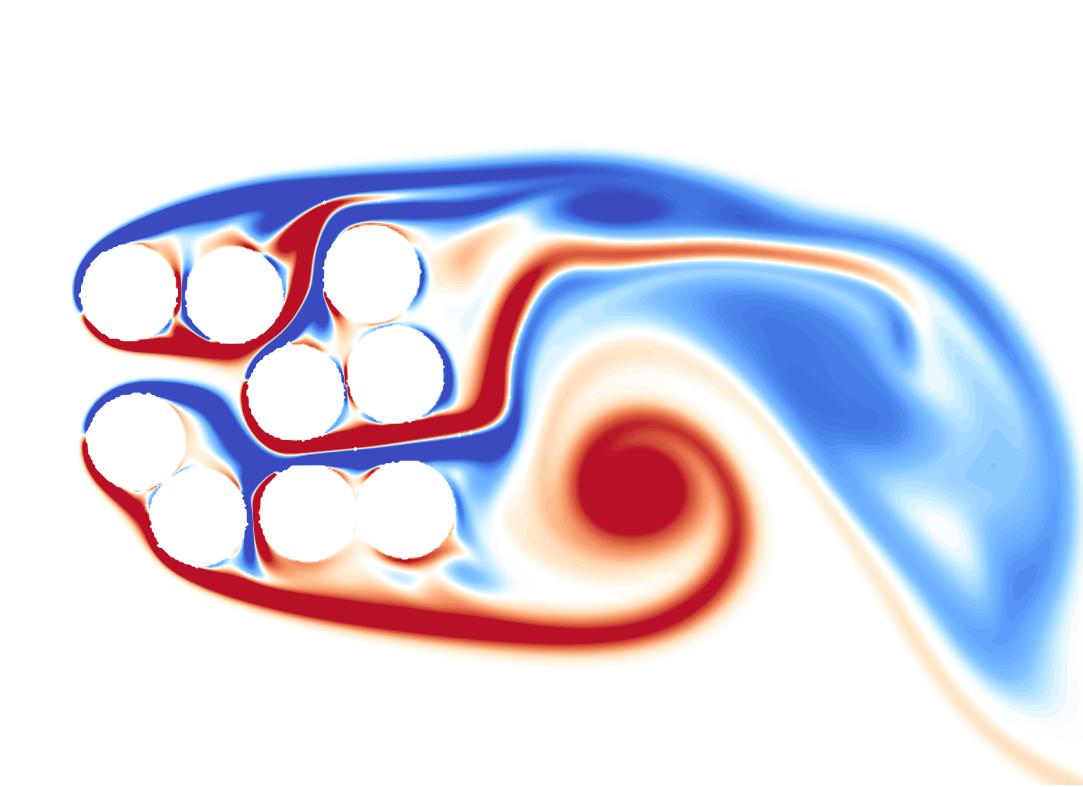}
  \hfill
  \includegraphics[width=0.24\textwidth]{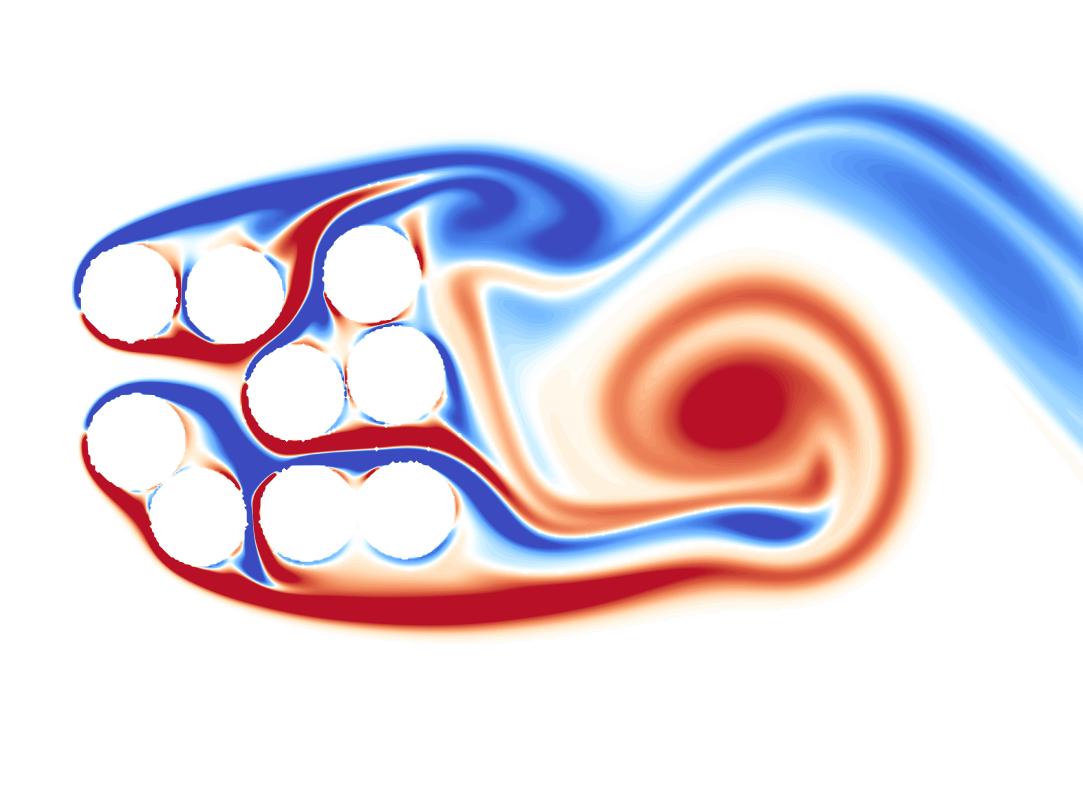}
  \hfill
  \includegraphics[width=0.24\textwidth]{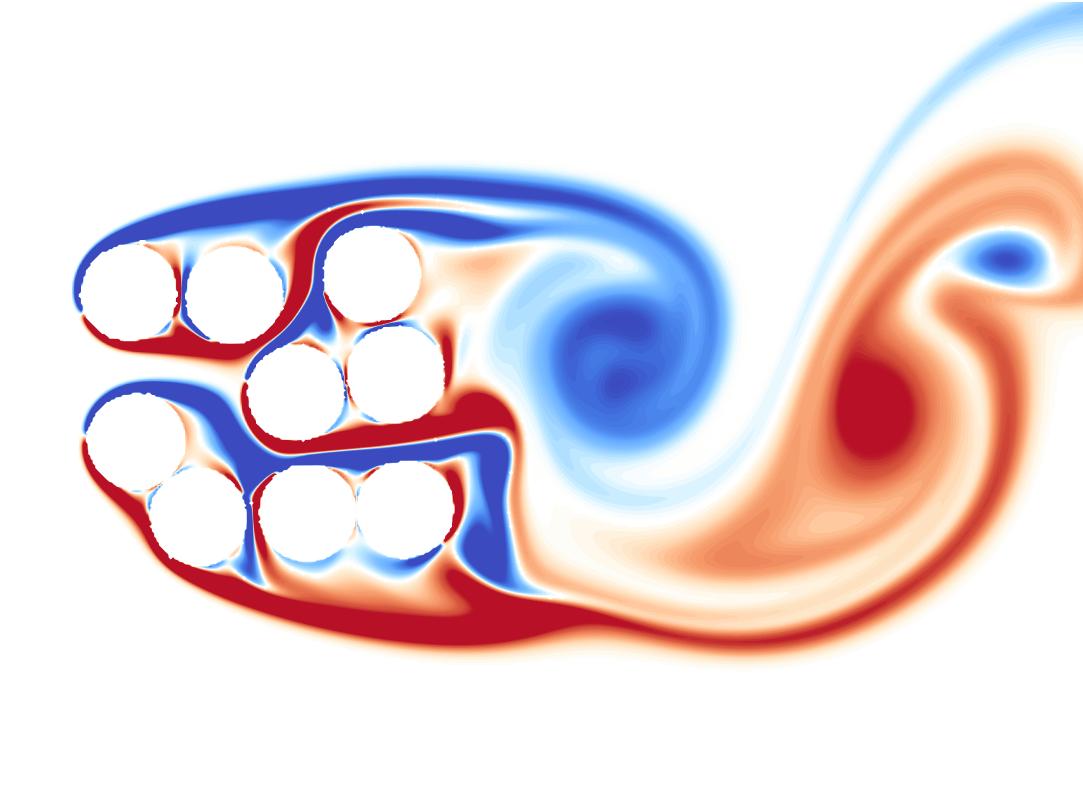}
  \end{minipage}
}
  \subcaption{}
  \label{proj1}
\end{subfigure}
\begin{subfigure}{\textwidth}
\fbox{
  \centering
\begin{minipage}{0.96\textwidth}
  \centering
  \includegraphics[width=0.24\textwidth]{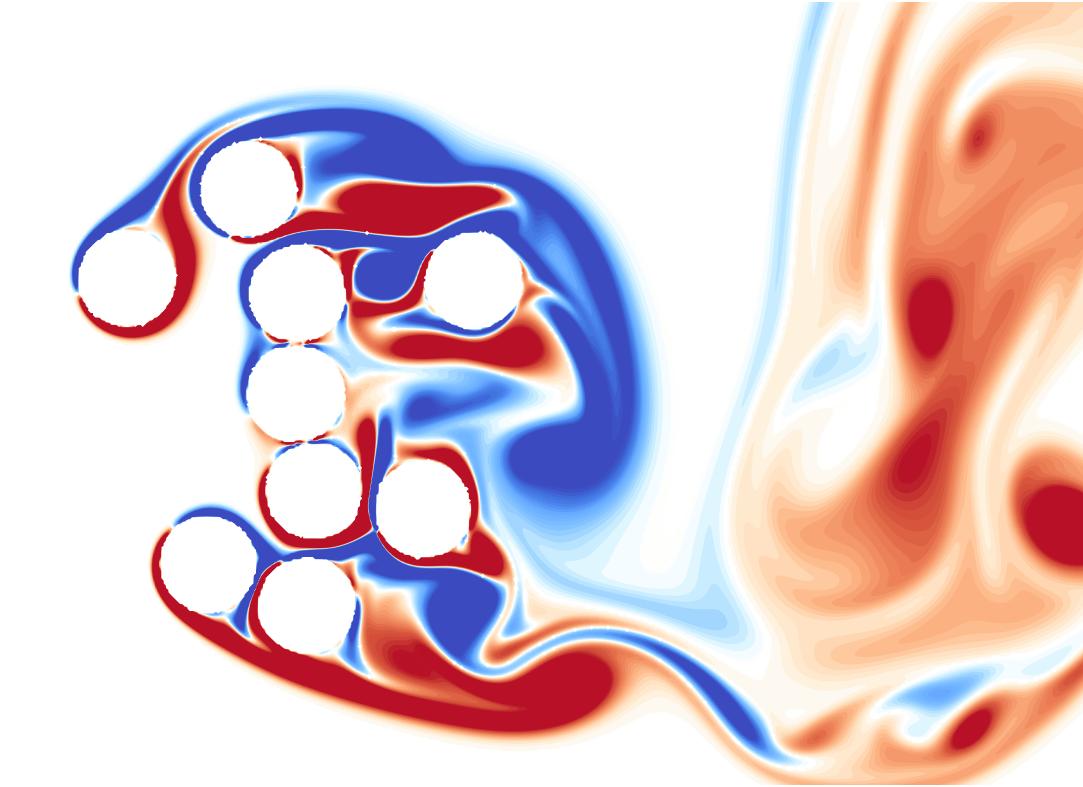}
  \hfill
  \includegraphics[width=0.24\textwidth]{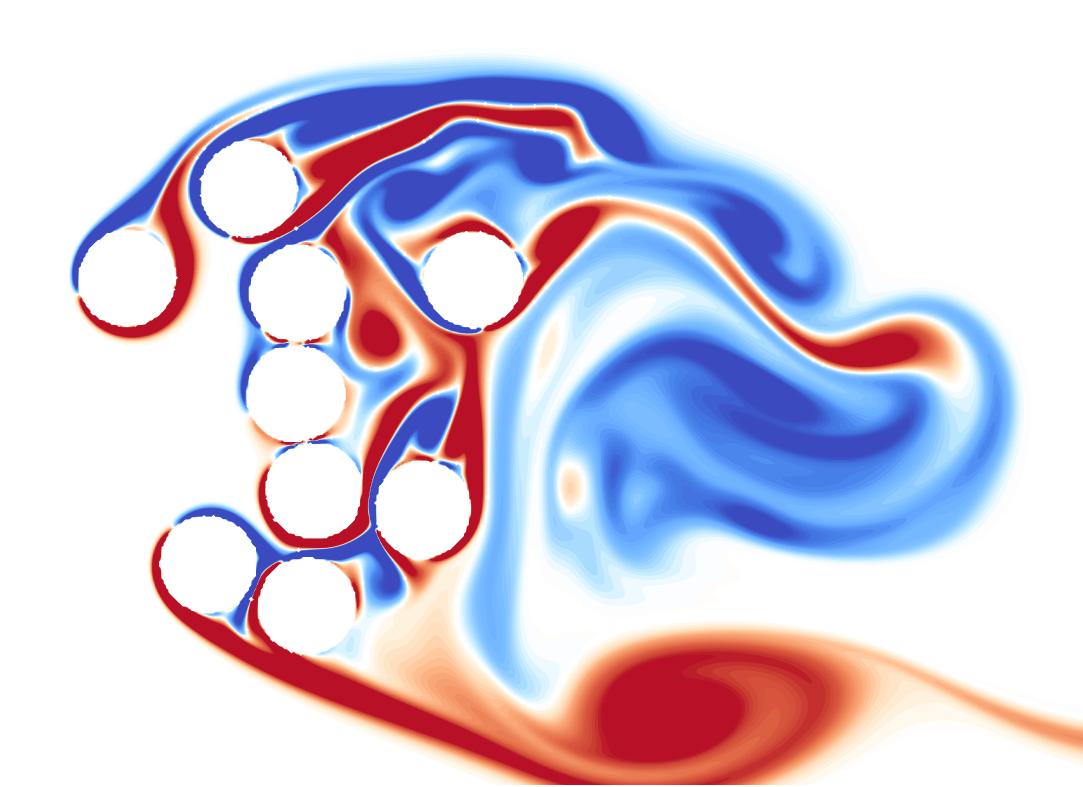}
  \hfill
  \includegraphics[width=0.24\textwidth]{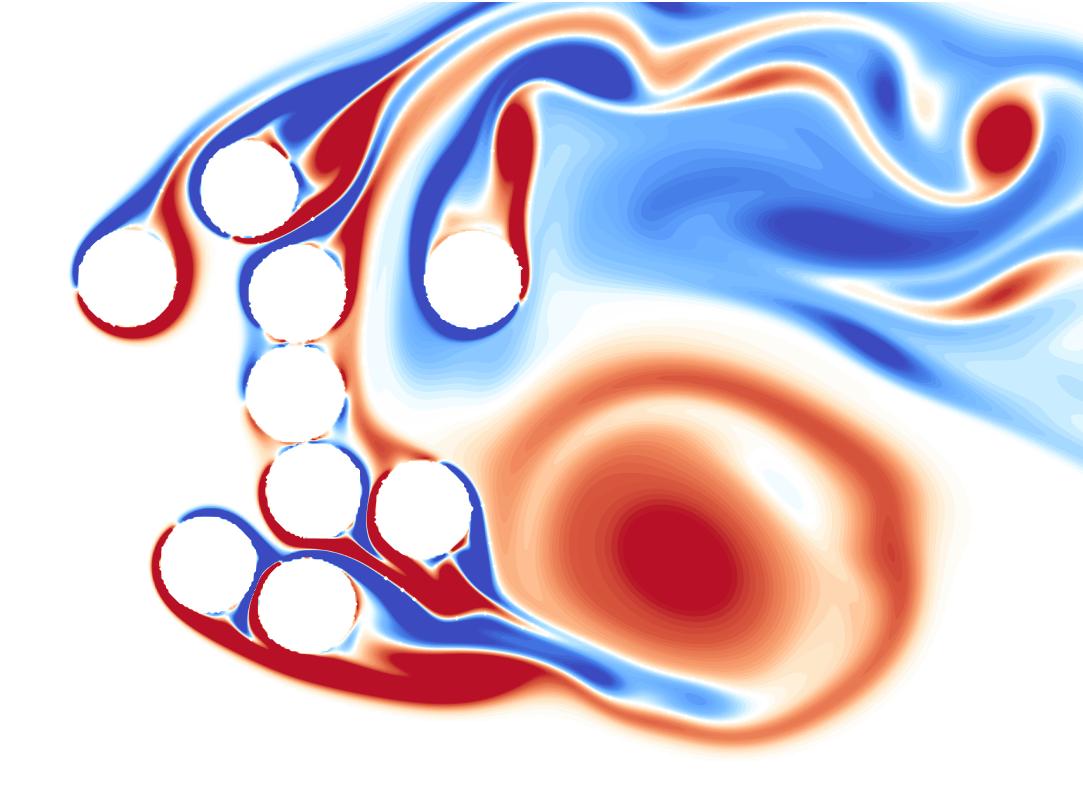}
  \hfill
  \includegraphics[width=0.24\textwidth]{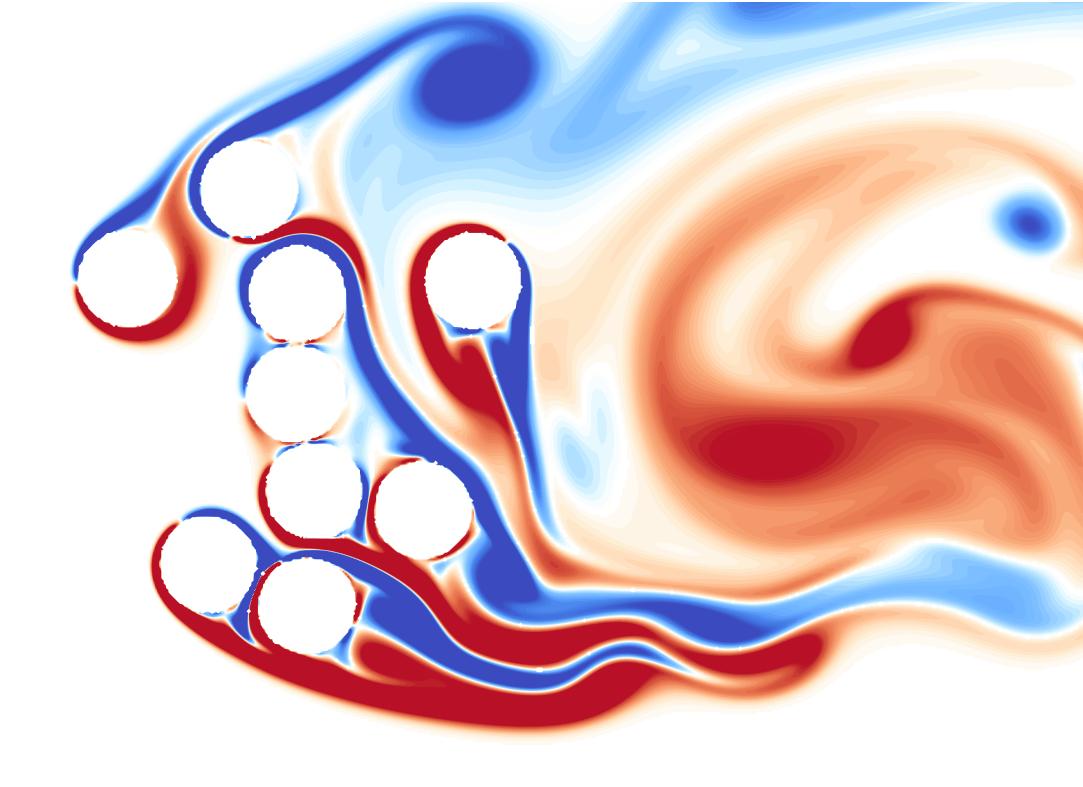}
  \end{minipage}
}
  \caption{}
  \label{proj2}
\end{subfigure}
\begin{subfigure}{\textwidth}
\fbox{
\begin{minipage}{0.96\textwidth}
  \centering
  \includegraphics[width=0.24\textwidth]{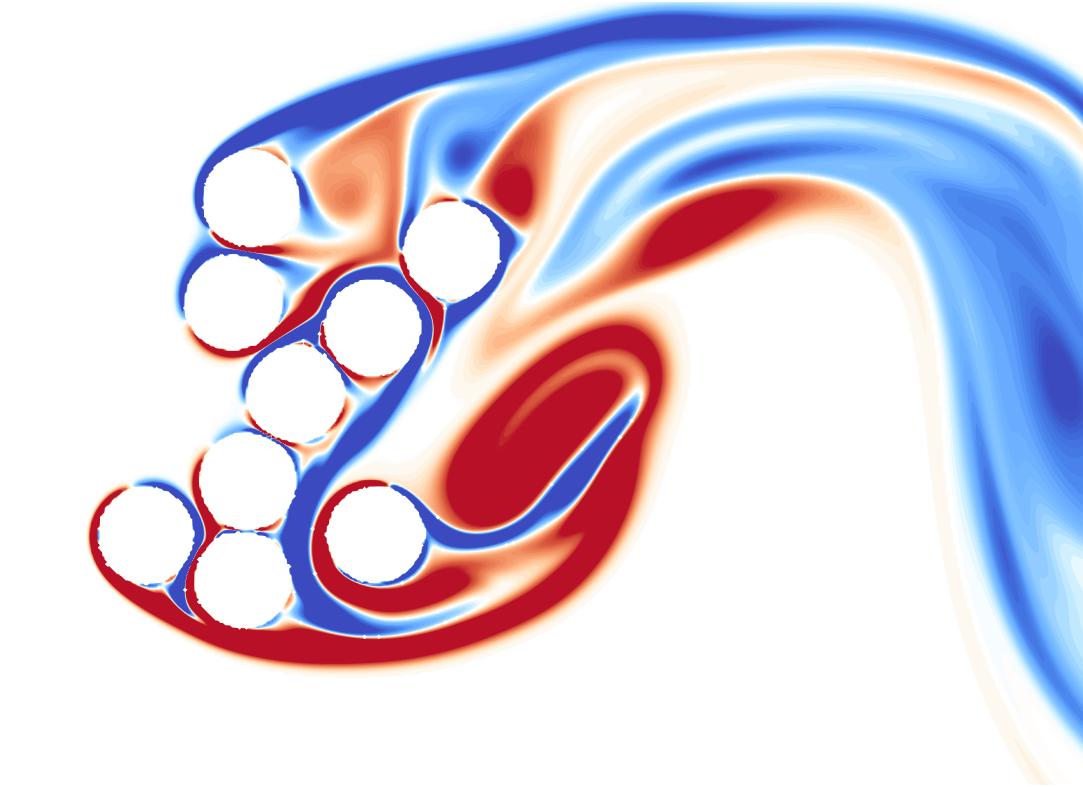}
  \hfill
  \includegraphics[width=0.24\textwidth]{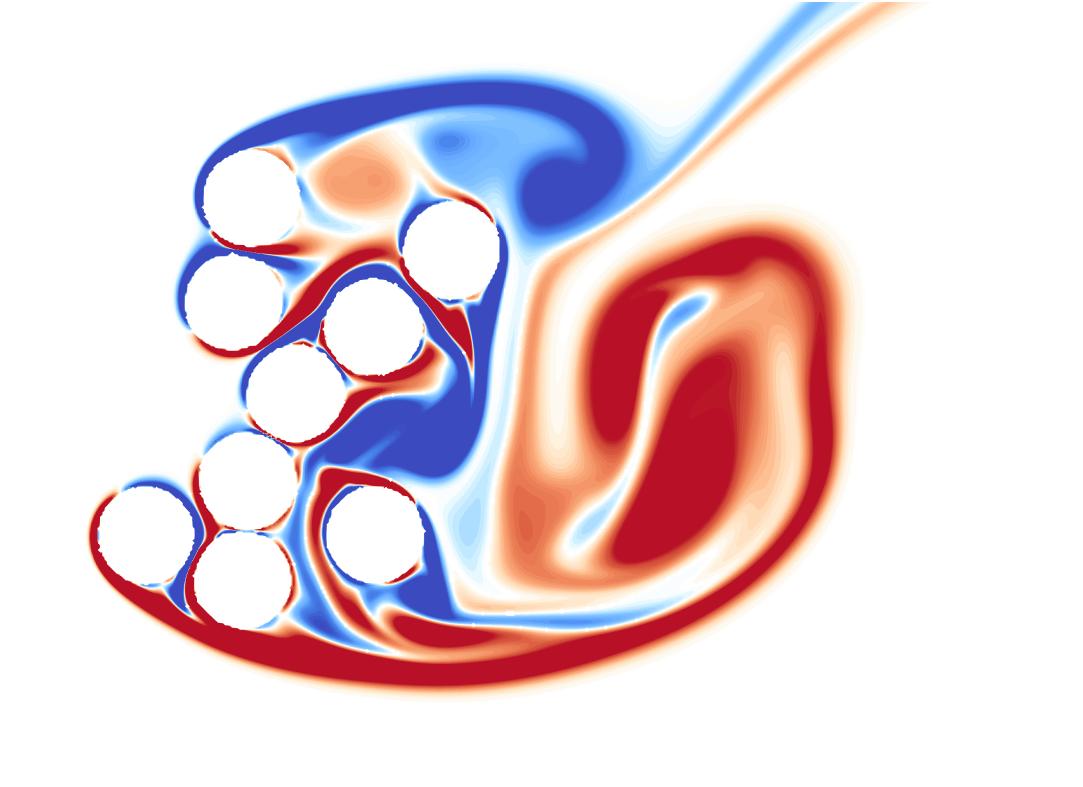}
  \hfill
  \includegraphics[width=0.24\textwidth]{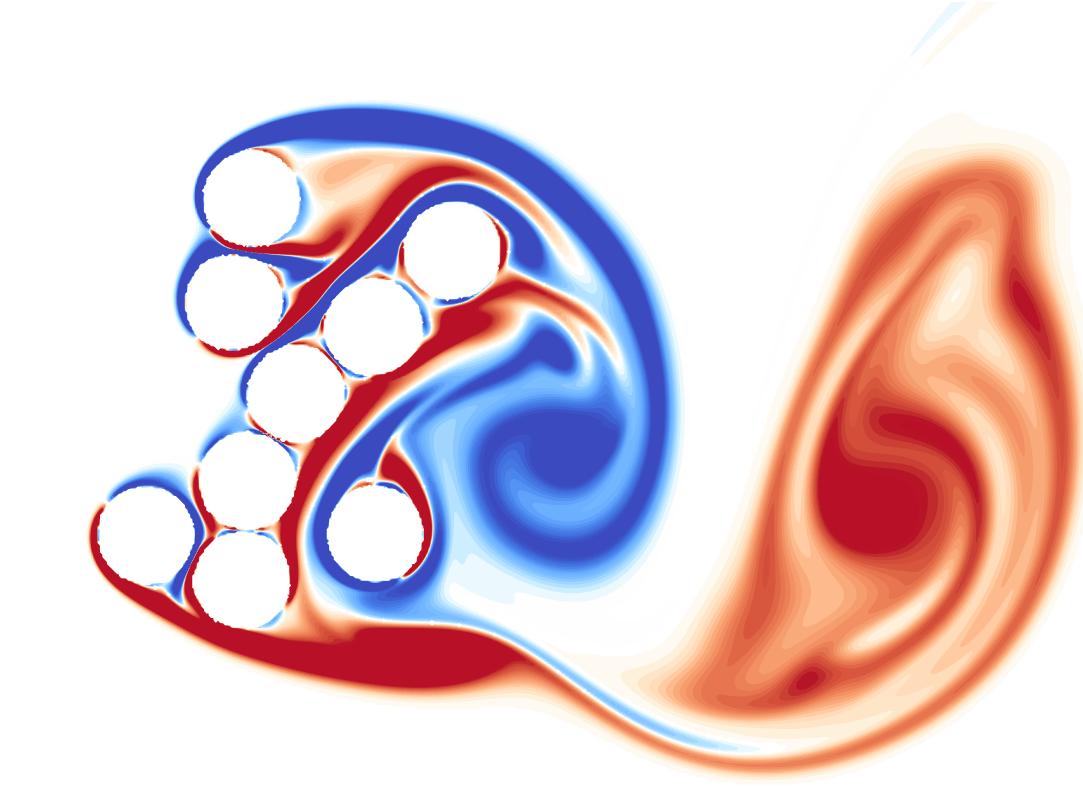}
  \hfill
  \includegraphics[width=0.24\textwidth]{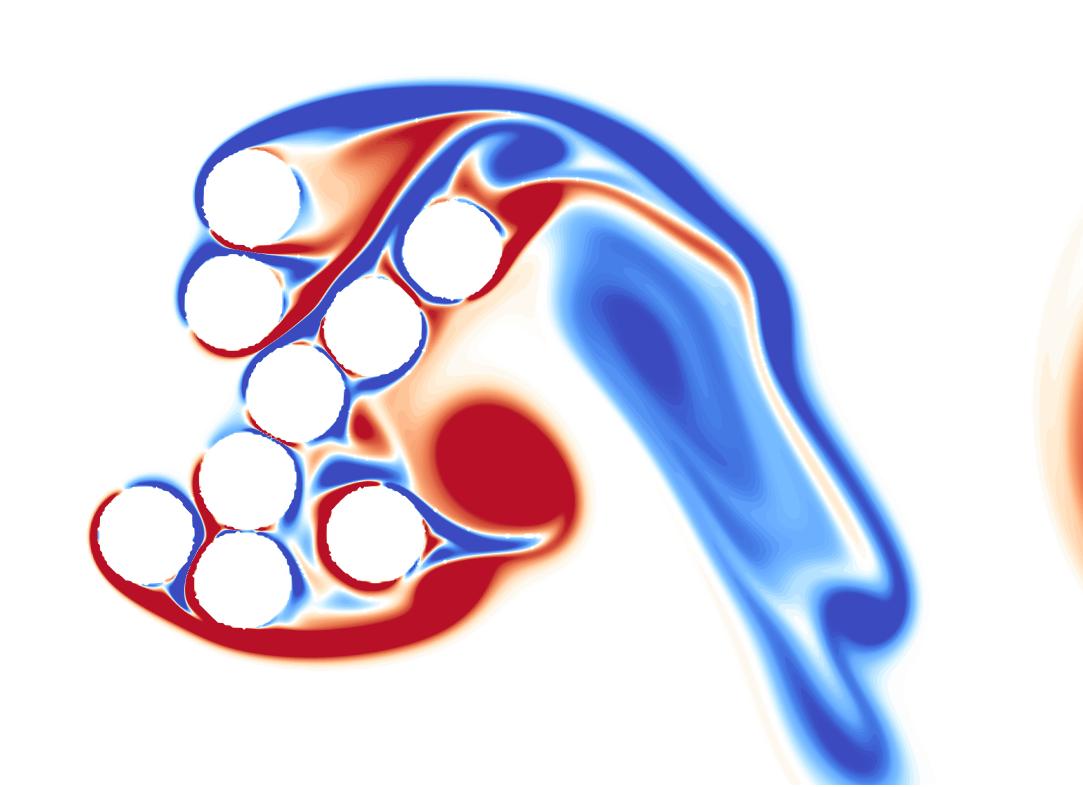}
  \end{minipage}
}
  \caption{}
  \label{proj3}
\end{subfigure}
\begin{subfigure}{\textwidth}
\fbox{
\centering
\begin{minipage}{0.96\textwidth}
\centering
  \includegraphics[width=0.24\textwidth]{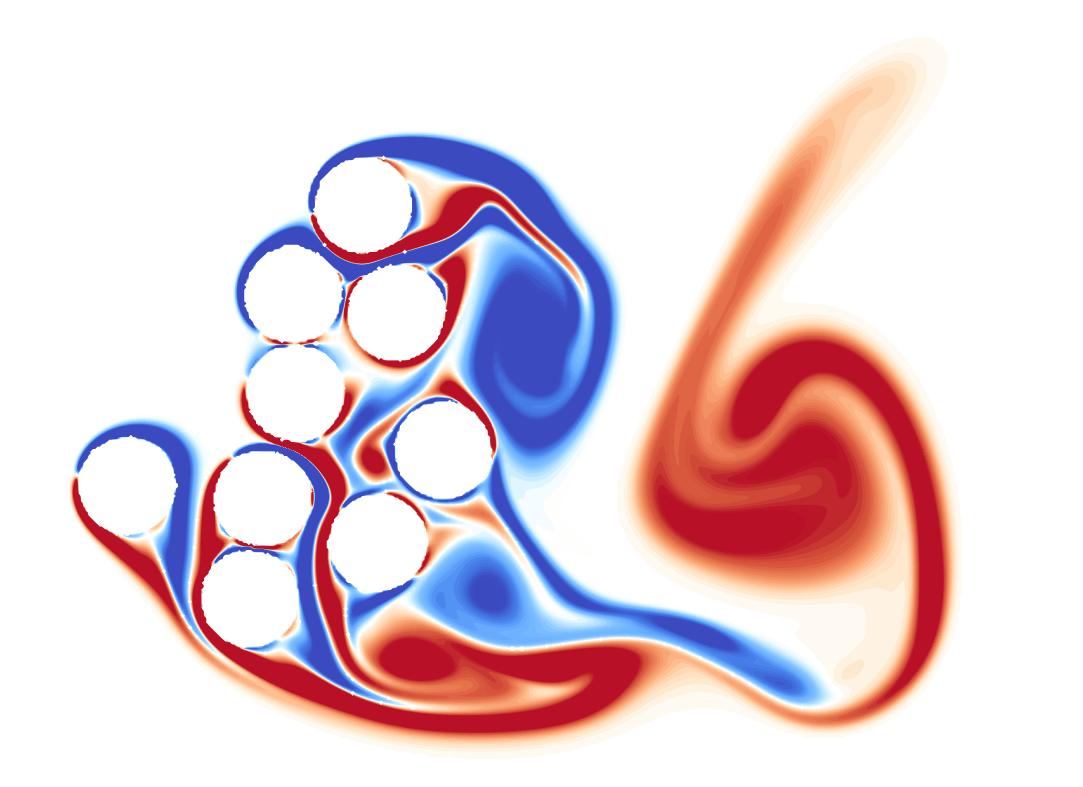}
  \hfill
  \includegraphics[width=0.24\textwidth]{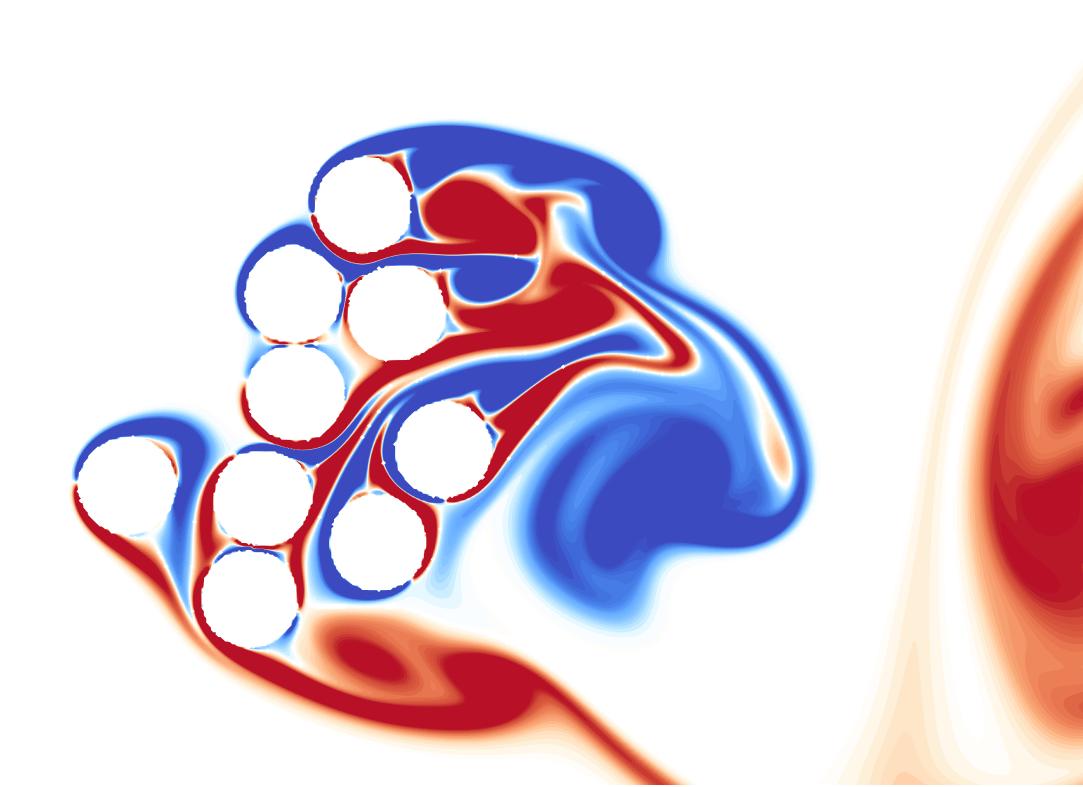}
  \hfill
  \includegraphics[width=0.24\textwidth]{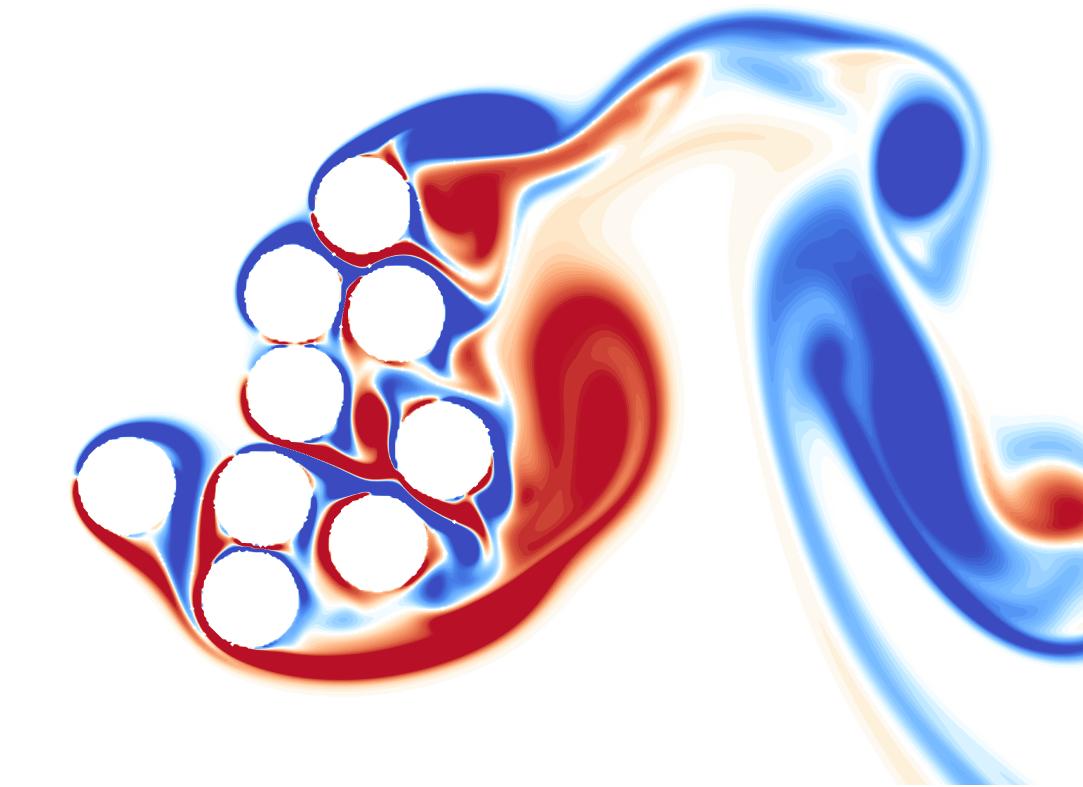}
  \hfill
  \includegraphics[width=0.24\textwidth]{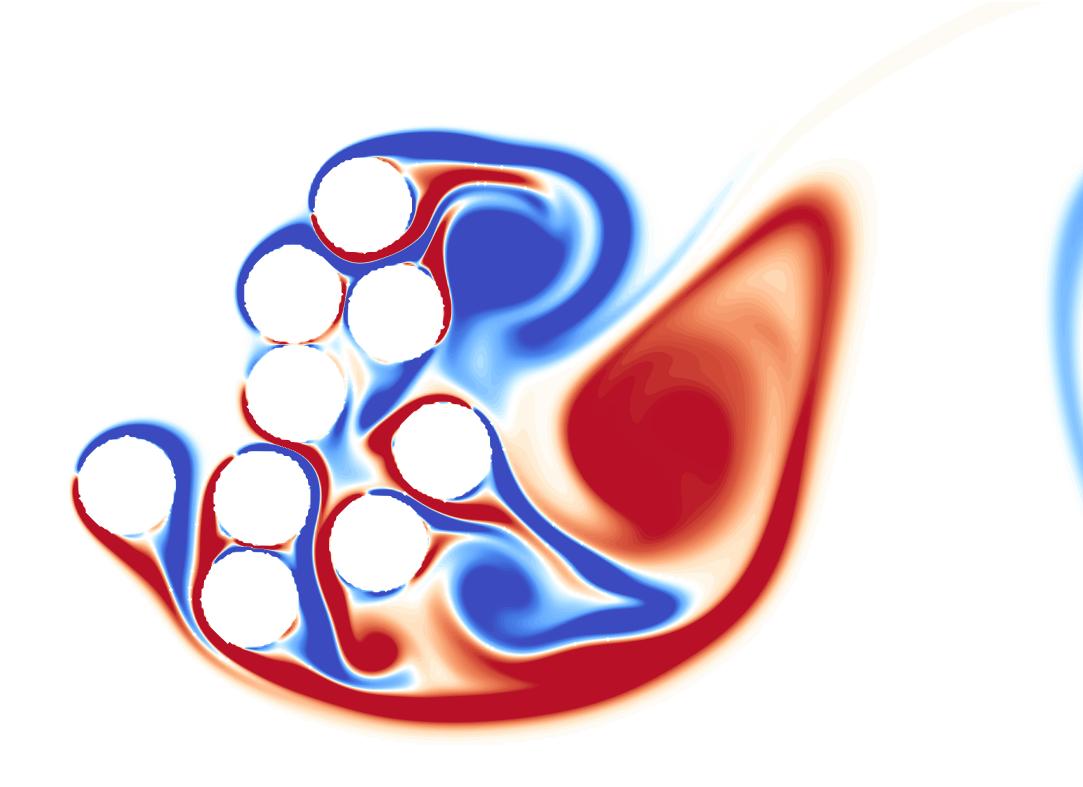}
  \end{minipage}
}
  \caption{}
  \label{proj4}
\end{subfigure}
\caption{\small{Time evolution of vorticity over one shedding time period from DNS data, with the corresponding colourbar provided in Figure \ref{fig:4wakes}. Panels \subref{proj1} through \subref{proj4} represent the Pareto-optimal individuals A through D, respectively, in order of increasing drag and enstrophy.}}
\label{fig:vorticitySnapshots}
\end{figure}

The differences in drag and enstrophy values associated with each of the four selected individuals ultimately arise from differences in how the flow behaves around and within the clusters. As mentioned in some of the studies discussed earlier \citep{Ricardo2016}, vorticity cancellation due to the interference of neighbouring cylinders plays a prominent role in determining wake characteristics. Thus, we take a closer look at flow patterns that develop within the porous structures for each of the four selected Pareto-optimal individuals, and examine the resulting impact on enstrophy and drag. The evolution of the vorticity field over one shedding time period is shown in Figure~\ref{fig:vorticitySnapshots} for the four selected arrays.
One notable observation for individual A is the presence of cohesive structures (i.e., internal boundary layers) that occupy the array's interior, and remain persistent throughout the shedding period without undergoing notable changes in form and strength. These structures exhibit a tendency to align in the streamwise direction and appear to `attach' to neighbouring cylinders. This is likely related to the relative closeness of adjacent cylinders and their specific arrangement within the array. Such internal cohesive structures are found less frequently as we move from individual A to D, where we instead observe structures that are less persistent and undergo rapid changes in shape, direction, and intensity over time. The changes are most prominently observed for individual D, i.e., the high-drag high-enstrophy configuration.

\begin{figure}
\begin{subfigure}{0.245\textwidth}
  \centering
	\includegraphics[width=\textwidth]{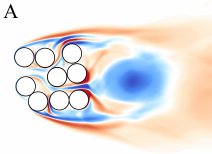}
   \subcaption{}
   \label{fig:1-modes}
  \end{subfigure}
\begin{subfigure}{0.245\textwidth}
  \includegraphics[width=\textwidth]{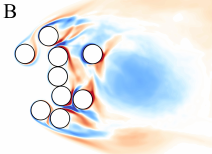}
  \subcaption{}
  \label{fig:2-modes}
\end{subfigure}
\begin{subfigure}{0.245\textwidth}
  \includegraphics[width=\textwidth]{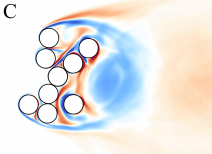}
  \subcaption{}
    \label{fig:3-modes}
\end{subfigure}
\begin{subfigure}{0.245\textwidth}
  \includegraphics[width=\textwidth]{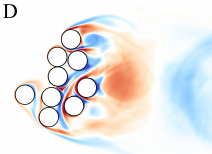}
  \subcaption{}
    \label{fig:4-modes}
\end{subfigure}
\begin{subfigure}{\textwidth}
	\centering
	\includegraphics[width=0.75\textwidth]{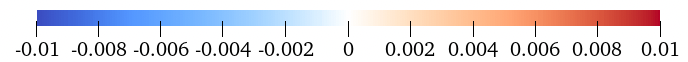}
\end{subfigure}
	\caption{The first POD mode of vorticity for the four selected arrangements obtained from DNS data, using columns of $\mathbf{V}$ from equation~\ref{eq:svdForPOD}. The first modes from all four cases were observed to account for at least 33\% of variation in the data. Please note that the colours do not necessarily correspond to positive or negative values of vorticity since they represent the eigenvector of the covariance matrix, with Euclidean norm  equal to 1.}
\label{fig:modes}
\end{figure}

For further analysis, vorticity data collected from the numerical simulations was decomposed into its constituent modes using Proper Orthogonal Decomposition (POD), also referred to as Principal Component Analysis. This allows us to isolate the most energetic flow structures (i.e., POD modes) in the interior and the wake of the cylinder arrays. We note that a POD mode's `energy' is not related to the concept of potential or kinetic energy, but it rather indicates how much variation in the data can be accounted for using a particular mode. Figure \ref{fig:modes} shows the most energetic modes which contain the highest variation in vorticity over time. Mathematically, these modes can be described as the first eigenvector (i.e., that associated with the largest eigenvalue) of the covariance matrix constructed using the flow snapshots. 

A total of 100 vorticity snapshots represented by $\omega(x,y,t)$, spanning one shedding time period in the steady state, were used for each of the individuals shown in Figure \ref{fig:modes}. Each such 2D snapshot was rearranged into a 1D row vector of length $M$ (the number of gridpoints) containing the vorticity values, and the 100 frames were arranged sequentially in rows. The time-averaged vorticity $\bar{\omega}(x,y)$ was subtracted from each frame in order to form the sample matrix $\mathbf{X}$:
\begin{equation}
\mathbf{X} = \begin{bmatrix} \mathbf{x}(t_{1}) \\ \mathbf{x}(t_{2}) \\ \cdots \\ \mathbf{x}(t_N) \end{bmatrix}_{N\times M}
\end{equation}
where the row vector $\mathbf{x}(t) = \omega(x,y,t) - \bar{\omega}(x,y)$, and $N=100$ is the number of snapshots.
The eigenvectors $\mathbf{v}_i$ of the covariance matrix $\mathbf{X^TX}$ correspond to the POD modes, while the eigenvalues $\lambda_i$ denote the contribution of each mode towards the total variation in the data about the time-mean. These were obtained using Singular Value Decomposition of the matrix $\mathbf{X}$, such that:
\begin{equation}
\mathbf{X} = \mathbf{U \bm{\Sigma} V^{T}} 
\label{eq:svdForPOD}
\end{equation}
where $\mathbf{U}$ and $\mathbf{V}$ are the left- and right-singluar matrices of $\mathbf{X}$, and $\bm{\Sigma}$ is a diagonal matrix consisting of the singular values. The columns of $\mathbf{V}$ are the eigenvectors $\mathbf{v}_i$ of $\mathbf{X}^T \mathbf{X}$. The elements of $\bm{\Sigma}$, $\sigma_{i}$, are arranged in descending order of magnitude, and are related to the corresponding eigenvalues as $\lambda_{i} = \sigma_{i}^{2}$. Thus, the POD modes are obtained in order of their contribution to variation in the data, from highest to lowest.

The first POD modes obtained in this manner for all four selected configurations are shown in figure~\ref{fig:modes}, and they highlight regions that correspond to the most significant variation in the vorticity field over one shedding time period. We note that any persistent vorticity structures that maintain a consistent form and strength in time are not expected to appear in these modes, since they get subtracted along with the time-averaged vorticity upon forming the sample matrix $\mathbf{X}$.
For the sake of convenience, we refer to the vortices shed by the arrays as a whole as primary vortices, and those shed by individual cylinders as secondary vortices. For individual A, we observe a prominent region of high variation just downstream of the array, which is caused by consistently shedding primary vortices of opposite sign, as observed in figure~\ref{proj1}. The other regions of high variation correspond to oscillatory internal jets (secondary vorticity) emerging from the rear and top of the array. These jets exist independently of the formation of the primary vortices to some extent, since they are driven by streamwise flow fed in through the array. At the other extreme, the first POD mode for individual D displays a more diffuse distribution, with fewer isolated regions of high variation. This indicates that a large spatial region of the flow for this array undergoes considerable variation in time. Moreover, it is evident from  figure~\ref{proj4} that the formation of the secondary vortices is driven primarily by the shedding of the two primary vortices along the top and bottom of the array. Minimal internal flow through array D does not allow for the formation of independently evolving secondary vortices (and the resulting jets), as in the case of array A. The POD modes for individuals B and C display behaviour that lies in between these two extremes, with array B displaying high variation streamwise jets forming via internal flow near the top and bottom of the array, and array C forming a combination of internal flow-driven and primary vortex shedding-driven secondary vortices.

The observations discussed here suggest that for a given number of cylinders, arrays that produce lower enstrophy have arrangements that allow flow to pass through the interior to some extent in the streamwise direction. Moreover, these cylinders are arranged such that the boundary layers generated by upstream cylinders attach more readily to downstream neighbours. On the other hand, the clusters with higher drag consist of arrangements that are aligned in the cross-stream direction, which increases the projected frontal area. The corresponding arrangements also allow minimal flow through the porous structure in the streamwise direction (Figure \ref{fig:avg_vel}), resulting in shedding patterns that more closely resemble those of rigid impervious objects.

\begin{figure}
    \centering
\begin{subfigure}{0.495\textwidth}
  \centering
  \includegraphics{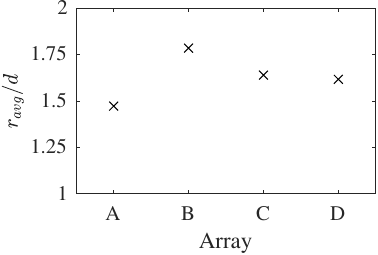}
  \caption{}
    \label{subfig:rAvg}
\end{subfigure}
\begin{subfigure}{0.495\textwidth}
    \includegraphics{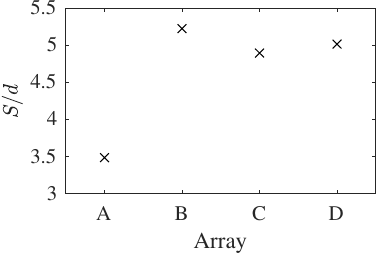}
    \caption{}
    \label{subfig:frontalS}
\end{subfigure}
\begin{subfigure}{0.495\textwidth}
    \centering
    \includegraphics{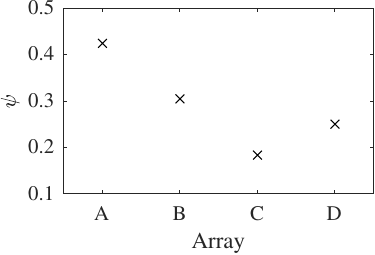}
    \caption{}
    \label{subfig:interiorFlux}
\end{subfigure}
	\caption{\label{fig:spacingFlux} \subref{subfig:rAvg} Measure of average array spacing, calculated as the mean radial distance of the cylinders from the central cylinder. \subref{subfig:frontalS} Projected surface area $S$ for the arrays normalized by cylinder diameter $d$ (assuming unit span). \subref{subfig:interiorFlux} Normalized internal flux through the arrays, calculated from DNS data using equation~\ref{eq:interiorFlux}.}
\end{figure}

To quantify the observations regarding the dependence of drag on internal flow through the arrays and cylinder alignment in the cross-stream direction, we compute the internal flux and the projected surface area for the four selected arrays. Additionally, the representative spacing for each array is also calculated as the average of the radial distances of the 8 cylinders from the central cylinder. The average internal flux was calculated from DNS data using the time-averaged speed over approximately 8 shedding time periods, and defining the convex hulls $\Omega$ manually for each array:
\begin{equation}
	\psi = \dfrac{\iint\limits_{\Omega} \lVert\bar{\bm{u}}(x,y)\rVert dA}{U_\infty(A_\Omega-9\pi d^2/4)}
	\label{eq:interiorFlux}
\end{equation}
where $A_\Omega$ is the area of the convex hull. The resulting plots are shown in figure~\ref{fig:spacingFlux} for the average spacing $r_{avg} = \sum_{i=1}^n r_{i}/n$ normalized by cylinder diameter $d$, projected frontal area $S$ normalized by $d$ (assuming unit span), and the non-dimensional internal flux $\psi$. We observe that there is no clear dependence of drag on average spacing, with $r_{avg}$ for individual D being smaller than those for B and C, although array D experiences the largest amount of drag (figure~\ref{fig:dragPIV}). On the other hand, the projected area seems to correlate better with drag, with drag increasing for higher values of $S/d$. We note that the projected area for individual D is slightly lower than that for individual B, although its drag is slightly higher than that of B. This will be examined further using the internal flux. The plot of internal flux indicates that configuration A allows the most flow to pass through the porous array, and amounts to approximately 42\% of the freestream flux through a comparable area. In general, internal flux tends to decrease with increasing drag, except for configuration C which was observed to have the minimum value of $\psi=0.18$ despite experiencing lower drag than D. This indicates that drag may be dependent on a combination of internal flux and the projected frontal area, since the value of $S/d$ is smaller for configuration C than for D. Similarly, individual B ($\psi=0.30$) is observed to allow more flow to pass through the interior than individual D ($\psi=0.25$), which may explain why the drag experienced by B is slightly lower than by D, despite having a slightly larger projected area.

\subsection{Sensitivity Analysis}
\label{subsec:sense}

\begin{figure}
\begin{subfigure}{0.32\textwidth}
  \centering
  \includegraphics[scale=0.8]{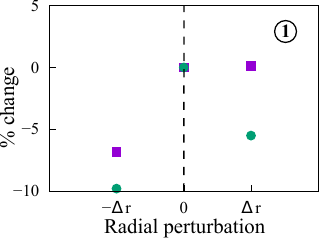}
   \subcaption{}
   \label{p4r1}
\end{subfigure}
\begin{subfigure}{0.32\textwidth}
    \centering
    \includegraphics[scale=0.8]{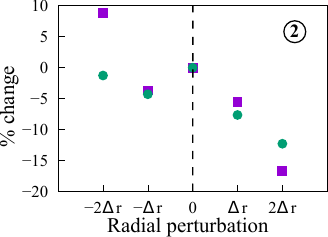}
       \subcaption{}
       \label{p4r2}
\end{subfigure}
\begin{subfigure}{0.32\textwidth}
    \centering
    \includegraphics[scale=0.8]{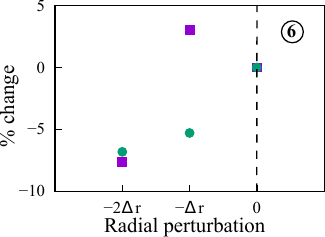}
       \subcaption{}
            \label{p4r6}
\end{subfigure}
\newline
\begin{subfigure}{\textwidth}
\begin{minipage}{0.5\textwidth}
\centering
\hfill
    \includegraphics[scale=0.8]{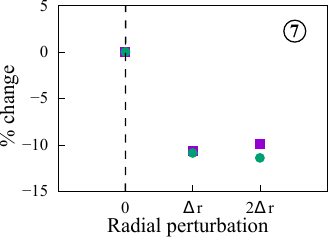}
       \subcaption{}
          \label{p4r7}
\end{minipage}
\begin{minipage}{0.5\textwidth}
\centering
    \includegraphics[scale=0.3]{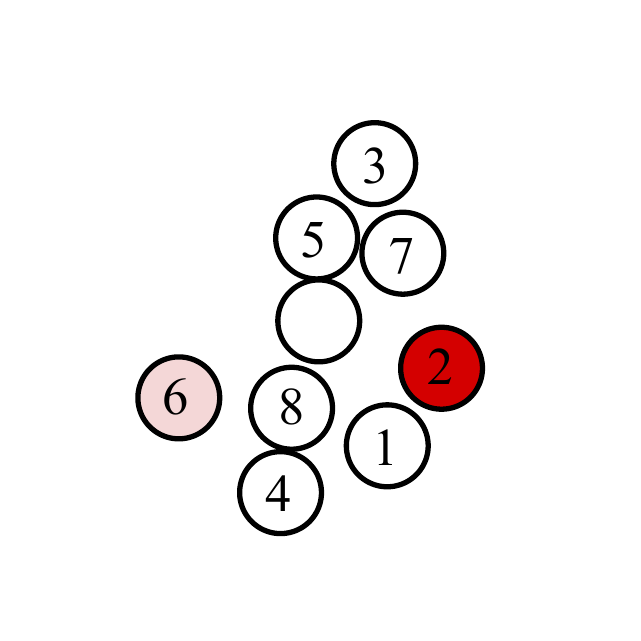}
       \subcaption{}
            \label{cyl-sense-radius}
       \hfill
\end{minipage}
\end{subfigure}
	\caption{Percentage change in enstrophy (purple squares) and drag (green circles) when the position of one cylinder in the high-drag arrangement (individual D) was changed with respect to $r$ ($\Delta r=d/4$) in DNS. Panels \subref{p4r1}, \subref{p4r2}, \subref{p4r6} and \subref{p4r7} correspond to data for cylinders 1, 2, 6 and 7 respectively, which are labelled in panel \subref{cyl-sense-radius}. The remaining cylinders occupy positions where they could not be perturbed in the radial direction without encountering collisions. The most sensitive (dark red) and least sensitive (light red) cylinders with respect to changes in $r$ are highlighted in panel \subref{cyl-sense-radius}.}
\label{fig:sensitivity-radius}
\end{figure}

\begin{figure}
\begin{subfigure}{0.32\textwidth}
  \centering
  \includegraphics[scale=0.8]{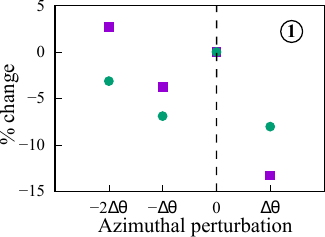}
   \subcaption{}
   \label{p4t1}
\end{subfigure}
\begin{subfigure}{0.32\textwidth}
    \centering
    \includegraphics[scale=0.8]{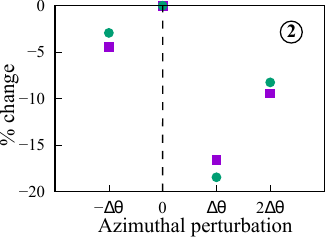}
   \subcaption{}
   \label{p4t2}
\end{subfigure}
\begin{subfigure}{0.32\textwidth}
    \centering
    \includegraphics[scale=0.8]{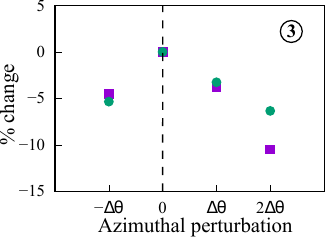}
   \subcaption{}
   \label{p4t3}
\end{subfigure}
\newline
\begin{subfigure}{0.32\textwidth}
    \centering
    \includegraphics[scale=0.8]{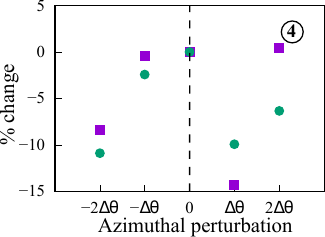}
   \subcaption{}
   \label{p4t4}
\end{subfigure}
\begin{subfigure}{0.32\textwidth}
    \centering
    \includegraphics[scale=0.8]{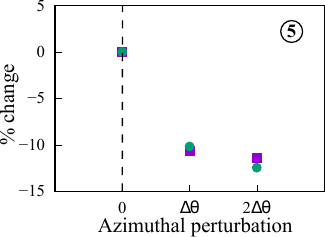}
   \subcaption{}
   \label{p4t5}
\end{subfigure}
\begin{subfigure}{0.32\textwidth}
    \centering
    \includegraphics[scale=0.8]{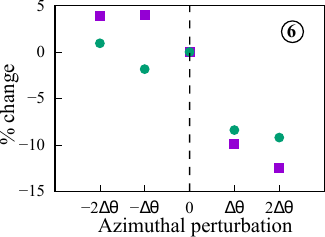}
   \subcaption{}
     \label{p4t6}
\end{subfigure}
\newline
\begin{subfigure}{0.32\textwidth}
    \centering
    \includegraphics[scale=0.8]{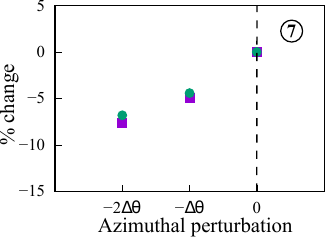}
   \subcaption{}
   \label{p4t7}
\end{subfigure}
\begin{subfigure}{0.32\textwidth}
    \centering
    \includegraphics[scale=0.8]{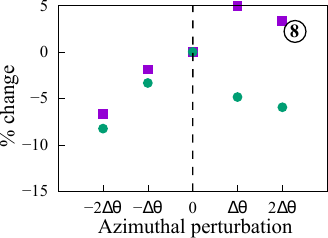}
   \subcaption{}
   \label{p4t8}
\end{subfigure}
\begin{subfigure}{0.32\textwidth}
    \centering
    \includegraphics[scale=0.3]{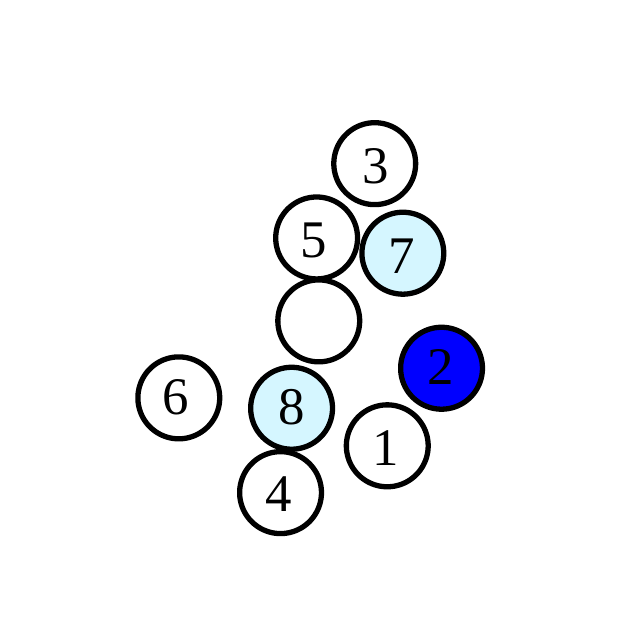}
       \subcaption{}
       \label{cyl-sense-theta}
\end{subfigure}
	\caption{Percentage change in enstrophy (purple squares) and drag (green circles) when the position of one cylinder in the high-drag array (individual D) was changed with respect to azimuthal angle $\theta$ ($\Delta \theta = \frac{\pi}{32}$) in DNS. Panels \subref{p4t1} through \subref{p4t8} correspond to data for cylinders 1 through 8, respectively, as shown in panel \subref{cyl-sense-theta}. The most sensitive (dark blue) and least sensitive (light blue) cylinders with respect to changes in $\theta$ are highlighted in panel \subref{cyl-sense-theta}.}
\label{fig:sensitivity-theta}
\end{figure}

\begin{figure}
\begin{subfigure}{0.245\textwidth}
  \centering
  \includegraphics[scale=0.35]{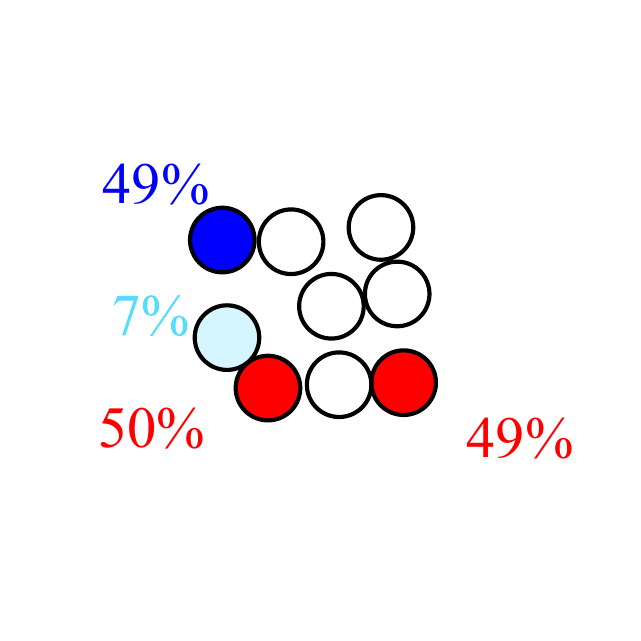}
   \subcaption{}
    \label{p1e}
\end{subfigure}
\begin{subfigure}{0.245\textwidth}
    \centering
    \includegraphics[scale=0.35]{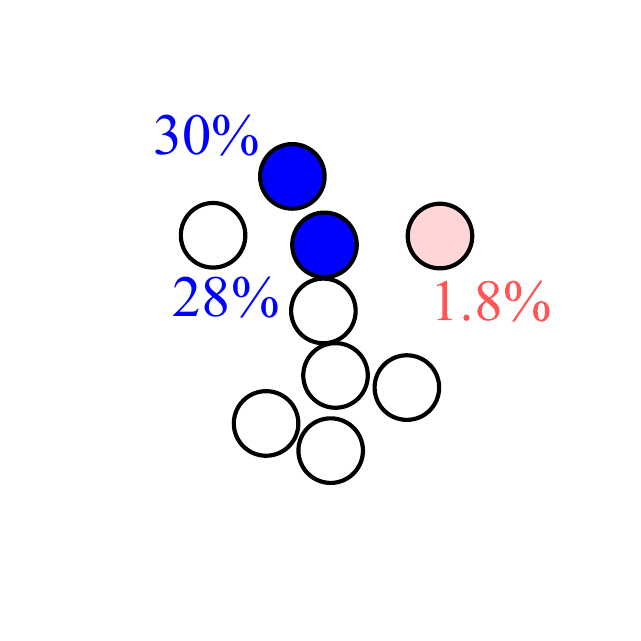}
       \subcaption{}
       \label{p2e}
\end{subfigure}
\begin{subfigure}{0.245\textwidth}
    \centering
    \includegraphics[scale=0.35]{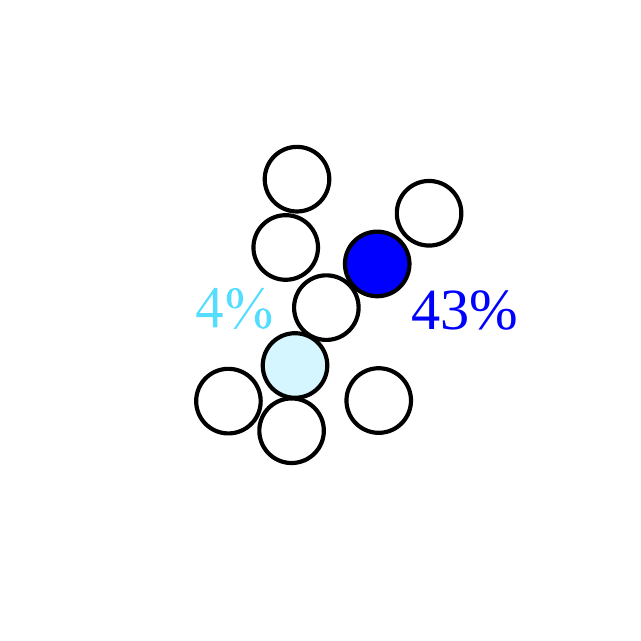}
       \subcaption{}
       \label{p3e}
\end{subfigure}
\begin{subfigure}{0.245\textwidth}
    \centering
    \includegraphics[scale=0.35]{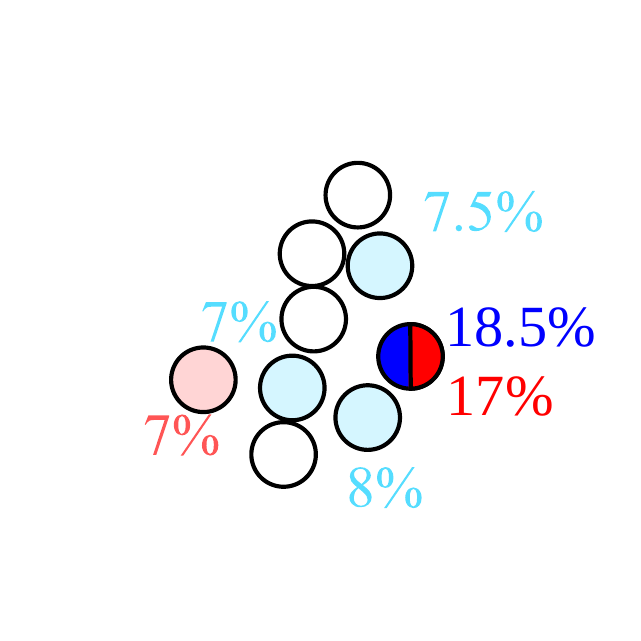}
       \subcaption{}
       \label{p4e}
\end{subfigure}
\newline
\begin{subfigure}{0.245\textwidth}
  \centering
  \includegraphics[scale=0.35]{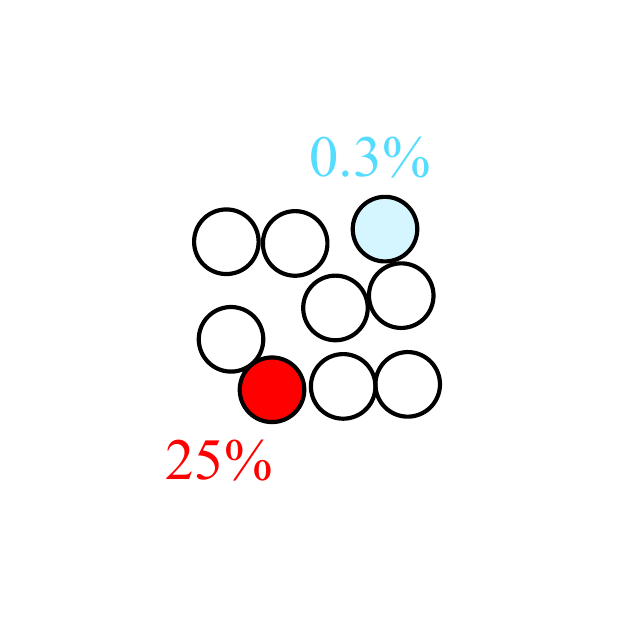}
   \subcaption{}
   \label{p1d}
\end{subfigure}
\begin{subfigure}{0.245\textwidth}
    \centering
    \includegraphics[scale=0.35]{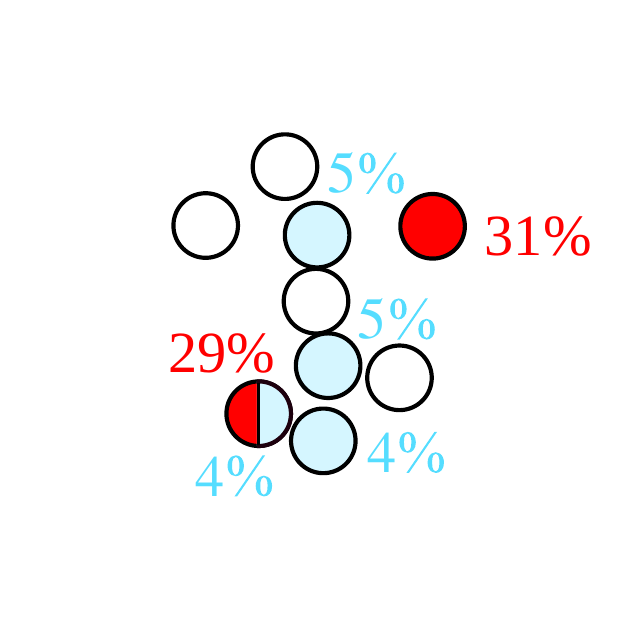}
       \subcaption{}
       \label{p2d}
\end{subfigure}
\begin{subfigure}{0.245\textwidth}
    \centering
    \includegraphics[scale=0.35]{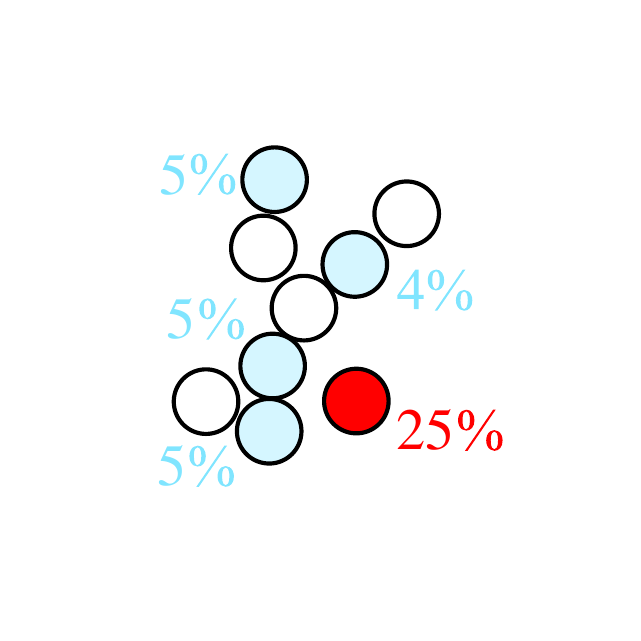}
    \subcaption{}
     \label{p3d}
\end{subfigure}
\begin{subfigure}{0.245\textwidth}
    \centering
    \includegraphics[scale=0.35]{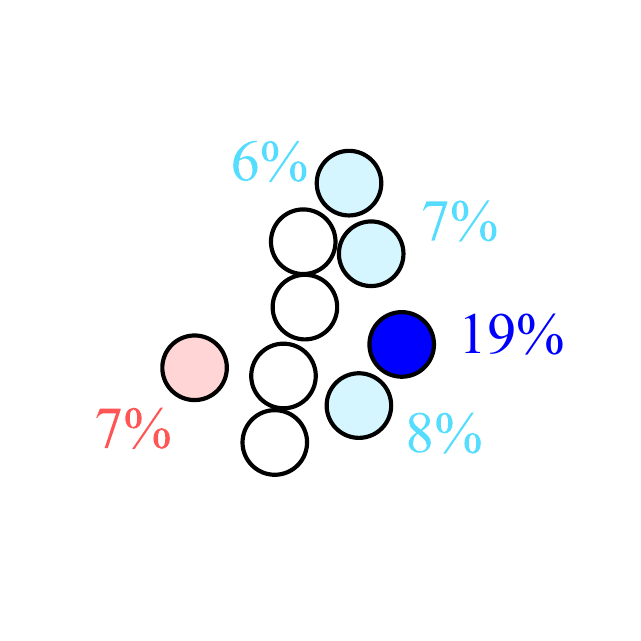}
       \subcaption{}
       \label{p4d}
\end{subfigure}
	\caption{The most and least sensitive cylinders for each of the four selected arrangements. Sensitivity with respect to enstrophy is examined in the top row (panels \subref{p1e} to \subref{p4e}), and that with respect to drag is examined in the bottom row (panels \subref{p1d} to \subref{p4d}). The dark red cylinders are most sensitive to changes in r (radial direction), and the dark blue cylinders are most sensitive to changes in $\theta$ (azimuthal direction). The lighter coloured cylinders are least sensitive with respect to r (light red) and $\theta$ (light blue). The values next to the cylinders indicate the absolute percentage change in drag or enstrophy upon perturbation.}
\label{fig:sense-highlight}
\end{figure}

To gain further insight into the role that the positioning of the cylinders plays in determining drag and enstrophy, we conduct a sensitivity analysis for the four selected Pareto-optimal cases. The position of each cylinder within an array was perturbed in the radial and azimuthal directions by small amounts, hence changing the corresponding $r_{i}$ and $\theta_{i}$ values one cylinder at a time. Cases where this resulted in a collision with neighbouring cylinders, or in exceeding the maximum patch diameter $D$, were not considered. The enstrophy and drag were calculated for all the perturbed cases, and detailed results for one of the scenarios are shown in Figures \ref{fig:sensitivity-radius} and \ref{fig:sensitivity-theta}. Additionally, the most- and least-sensitive cylinders with regard to change in enstrophy and drag are shown in Figure \ref{fig:sense-highlight} for each of the four arrays. 

Figure \ref{fig:sensitivity-radius} shows the percentage change in drag and enstrophy for individual D when each cylinder's position was changed slightly by $\pm \Delta r$ and $\pm 2\Delta r$ (where $\Delta r = d/4$) in the radial direction with respect to the central cylinder. Similarly, Figure \ref{fig:sensitivity-theta} shows the percentage change in drag and enstrophy when each cylinder's position was changed by $\pm \Delta \theta$ and $\pm 2\Delta \theta$ (where $\Delta \theta = \frac{\pi}{32}$) in the azimuthal direction about the center of the array. The cylinders showing the highest and lowest sensitivity overall (i.e., for drag as well as enstrophy) have been highlighted in the respective figures. As observed in Figure \ref{fig:sensitivity-radius}, only the cylinders labelled 1, 2, 6 and 7 were able to be perturbed in the radial direction. The largest change in enstrophy was observed for cylinder 2 when its radial distance was increased by $2\Delta r$. The largest change in drag was also observed for cylinder 2 at this perturbation distance. Hence, cylinder 2 is labelled as the most sensitive cylinder overall for individual D with respect to radial perturbation (Figure \ref{cyl-sense-radius}). On the other hand, cylinder 6 exhibits the smallest maximum change in drag, but has a maximum change in enstrophy comparable to that of cylinder 1. Hence, cylinder 6 is labelled as the overall least sensitive cylinder for individual D with respect to radial perturbations. Similarly, in Figure \ref{fig:sensitivity-theta}, cylinder 2 was found to have the largest maximum change in enstrophy as well as drag when perturbed in the azimuthal direction, whereas cylinder 7 and 8 both showed similar and overall smallest changes in enstrophy and drag (Figure \ref{cyl-sense-theta}). We note that none of the perturbed array configurations for the four selected individuals resulted in higher drag and lower enstrophy simultaneously, compared to their respective base unperturbed configurations. More specifically, none of the perturbed individuals were better than the original Pareto-optimal individuals in both fitness metrics, which indicates that the Pareto-optimal solutions obtained using the multi-objective optimization are robust.

In Figure \ref{fig:sense-highlight}, the sensitivity of the four selected Pareto-optimal individuals is considered separately with regard to drag and enstrophy. The cylinders that have the highest and least impact on each of these two aspects upon perturbation are highlighted. The top row (Figures \ref{p1e} - \ref{p4e}) depicts sensitivity with regard to enstrophy, and the bottom row (Figures \ref{p1d} - \ref{p4d}) depicts sensitivity with respect to drag. One notable observation is that across all the examined cases, the perturbations tended to have a greater influence on wake enstrophy than on drag. This indicates that enstrophy is more sensitive than drag to specific cylinder placement within porous arrays.

For individual A in Figure \ref{p1e}, we observe that the cylinders most sensitive with respect to enstrophy are the ones whose perturbation disrupts boundary layer attachment among neighbouring cylinders (as observed in Figure \ref{proj1}), leading to an increase in wake enstrophy. The dark blue cylinder in Figure \ref{p1e} blocks flow from passing through the interior of the array when perturbed in the azimuthal direction, which also increases enstrophy due to weaker secondary vortices (and hence, weaker vorticity cancellation). The phenomenon of vorticity cancellation, especially at the low local Reynolds number in the interior of the arrays, is similar to that observed by \citet{Moulinec2004}, who found that oppositely signed vorticity in close proximity led to cancellation via diffusion. On the other hand, the least sensitive cylinder in Figure \ref{p1e}, when perturbed in the azimuthal direction, was observed to retain its tendency to generate boundary layers that remain attached to neighbouring cylinders. Examining the drag sensitivity for individual A, we observe in Figure \ref{p1d} that perturbing the dark red cylinder in the radial direction increases drag by $25\%$. This may be related to an overall increase in the effective frontal surface area as the cylinder in question is moved radially outward. Moreover, increasing its distance from the center disrupts the continuity of the attached boundary layer observed in Figure \ref{proj1}, thereby increasing drag. The same cannot be said for the rest of the cylinders for individual A, since even after perturbation they remained sufficiently close to their neighbours so as not to disrupt the continuous boundary layer. Meanwhile, the light blue cylinder caused no significant variation in drag when perturbed in the azimuthal direction and it did not induce any major changes to the overall flow pattern, presumably because the internal channels that allow flow to pass through the array remain mostly unchanged.

For individual B in Figures \ref{p2e} and \ref{p2d}, we note that a majority of the cylinders, especially the outermost ones, are restricted in perturbations with respect to $r$; positive perturbations would cause cylinders to exceed the outer patch diameter $D$, whereas negative perturbations would lead to collisions. With respect to drag (Fig. \ref{p2d}), one cylinder in particular affects drag significantly when perturbed radially ($29\%$), and minimally when perturbed in the azimuthal direction ($4\%$). This can be explained by the influence the respective perturbations have on internal flow through the array. Upon decreasing the radial distance of the cylinder in question, it completely blocks off internal flow through the lower half of the array. As a result, the freestream gets redirected around the lower half, behaving more closely to flow around a cylinder. This reduces the interaction of the oncoming flow with the internal cylinders, most of which are placed in the cross-stream direction, resulting in lower drag. We note that azimuthal perturbations for this particular cylinder do not have as significant an impact on the internal flow, resulting in only a $4\%$ change. The remaining light blue cylinders in Figure \ref{p2d} obstruct the incoming freestream, and the extent of their blockage is not affected by perturbations in $\theta$. We also note that one of the cylinders in the array is minimally relevant for enstrophy but highly relevant for drag ($1.8\%$ vs $31\%$), and another cylinder for which the reverse is true ($28\%$ vs $5\%$). If we consider the vorticity shown in Figure~\ref{proj2}, the cylinder coloured light red in Figure \ref{p2e} is isolated from the other cylinders in the array, and generates a highly unsteady shedding pattern of its own. Hence, even though it changes the flow structure slightly upon perturbation, the structures it affects have a small impact on the overall enstrophy ($1.8\%$).

For individual C in Figures \ref{p3e} and \ref{p3d}, we observe one particular cylinder which has a large impact on enstrophy ($43\%$) and a minimal impact on drag ($4\%$). This is because the cylinder is in the vicinity of the most consistent internal flow structures that form for this particular array. Increasing the azimuthal angle will hinder the formation of the opposing-vorticity boundary layers found at an angle of approximately $\pi/4$ near the top of the array in Figure \ref{proj3}. However, since the dominant direction of local flow is tangential to the cylinder, its overall contribution to drag does not change significantly upon azimuthal perturbation. In Figure \ref{p3d}, we observe that the light blue cylinders do not affect drag significantly upon perturbation in the $\theta$ direction for the same reason as that for individual B, i.e., perturbations in the azimuthal direction do not result in significant changes to the overall flow. The dark red cylinder in \ref{p3d} causes a decrease in drag as its distance from the center increases. This may be related to the fact that it gets better aligned with the wakes of the cylinders upstream of it, which is also accompanied by a significant reduction in enstrophy (30\% reduction, not annotated on the figure). This reinforces an earlier observation that boundary layers of neighbouring cylinders staying attached is a significant contributor to reducing enstrophy.

Individual D (Figures \ref{p4d} and \ref{p4e}) appears overall to be less sensitive to perturbations than the other 3 individuals, since it has a comparatively large number of lighter coloured cylinders, and the absolute percent change values are smaller. This may be because the localized flow structures are highly unsteady (Figure \ref{proj4}), so that perturbations to any one specific cylinder will play a smaller role in modifying the overall flow structure. We observe that the same cylinder is most sensitive towards drag (azimuthal direction, 19\%) and towards enstrophy (both radial and azimuthal directions, 17\% and 18.5\%). Decreasing the azimuthal angle for this cylinder would connect it to the lower neighbouring cylinder such that their boundary layers remain attached, thereby reducing both drag and enstrophy. In Figure \ref{p4d}, the single light red cylinder has a minimal impact on drag as well as on enstrophy (7\%) upon radial perturbation, since it is isolated and does not impact the resultant flow significantly.

We note that while we aimed to perform sensitivity analysis for all of the cylinders in each array, perturbing certain cylinders would have resulted in an overlap or collision with neighboring cylinders, and thus, these cylinders were excluded. Furthermore, Figure~\ref{fig:sense-highlight} highlights only those cylinders that displayed the highest and lowest sensitivity for a given configuration, although several of the unmarked cylinders were also examined. Overall, the observations presented here indicate that the internal arrangement of cylinders in an array is an important consideration for controlling drag and enstrophy. The sensitivity analysis reveals that enstrophy is minimized by directing incoming flow to pass within the array, which leads to vorticity cancellation \citep{Moulinec2004,Ricardo2016} and the suppression of individual vortex shedding due to the attachment of boundary layers among neighbouring cylinders. We also observe that drag appears to be dependent on the projected frontal area, as well as on boundary layer attachment among neighbouring cylinders to a smaller extent. Variation in drag based on cylinder placement was also observed in the study by \citet{Shan2019}, who found that random physical arrangements resulted in higher drag compared to uniform grid arrangements. These observations indicate that a high-level metric such as porosity may not be sufficient on its own to predict wake characteristics for porous arrays.

\subsection{Particle Tracking and Lagrangian Coherent Structures}
\label{subsec:ftle}

We now use Lagrangian Particle tracking to examine the trajectories of tracer particles seeded into the flow to determine the behavior of each of the selected Pareto-optimal arrays with regard to particle sedimentation and erosion. We note that sediment transport is a complex process that depends on several flow and sediment properties, particularly the bed shear stress \citep{Lopez1998, Kothyari2009}. However, tracers can provide a useful approximation of particle transport around and within the porous arrays. We consider the initial transient state separately from the steady vortex shedding state, to explore how particle transport may differ in systems involving intermittent flow (e.g., wave-dominated coastal regions) from those involving continuous flow (e.g., rivers). In both scenarios, particles were initialized separately upstream of the porous arrays and in the cylinders' immediate vicinity and wake (Appendix Figure~\ref{fig:particles_tr}). Apart from particle tracking, Lagrangian Coherent structures (LCS) associated with steady state shedding were computed using the Finite time Lyapunov Exponent (FTLE) \citep{Haller2000, Shadden2005, Lekien2007}. \cite{Ku2018} used FTLE analysis to study the transport and dispersion of small particles in stratified flows, demonstrating that the particles could not pass through isosurfaces of the FTLE field. Moreover, \cite{Giudici2021} used the concept of LCSs to successfully identify areas in the Gulf of Finland where floating items tended to accumulate in large quantities.

\subsubsection{Particle tracking during the startup transient state}

\begin{figure}
\begin{subfigure}{0.5\textwidth}
  \centering
  \includegraphics[width=\textwidth]{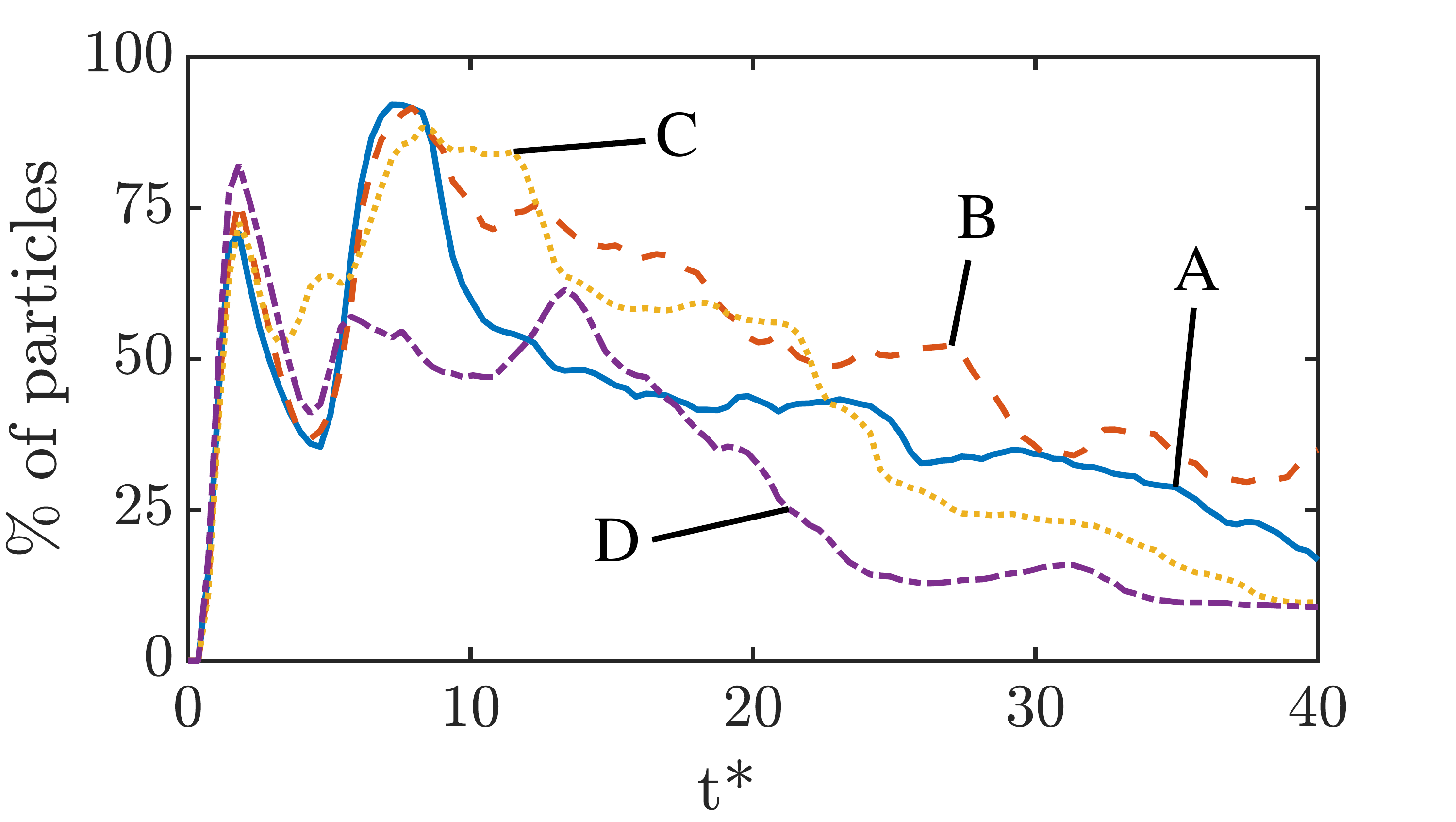}
  \subcaption{}
  \label{fig:graph_tr_in}
\end{subfigure}
\begin{subfigure}{0.5\textwidth}
  \centering
  \includegraphics[width=\textwidth]{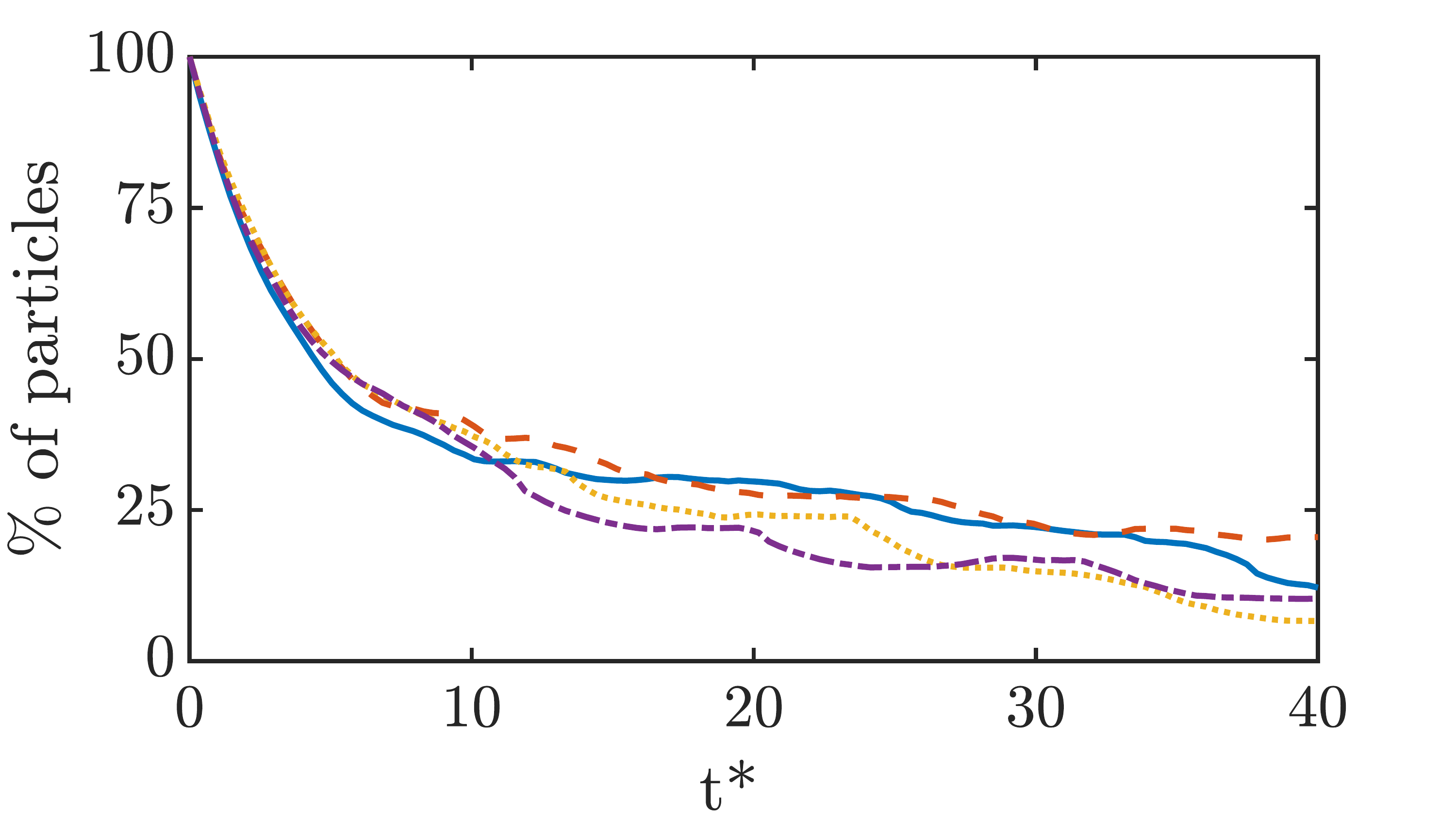}
  \subcaption{}
  \label{fig:graph_tr_surr}
\end{subfigure}
	\caption{Percentage of seeded particles that are present in the arrays' vicinity, i.e., in a prescribed observation window, over the first $40t^*$ of transient flow in DNS. \subref{fig:graph_tr_in} The plots represent the time-evolution of the upstream-seeded particles, whereas the plots in \subref{fig:graph_tr_surr} represent the time-evolution of the vicinity-seeded particles.}
\label{fig:graph_tr}
\end{figure}

In order to study particle entrainment, two groups of particles were seeded in the flow field. The first group was initialized in the incoming flow upstream of the array in a region of size $\sim 0.15D \times 1.1D$  (upstream-seeded - Appendix Figure \ref{fig:particles_tr}). The second group was initialized in the areas surrounding the cylinders and in the near-wake region with dimensions $\sim 1.6D \times 1.1D$  (vicinity-seeded). The cross-stream width of these regions was varied according to the  width of the arrays, which can vary slightly between the selected individuals. The particles' motion can be observed in the animations provided in Supplementary Movies 2 and 3, and corresponding snapshots are provided in Appendix Figure \ref{fig:particles_tr} at non-dimensional times $t^{*}=0$ and $t^{*}=40$ (where $t^{*} = tU_{\infty}/d$). The plots in Figure \ref{fig:graph_tr} show the percentage of particles found in the vicinity of the cylinders (i.e., in a prescribed observation window described by the region where the green-coloured particles are initialized in Appendix Figure \ref{fig:particles_tr}), separately for the upstream-seeded particles in Figure \ref{fig:graph_tr_in} and the vicinity-seeded particles in Figure \ref{fig:graph_tr_surr}. The initial results are identical across all four cases up until $t^*=2$, since distinct flow features have not yet had time to develop. In Figure \ref{fig:graph_tr_in}, the sudden dip after $t^{*} = 2$ corresponds to the upstream-seeded particles exiting and re-entering the spatial window. Individual D ends up losing a significant percentage of the particles after this stage, which is related to most of the flow being directed around the array instead of through the interior. Individuals B and individual C retain the highest proportion of particles initially ($10 \leq t^{*} \leq 20$), after which individual A out-performs individual C. Similarly, in Figure \ref{fig:graph_tr_surr} individual B retains almost twice as many vicinity-seeded particles in the spatial window as the other three arrays. This may be because in the initial transient state, i.e., before vortex shedding begins, the re-circulation region is a prominent factor in determining the extent of particle retention. In Figure \ref{fig:vorticitySnapshots}, we observe that individuals B and C generate large recirculation regions in their wakes. Overall, individual B is most effective in retaining both the upstream-seeded and the vicinity-seeded particles in the initial transient state, with individual A being a close second.

\subsubsection{Particle tracking during the steady vortex shedding state}
\label{sssec:LPT-ss}

\begin{figure}
\begin{subfigure}{0.5\textwidth}
  \centering
  \includegraphics[width=\textwidth]{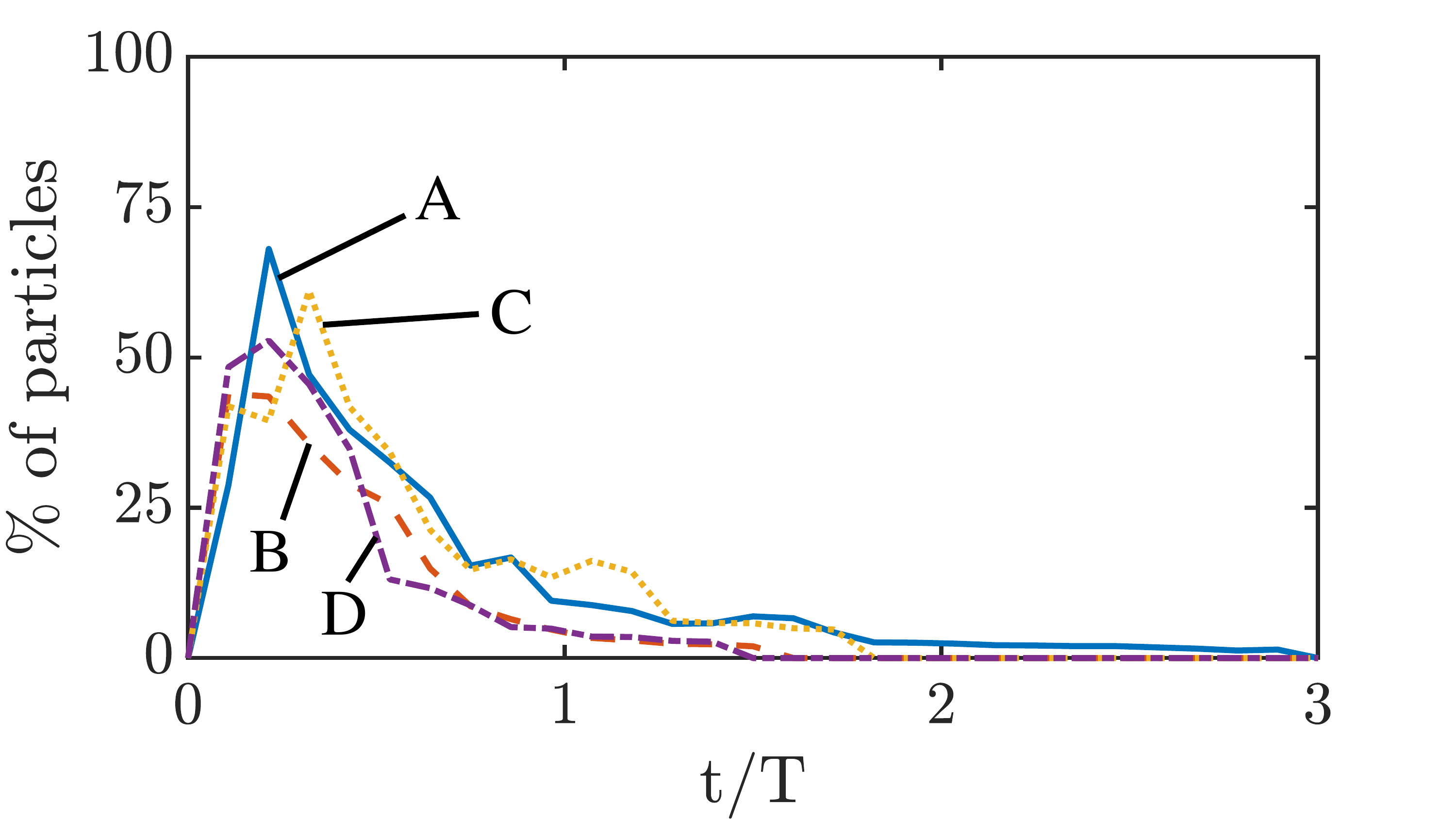}
  \subcaption{}
  \label{fig:graph_ss_in}
\end{subfigure}
\begin{subfigure}{0.5\textwidth}
  \centering
  \includegraphics[width=\textwidth]{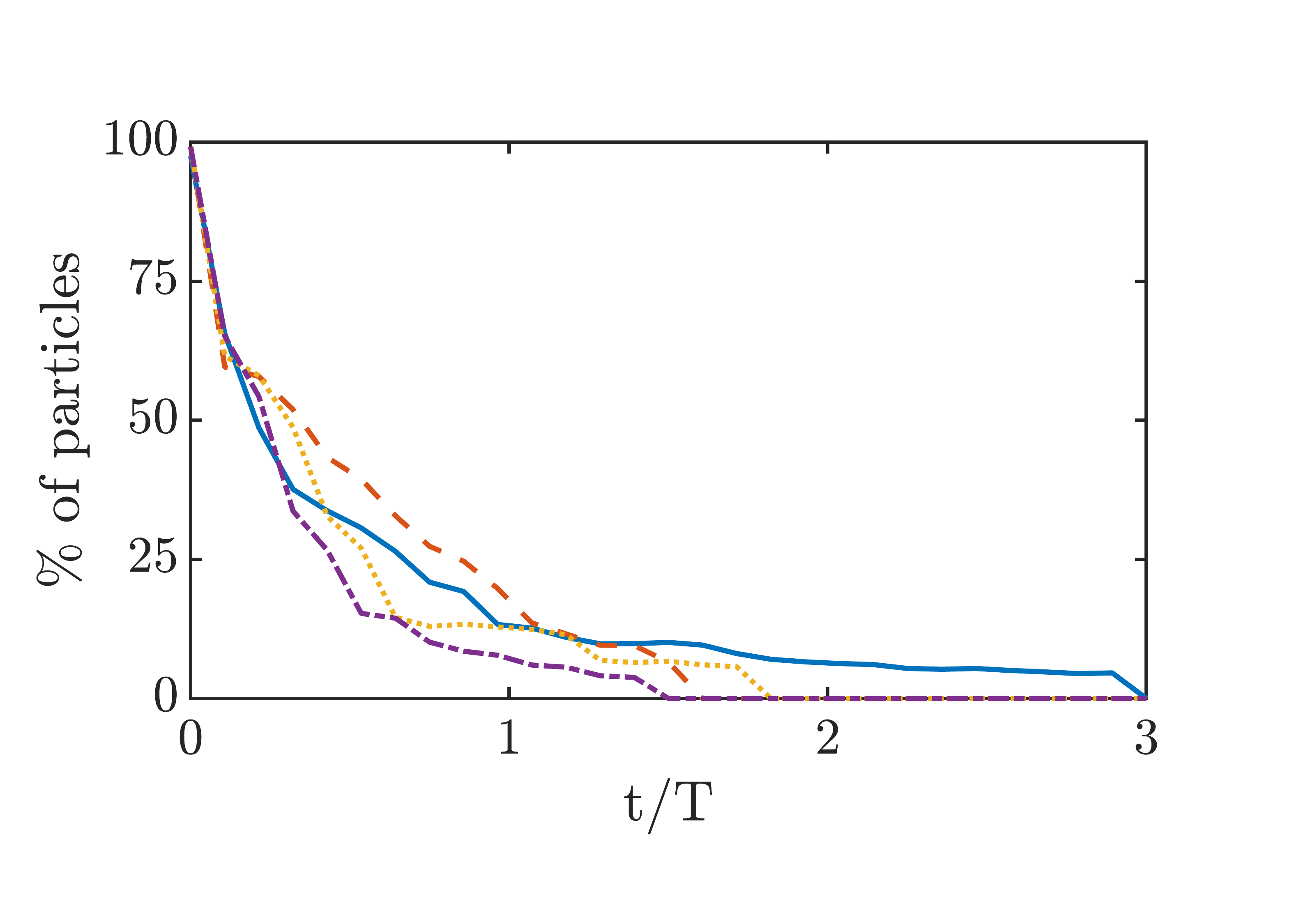}
  \subcaption{}
  \label{fig:graph_ss_surr}
\end{subfigure}
	\caption{The percentage of initialized particles found in the prescribed observation window over three shedding time periods in the quasi-steady state, for each of the four arrangements using DNS data. \subref{fig:graph_ss_in} Retention of upstream-seeded particles. \subref{fig:graph_ss_surr} Retention of vicinity-seeded particles.}
\label{fig:graph_ss}
\end{figure}

Appendix Figure \ref{fig:particles_ss} shows particle tracers for the four selected individuals, through 3 vortex shedding cycles once the wake flow reaches a steady shedding state. The plots in Figure \ref{fig:graph_ss} show the percentage of the initialized particles that remain in the vicinity of the cylinders, i.e., in the observation window described earlier, separately for the upstream-seeded (Figure \ref{fig:graph_ss_in}) and the vicinity-seeded particles (Figure \ref{fig:graph_ss_surr}). In Figure \ref{fig:graph_ss_in}, we observe that individual C retains the largest proportion of the upstream-seeded particles until about 1.5 time periods, after which individual A takes over. Similarly, in Figure \ref{fig:graph_ss_surr}, individuals B and C have the highest retention until one time period, and by the end of three time periods, individual A is the only arrangement to retain approximately 5\% of the vicinity-seeded particles within the observation window, while the other arrangements retain a negligible amount of particles. We will examine the relationship between particle retention and the flow field further with the help of Lagrangian coherent structures in \S\ref{sssec:LCS}.

\subsubsection{Lagrangian Coherent Structures (LCSs)}
\label{sssec:LCS}

\begin{figure}
\begin{subfigure}{\textwidth}
\centering
  \includegraphics[width=0.24\textwidth]{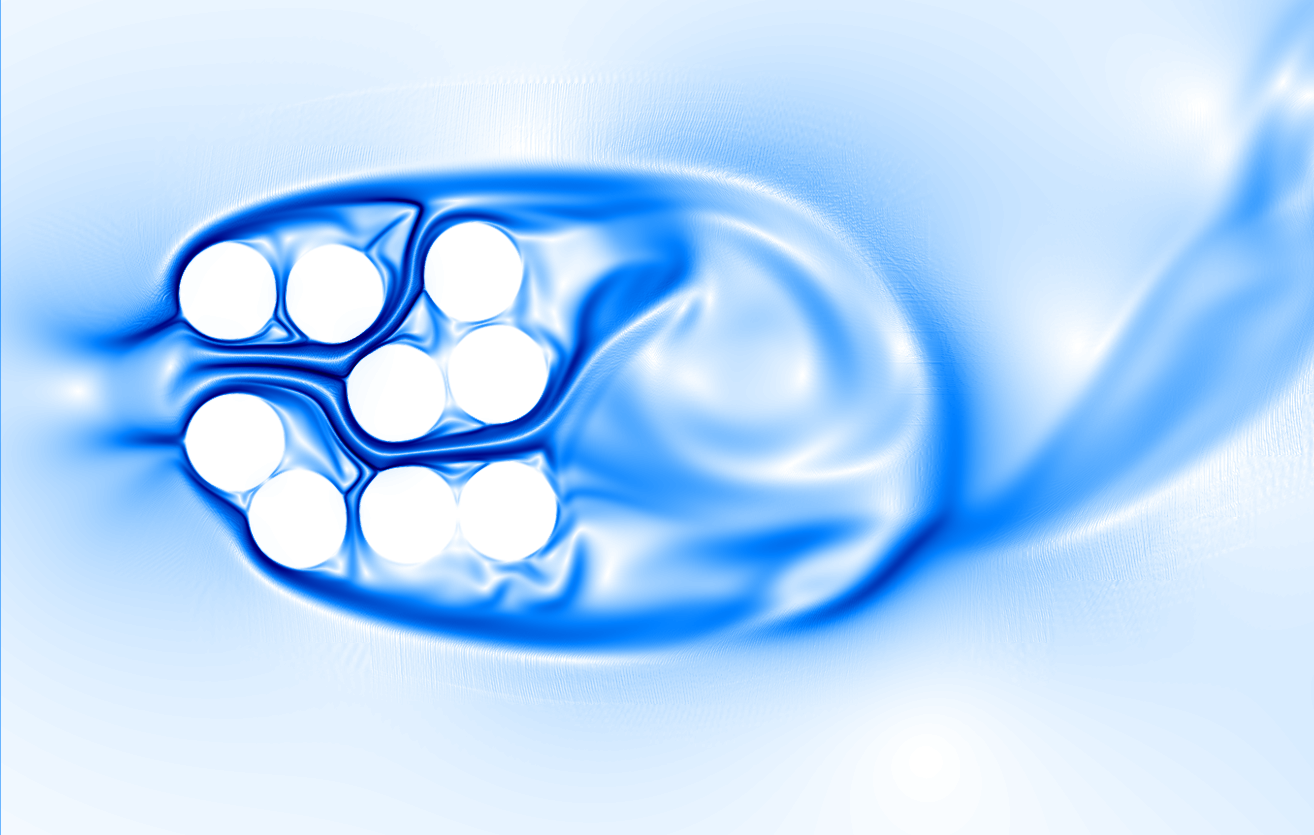}
  \includegraphics[width=0.24\textwidth]{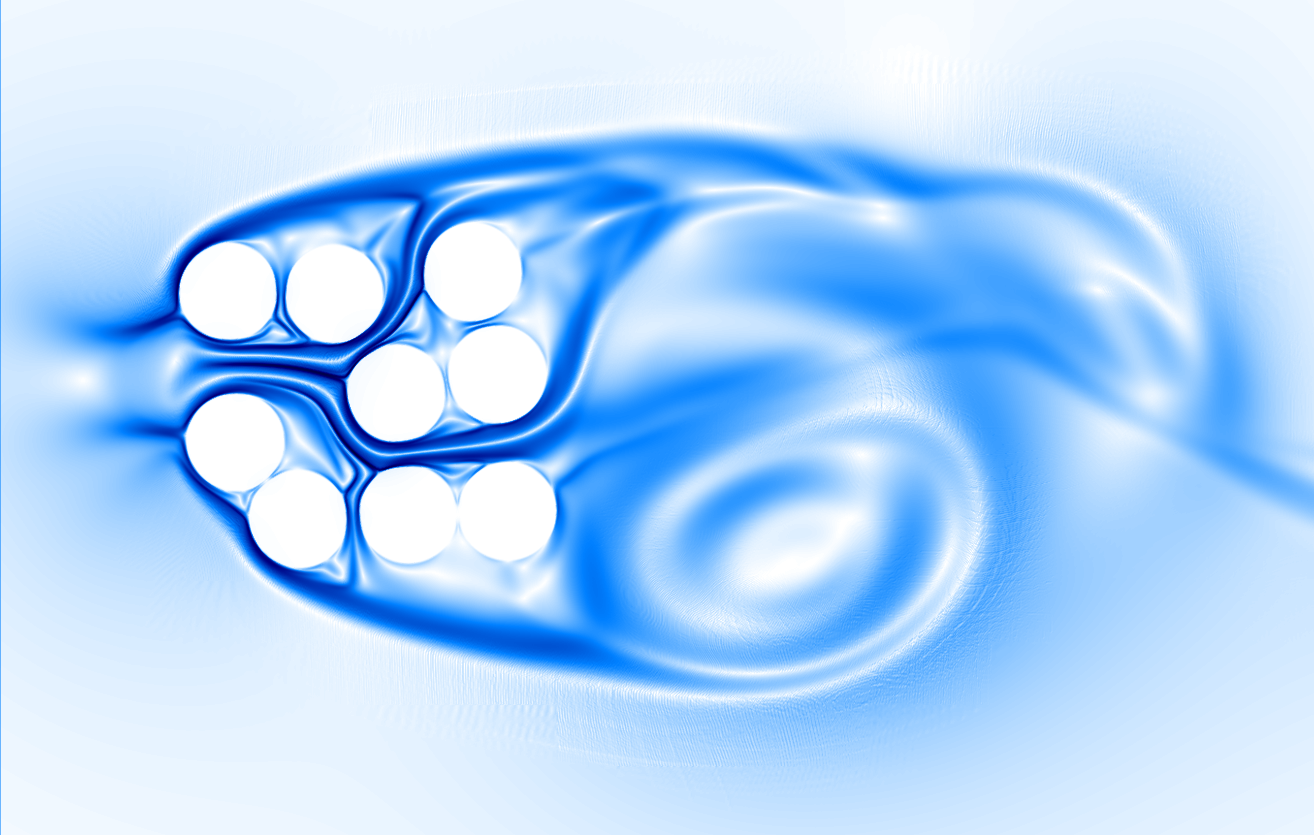}
  \includegraphics[width=0.24\textwidth]{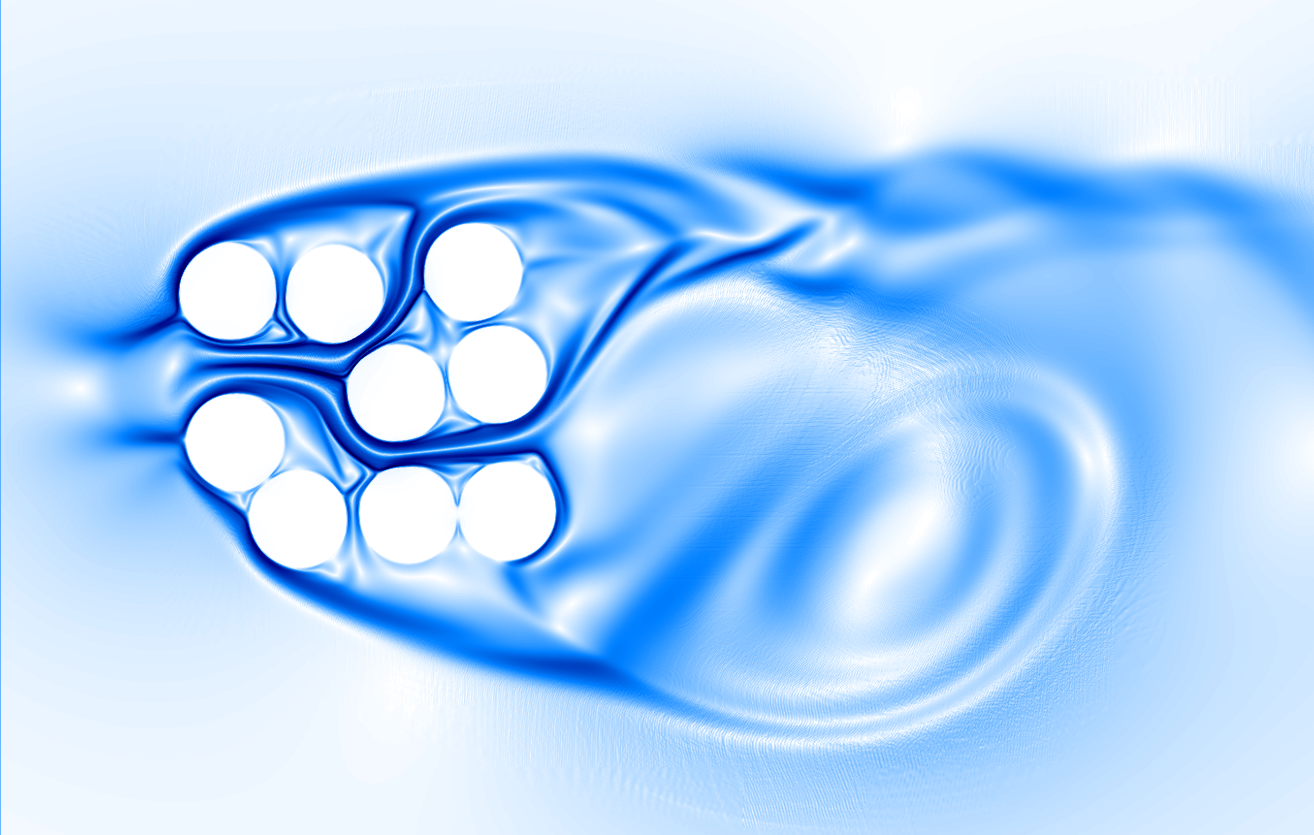}
  \includegraphics[width=0.24\textwidth]{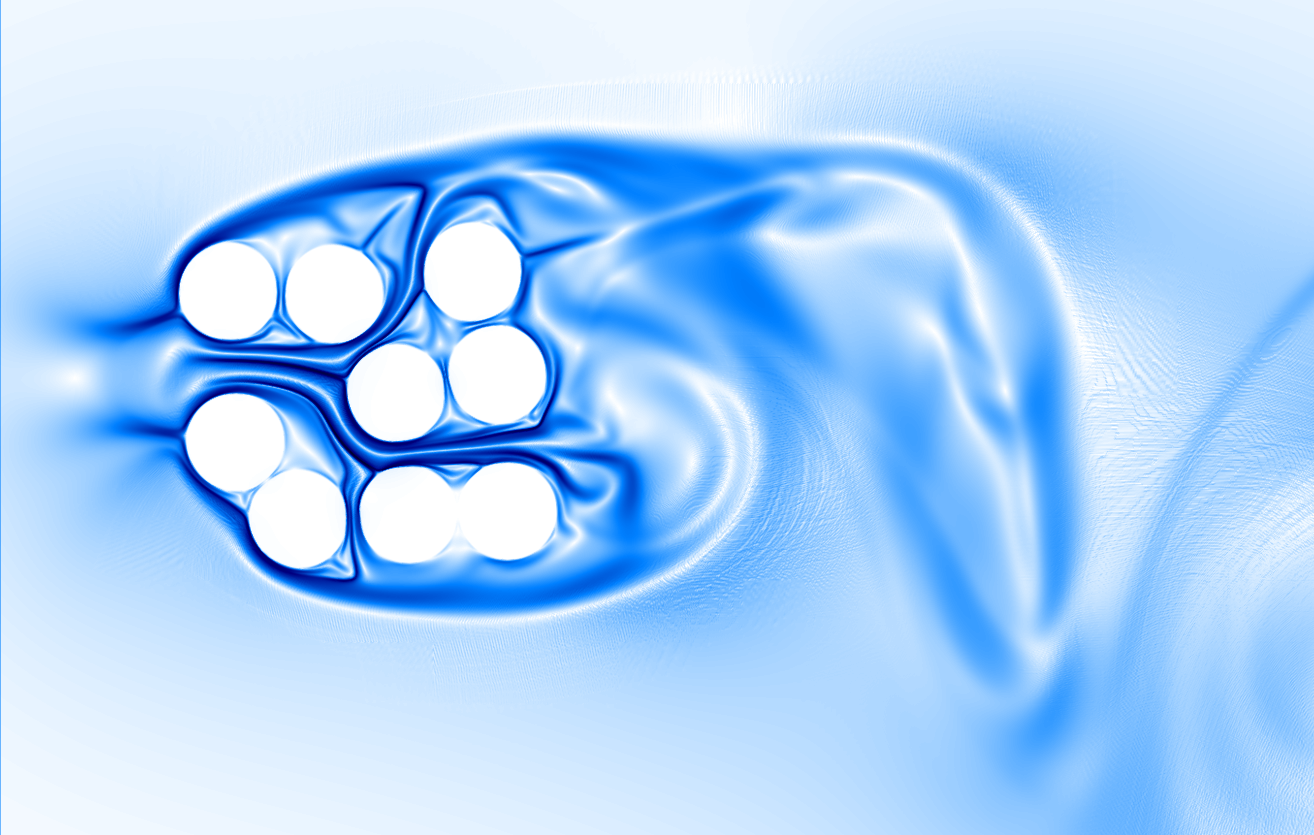}
  \subcaption{}
  \label{fig:lcs1}
\end{subfigure}

\begin{subfigure}{\textwidth}
\centering
  \includegraphics[width=0.24\textwidth]{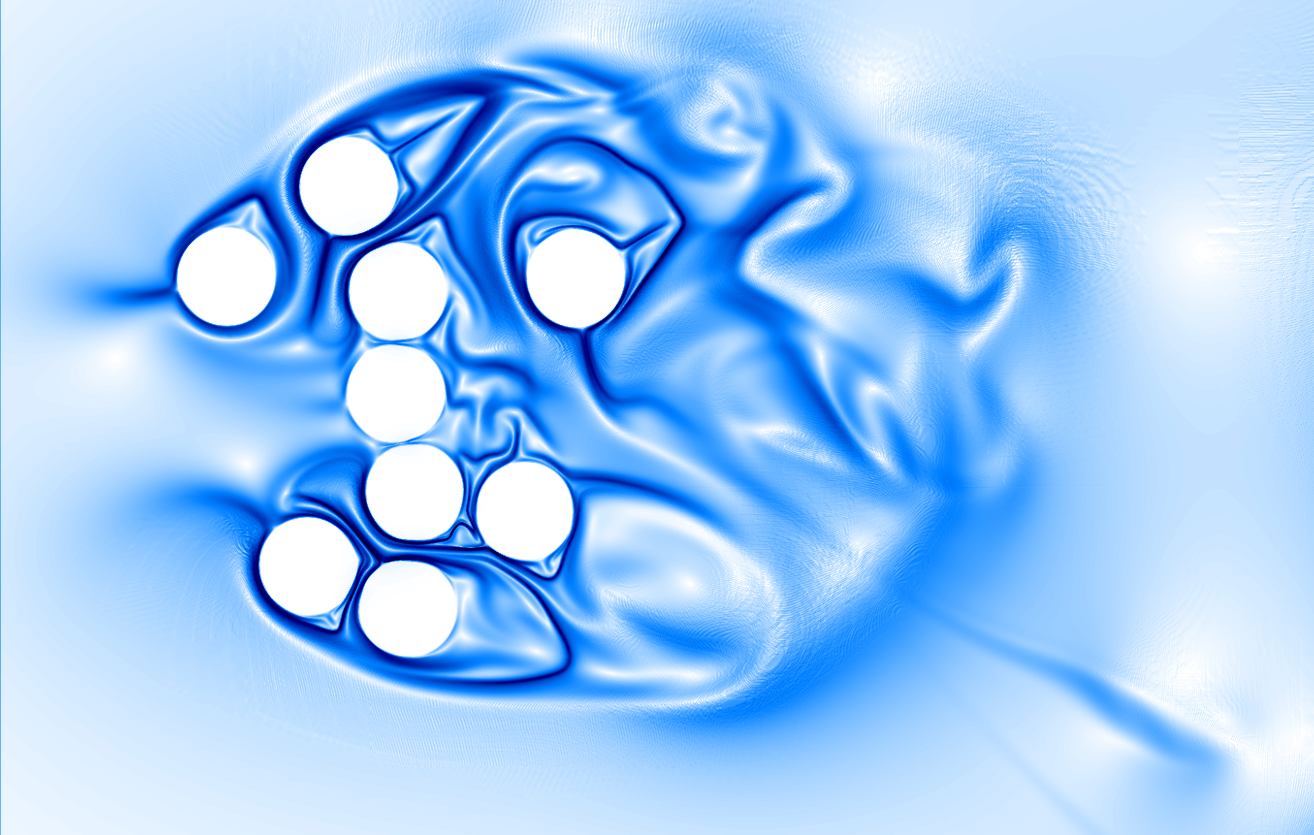}
  \includegraphics[width=0.24\textwidth]{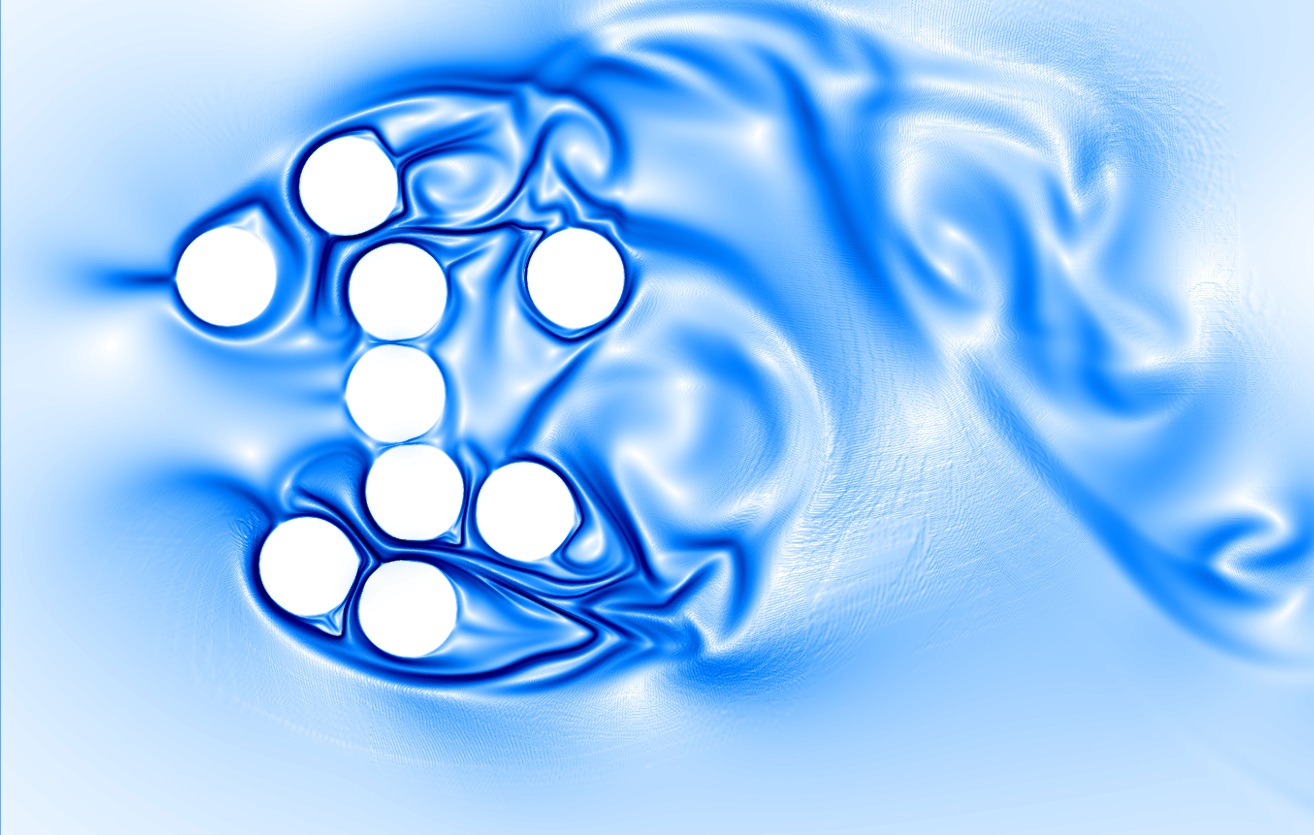}
  \includegraphics[width=0.24\textwidth]{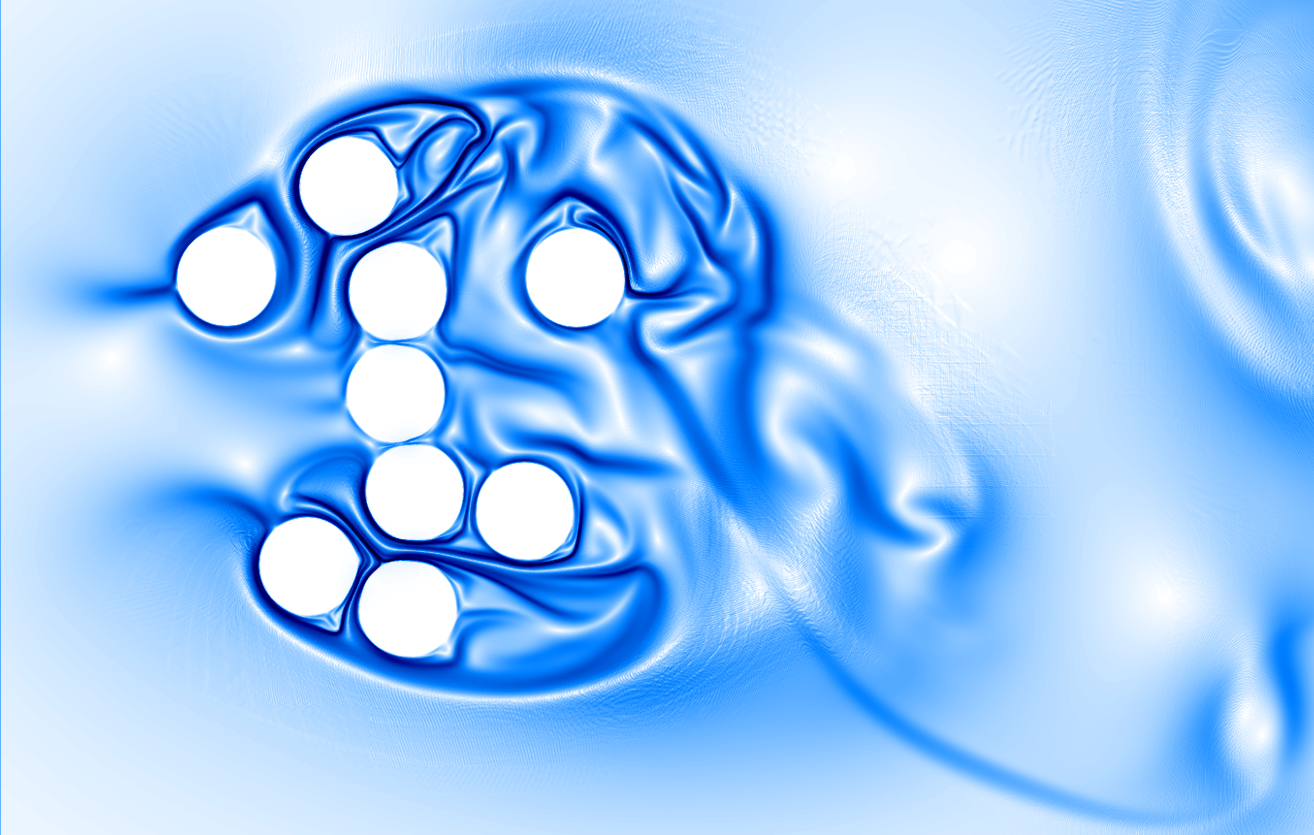}
  \includegraphics[width=0.24\textwidth]{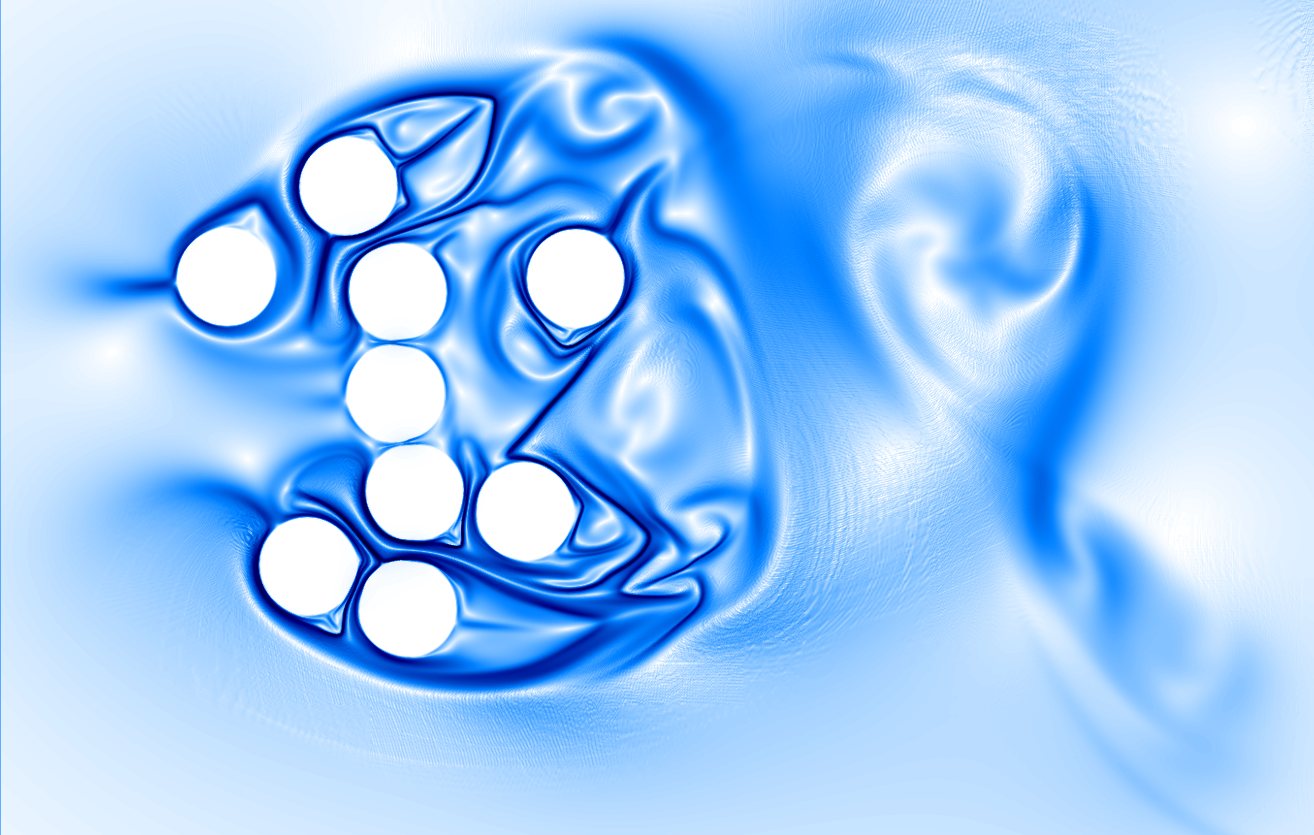}
  \subcaption{}
  \label{fig:lcs2}
\end{subfigure}

\begin{subfigure}{\textwidth}
\centering
  \includegraphics[width=0.24\textwidth]{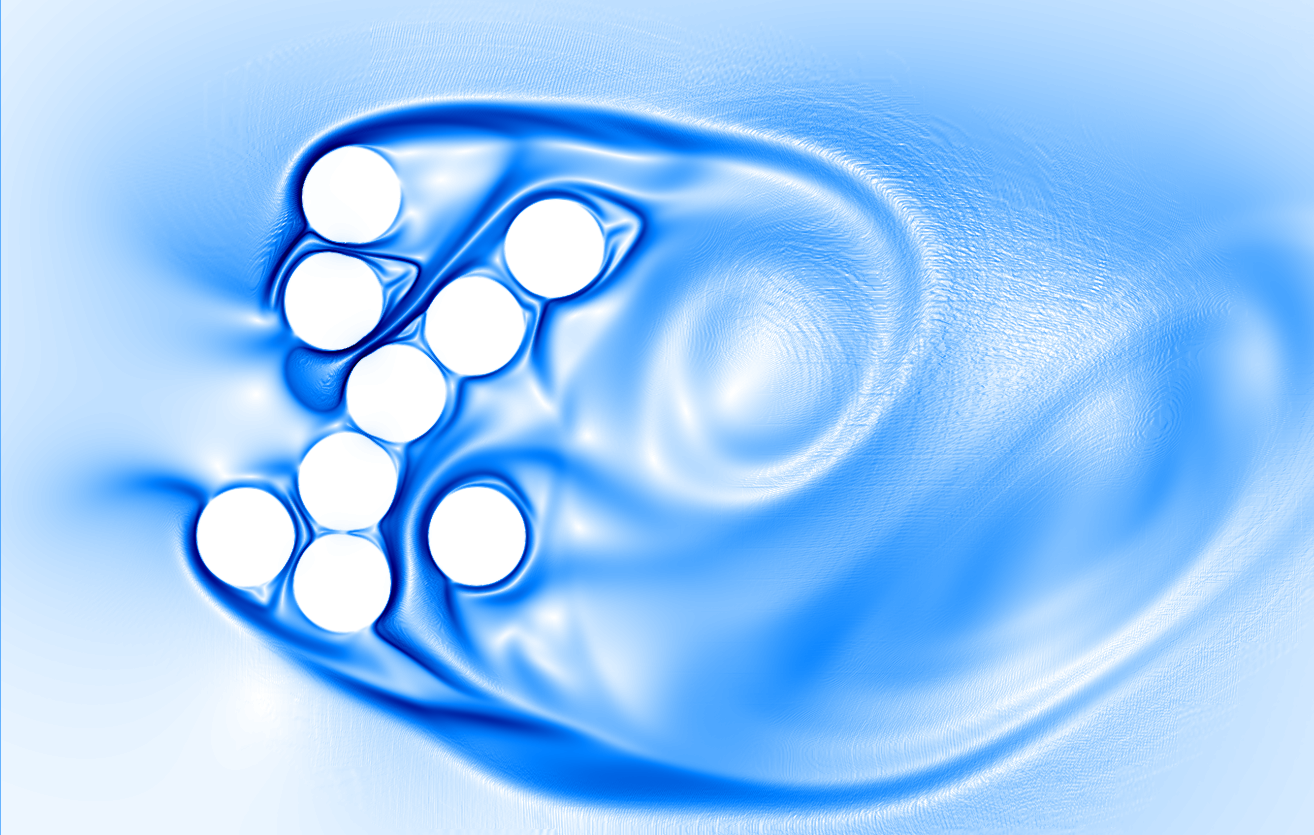}
  \includegraphics[width=0.24\textwidth]{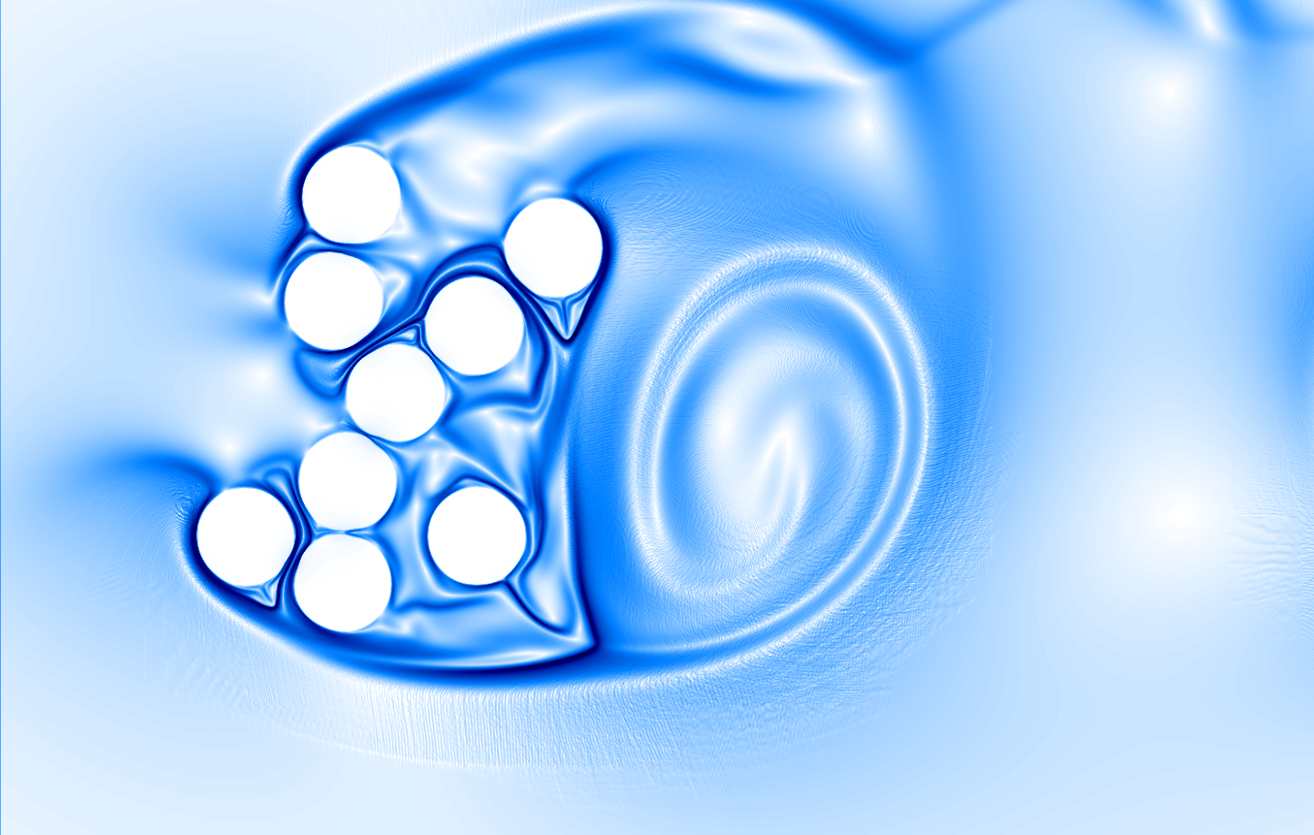}
  \includegraphics[width=0.24\textwidth]{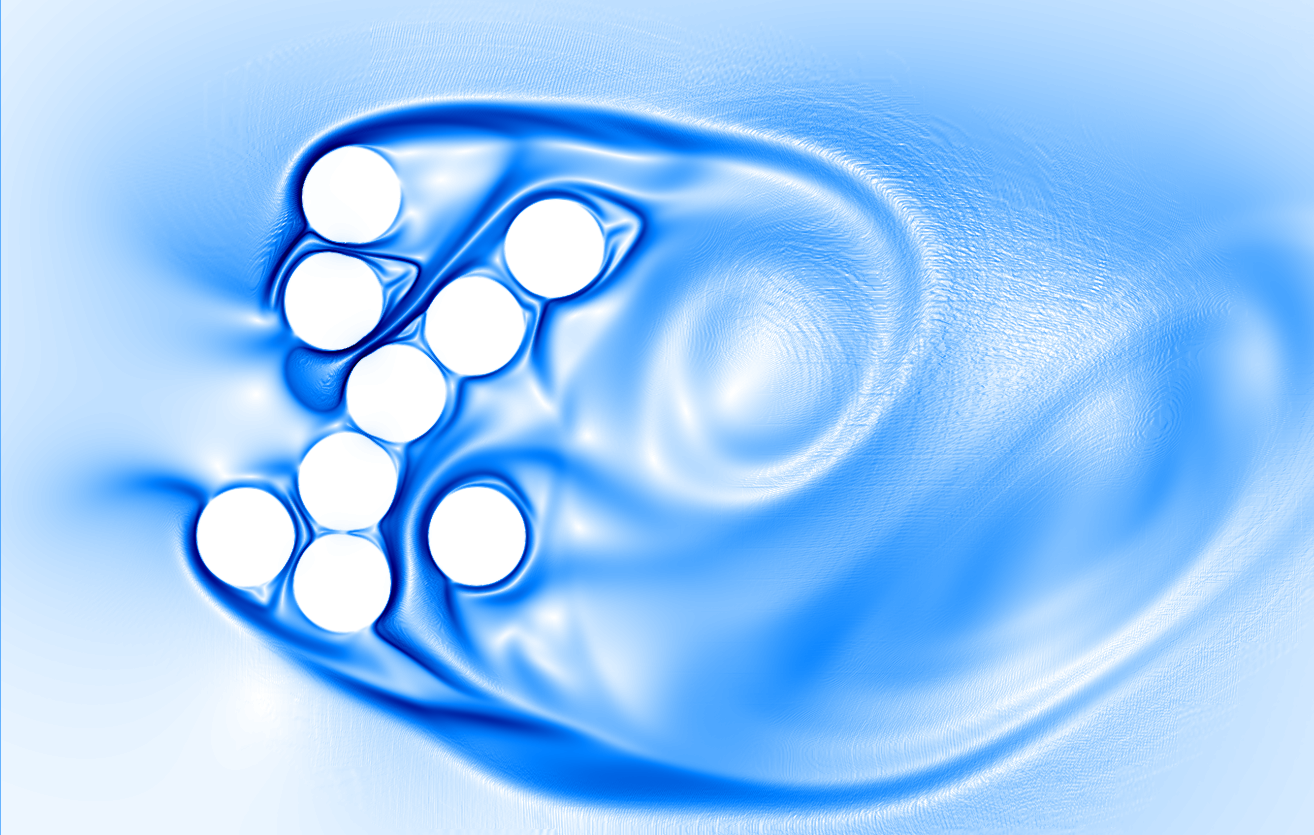}
  \includegraphics[width=0.24\textwidth]{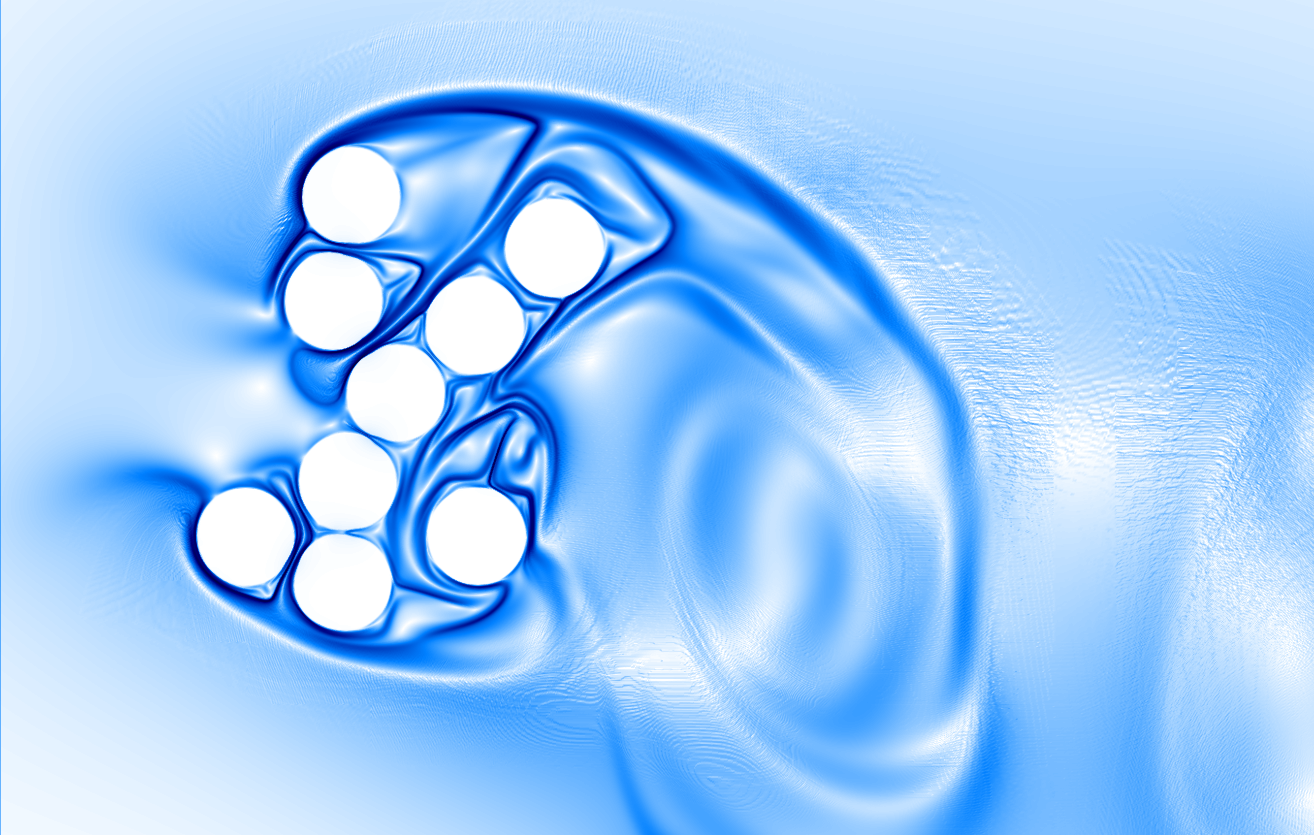}
  \subcaption{}
  \label{fig:lcs3}
\end{subfigure}

\begin{subfigure}{\textwidth}
\centering
  \includegraphics[width=0.24\textwidth]{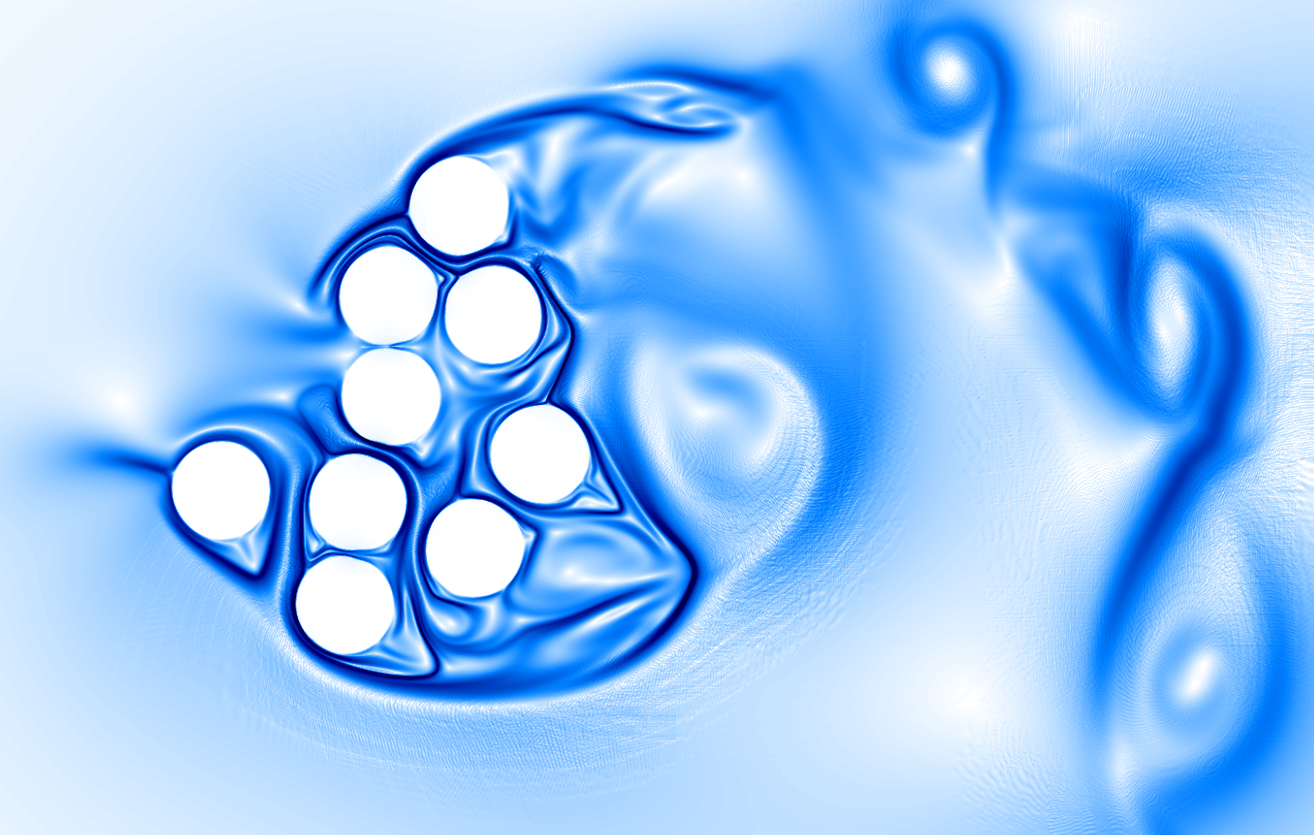}
  \includegraphics[width=0.24\textwidth]{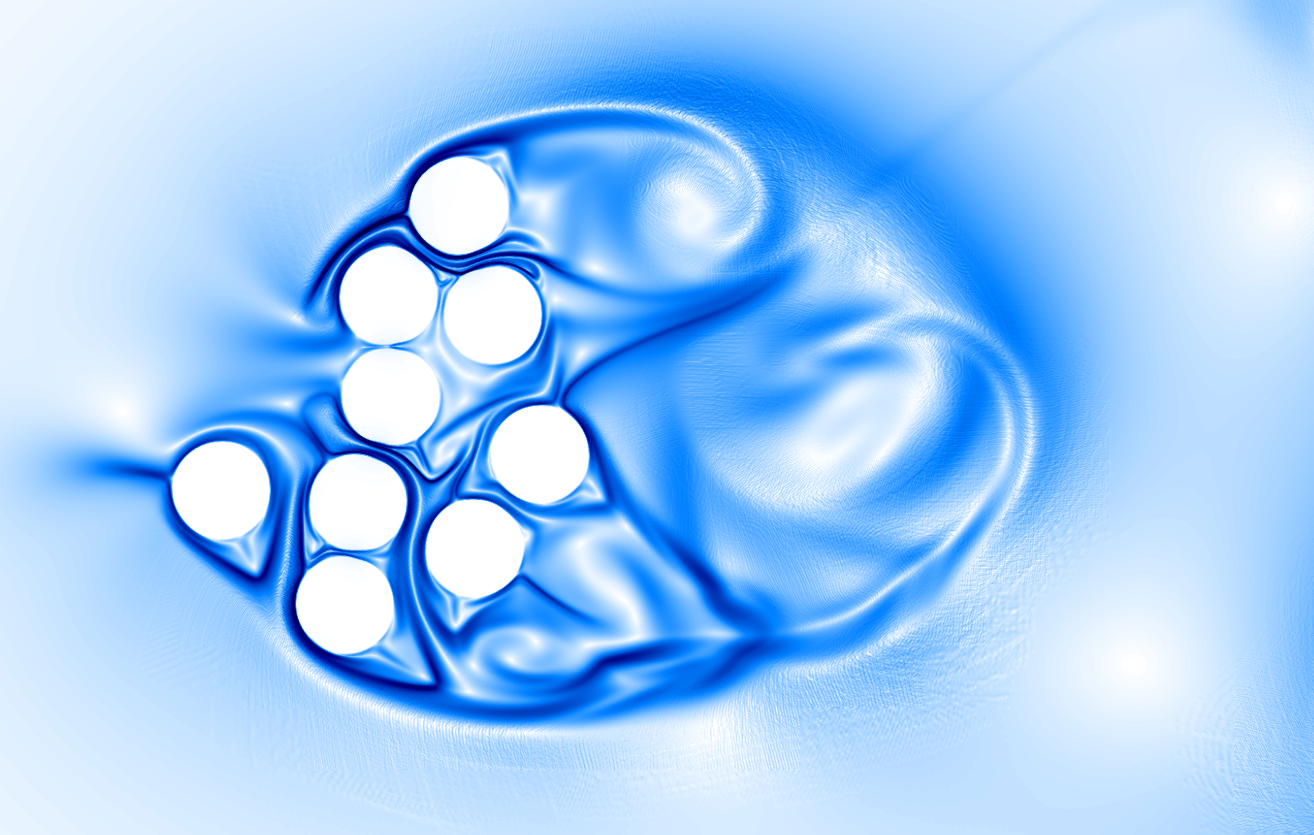}
  \includegraphics[width=0.24\textwidth]{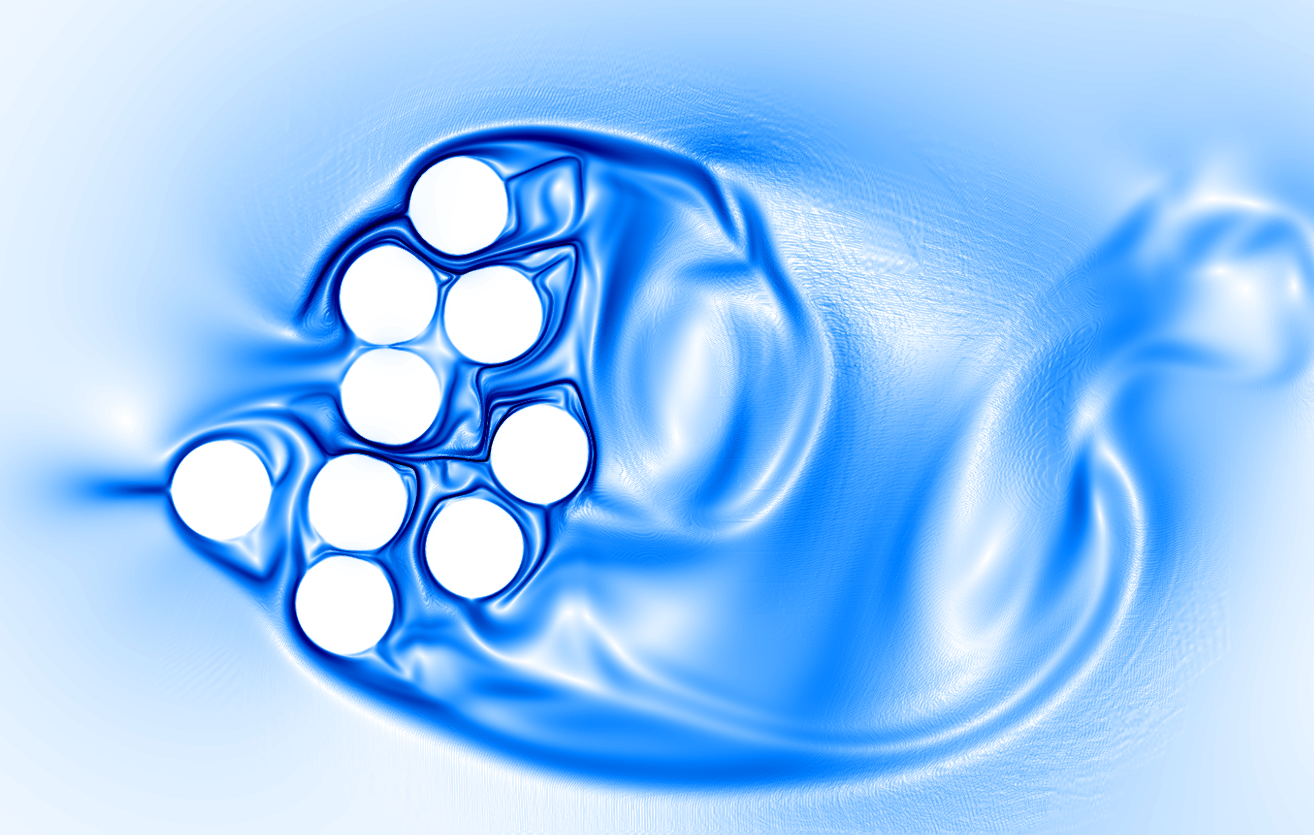}
  \includegraphics[width=0.24\textwidth]{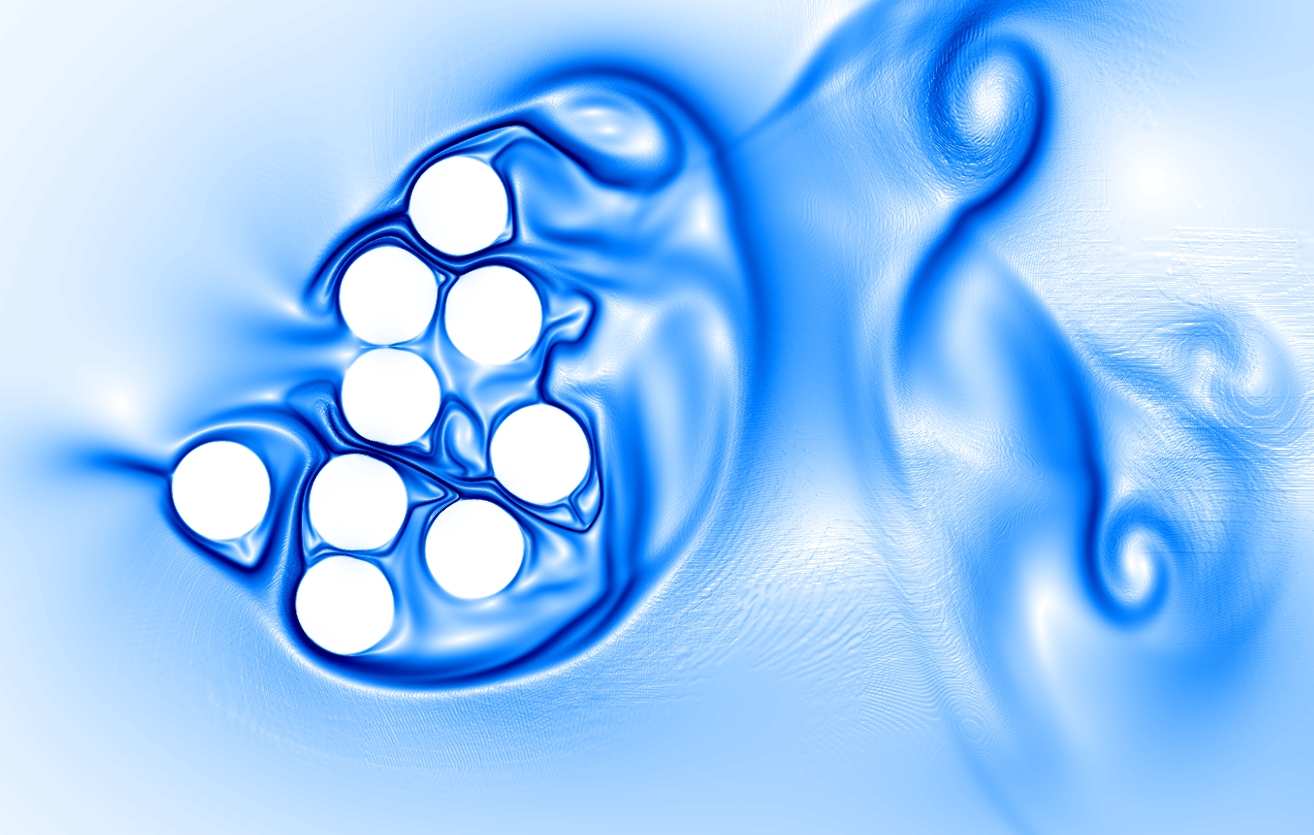}
  \subcaption{}
  \includegraphics[width=\textwidth]{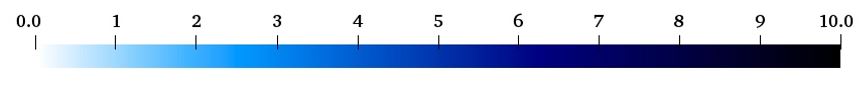}
  \label{fig:lcs4}
\end{subfigure}
	\caption{Forward FTLE field computed from DNS for individuals A through D at $t = t_{o}$, $t_{o} +0.25T$, $t_{o} + 0.5T$ and $t_{o} + 0.75T$, where $T$ is the shedding time period for each of the arrays. The darkest regions represent the ridges of the Lagrangian Coherent Structures.}
\label{fig:LCS}
\end{figure}

We now examine LCSs computed at various times during a complete shedding period for each of the four selected arrangements. The Finite Time Lyapunov Exponent (FTLE) field was computed for each of the four arrays in the steady shedding state, and the resulting images are shown in Figure \ref{fig:LCS}. An integration time window of approximately 1/10th of the shedding time period was used at 4 different stages in a shedding cycle to capture the variation of the LCSs with time. The dark lines visible in the FTLE fields correspond to the LCSs, where the threshold for distinguishing a ridge as a LCS is taken to be approximately 70\% of the maximum \citep{Shadden2011,Giudici2021}. The most consistent LCSs are found in Figure \ref{fig:lcs1}, where the coherent structures maintain their shape throughout the shedding period. This is especially evident in the interior of the array, where the ridges visible in the FTLE field do not vary with time. In Figures \ref{fig:lcs2}, \ref{fig:lcs3}, and \ref{fig:lcs4}, the extent of time-variation in the LCSs keeps increasing as we go from B to D. Since there is minimal flux across an LCS, the scenarios with steady LCSs are more capable of trapping particles. This implies that the arrangement in \ref{fig:lcs1} should be highly effective in retaining particles in the steady shedding state, which agrees well with the observations from \S\ref{sssec:LPT-ss}.

{\subsection{Manually designed arrangements}
\label{subsec:manual}

Based on the characteristics observed for the four selected Pareto-optimal arrays, we now examine several manually-designed arrays and assess their performance in relation to the designs obtained via optimization. Since drag and enstrophy display a general monotonic relationship with each other (Figure~\ref{fig:gen33results}), many of the array characteristics that maximize drag conflict with those that minimize enstrophy. This makes it difficult to design arrays that attain high drag and low enstrophy simultaneously, which are criteria that can benefit coastal-protection designs as discussed earlier. One of the main observations in the previous sections has been the existence of internal boundary layers within the porous arrays, which remain attached to neighbouring downstream cylinders. Based on experimental observations by \citet{Sumner2000}, this phenomenon occurs when the individual cylinders are separated by a relatively small distance (i.e., $P/d = 1 \text{ to } 1.25$), and when the angle they make with each other relative to the direction of the freestream velocity is small ($\alpha = 0 \text{ to } 20\degree$). 

The manually-designed arrays examined here are shown in Figure~\ref{fig:manual_arrangement}, and the resulting drag and enstrophy values are shown in Figure~\ref{fig:manual} along with a few regular arrangements from the initial optimization generation for comparison. The array in Figure \ref{fig:manSquare} includes a regularly-spaced grid arrangement of individual cylinders in a square shape. The spacing is kept within the $P/d < 1.25$ limit described by \citet{Sumner2000} in both the streamwise and cross-stream directions. Based on our prior observations, we expect that this base configuration will lead to low enstrophy, given the existence of internal channels that allow for the formation of internal boundary layers that remain attached to neighbouring cylinders. Furthermore, the relatively small projected frontal area as well as considerable internal flux through the array can be expected to lead to low drag. Both these expectations are confirmed in Figure~\ref{fig:manual}, which indicates that the performance of the array shown in Figure~\ref{fig:manSquare} closely resembles that of Pareto-optimal individual A, which is located at the left extreme of the Pareto front.

The second array, shown in Figure~\ref{fig:manSquareWonky} is a slight modification of the base square array, where the cylinders are spaced farther apart in the cross-stream direction. This has two effects: 1) increased flux through the interior results in decreased drag; and 2) the internal boundary layers of oppositely signed vorticity are farther apart, which reduces the extent of vorticity-cancellation via diffusion, resulting in increased wake enstrophy. Both these expectations are confirmed from Figure~\ref{fig:manual}, which indicates that this design entails lower drag and higher enstrophy than the base square array. The third design shown in Figure~\ref{fig:manSquareDiamond} was obtained by rotating the base square array by $45\degree$. The increase in projected frontal area can be expected to lead to increased drag. Additionally, the internal boundary layers no longer remain attached to neighboring cylinders and exhibit unsteady behaviour, since $\alpha=45\degree$ now exceeds the range prescribed by \citet{Sumner2000} ($\alpha < 20\degree$). Thus we expect a notable increase in both drag and enstrophy compared to the base square array, which is confirmed from Figure~\ref{fig:manual}. 

While the first design is relatively simple and attains the expected low-drag low-enstrophy performance, it does not meet the requirement of high-drag low-enstrophy that a coastal-protection design might benefit from. The two modifications to the base square array also do not exhibit the required characteristics, since either the drag remains too low or the enstrophy increases noticeably. The fourth design shown in Figure~\ref{fig:manDcyl} was based on a symmetric rearrangement of Pareto-optimal Individual D. We observe that while it leads to the expected increase in drag on account of the larger projected frontal area, it also leads to a considerable increase in enstrophy compared to the base square array. 

To counter this rise in enstrophy, the design shown in Figure~\ref{fig:manSigma} was developed with the aim of achieving both high drag and low wake enstrophy simultaneously. The projected frontal area for this configuration is similar to that of the D-shaped array described previously, which is beneficial with regard to high drag. However, this configuration no longer allows flow to pass through the array, which removes the possibility of enstrophy reduction via interactions among internal boundary layers. Thus, an alternate means of reducing enstrophy was devised in the form of slightly curved boundaries along the top and bottom edges of the array, and the large posterior cavity visible in Figure~\ref{fig:manSigma}. The curved boundaries allow the outermost boundary layers to remain attached and give rise to strong primary vortices shed from the top and bottom of the array, which get redirected towards the posterior cavity. As these primary vortices interact with the posterior cavity, they generate strong secondary vortices as can be observed in Figure~\ref{fig:manSigma}. This interaction between the primary and secondary vortices contributes to vorticity diffusion in the same manner as achieved by internal boundary layers of opposing vorticity, and results in lower wake enstrophy. This particular design was observed to perform well with regard to both metrics, with drag being comparable to that of the rotated square array from Figure~\ref{fig:manSquareDiamond} and enstrophy being only slightly higher than that for the base square array, indicating that it may be a suitable candidate for coastal structure design. We note that the design lies on the Pareto-front in Figure~\ref{fig:manual}, indicating that it is difficult to exceed the performance of the designs obtained via optimization. 

In addition to the 9-cylinder arrays, an additional 18-cylinder design was devised as shown in Figure~\ref{fig:man18cyl} (with $C_d = F_{Drag}/(0.5\rho U_\infty^2 \cdot 18d)$), to determine whether the observations presented here may extend to larger arrays. The design of this porous array specifically aims to reduce wake enstrophy by promoting interactions among boundary layers of opposite vorticity that form in the interior. We note from Figure~\ref{fig:manual} that the enstrophy for this design is comparable to, and lower than that of several 9-cylinder arrays, including the two regular arrangements from the initial generation.

\begin{figure}
\centering
\begin{subfigure}{0.32\textwidth}
  \centering
	\includegraphics[width=\textwidth]{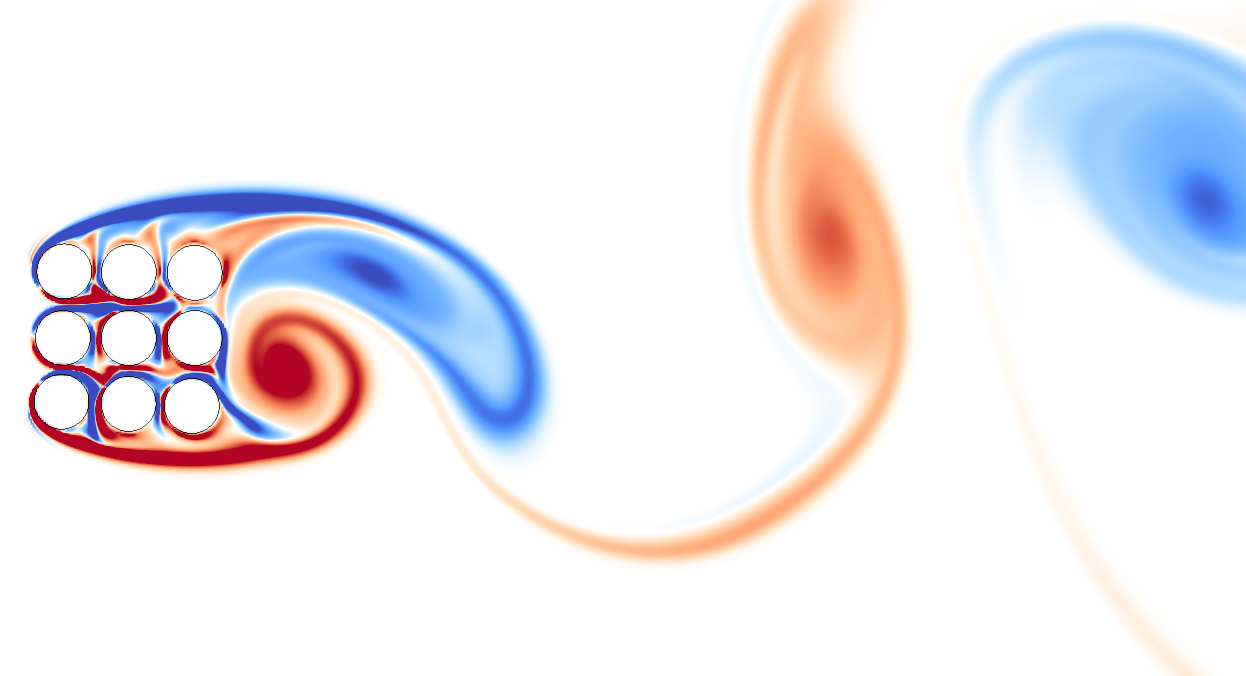}
  \caption{}
  \label{fig:manSquare}
\end{subfigure}
\begin{subfigure}{0.32\textwidth}
  \centering
	\includegraphics[width=\textwidth]{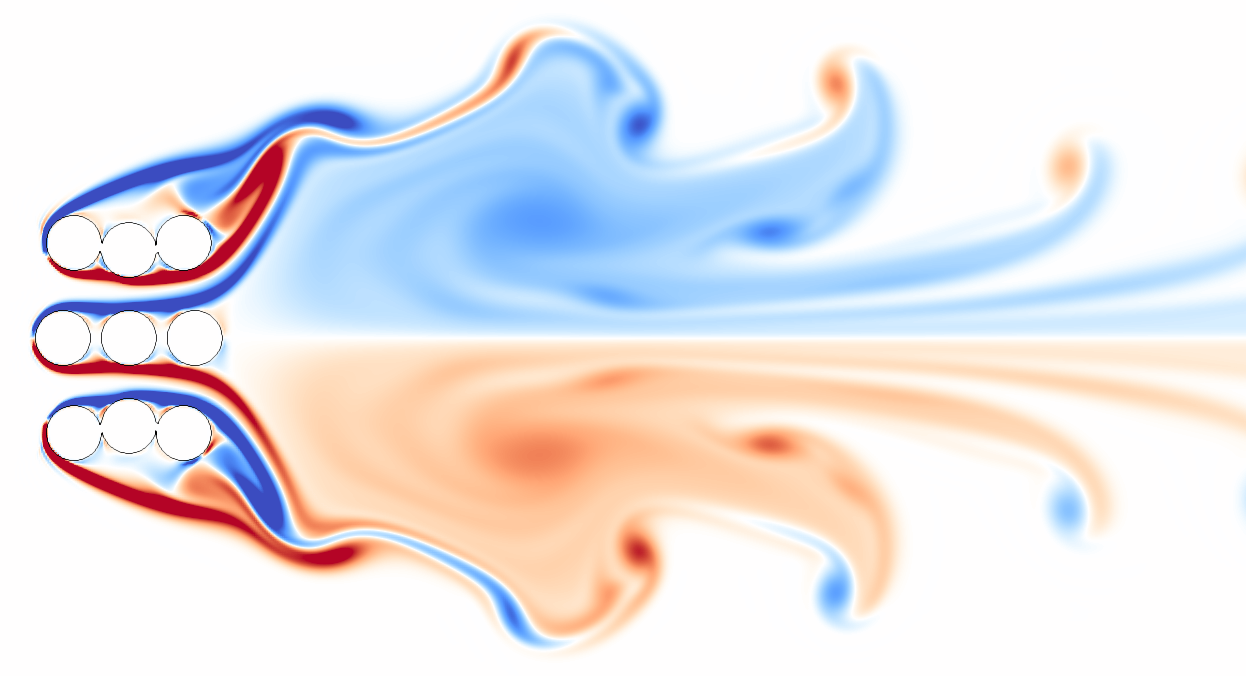}
  \caption{}
\label{fig:manSquareWonky}
\end{subfigure}
\begin{subfigure}{0.32\textwidth}
  \centering
	\includegraphics[width=\textwidth]{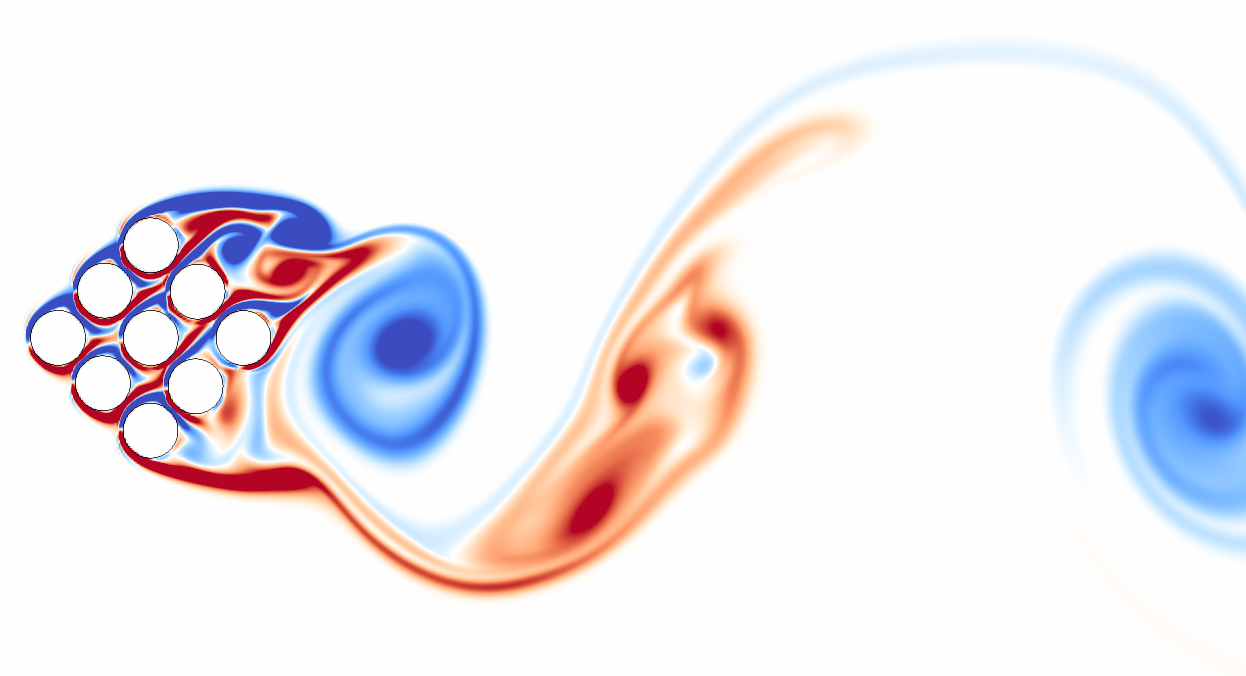}
  \caption{}
\label{fig:manSquareDiamond}
\end{subfigure}
\begin{subfigure}{0.32\textwidth}
  \centering
	\includegraphics[width=\textwidth]{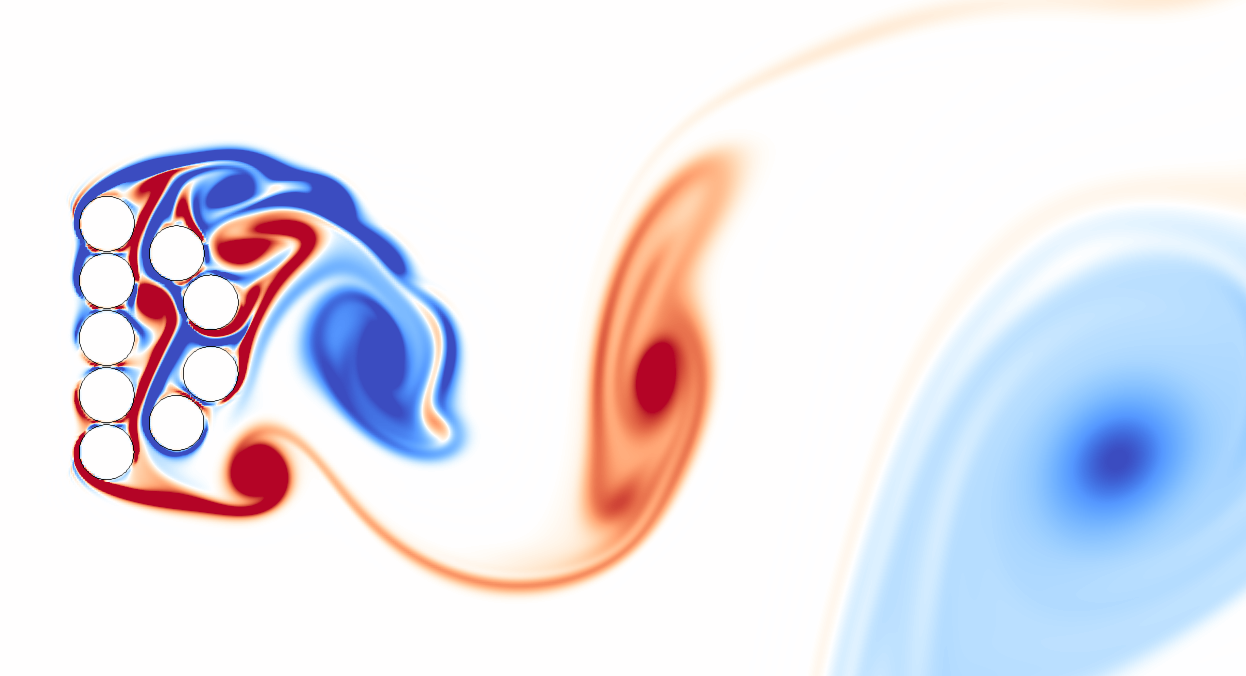}
    \caption{}
\label{fig:manDcyl}
\end{subfigure}
\begin{subfigure}{0.32\textwidth}
  \centering
	\includegraphics[width=\textwidth]{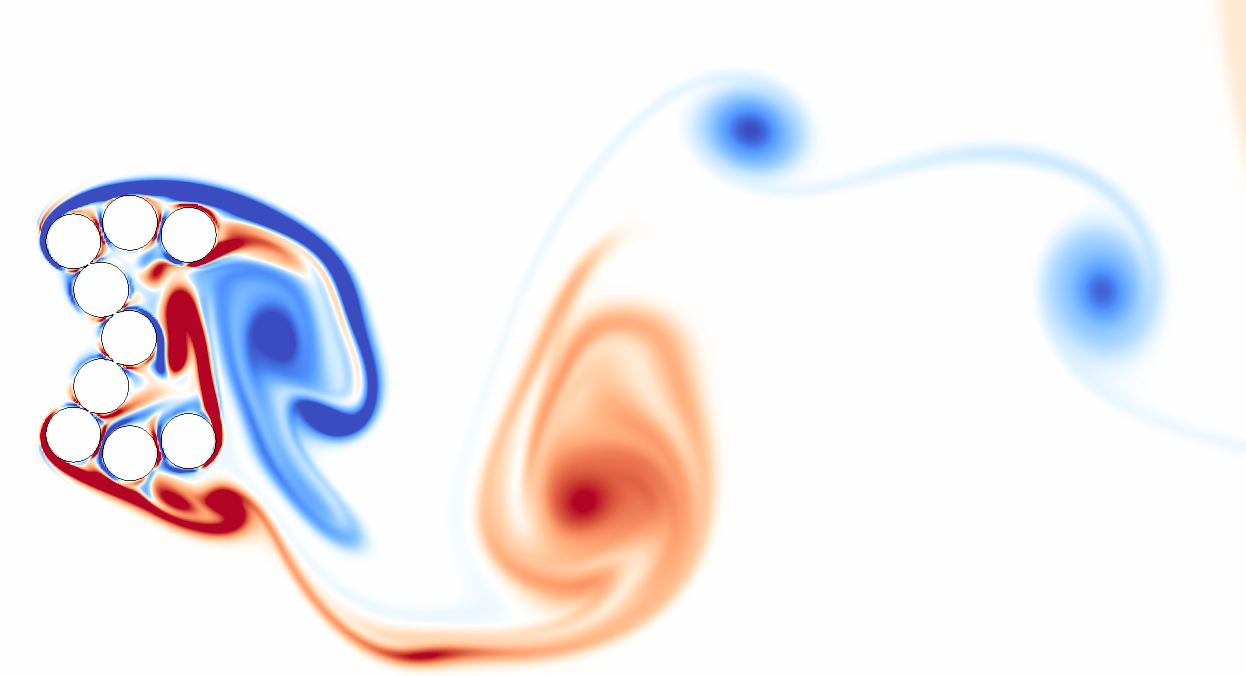}
    \caption{}
\label{fig:manSigma}
\end{subfigure}
\begin{subfigure}{0.32\textwidth}
  \centering
	\includegraphics[width=\textwidth]{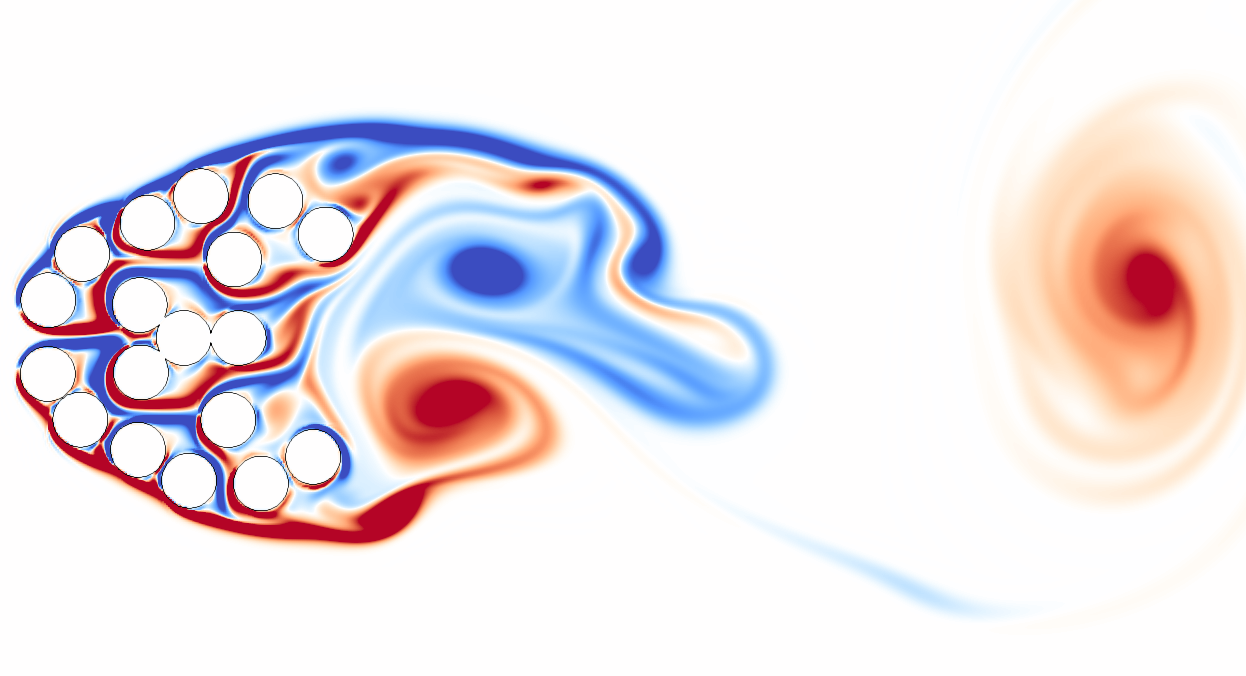}
    \caption{}
\label{fig:man18cyl}
\end{subfigure}
	\caption{Vorticity from DNS for several manually-designed arrays. The corresponding colourbar is provided in Figure~\ref{fig:4wakes}.}
\label{fig:manual_arrangement}
\end{figure}

\begin{figure}
    \centering
    \includegraphics{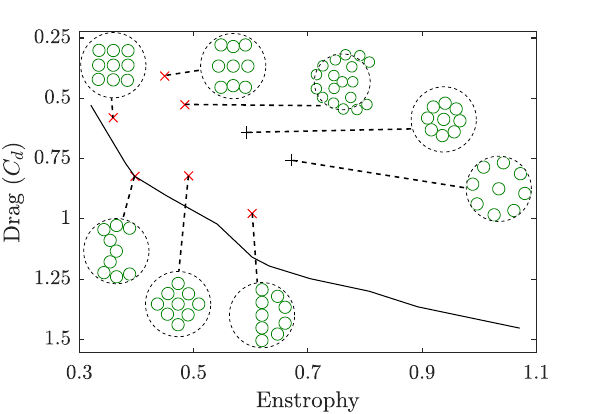}
	\caption{The drag and enstrophy values obtained from DNS for the manual arrangements shown in Figure~\ref{fig:manual_arrangement} ($\times$). The values for two of the configurations from the initial generation are also shown (+). The Pareto front from Figure~\ref{fig:gen33} is included for comparison.}
    \label{fig:manual}
\end{figure}


\section{Discussion}
\label{sec:discussion}

The present study extends earlier work by \cite{Kazemi2017} and \cite{Kazemi2018}, where the porosity and spacing between individual cylinders were varied for an array of 9 cylinders. In other experiments involving porous cylinder arrays, \cite{Chen2012} considered cylinders of various shapes, diameters, and patch densities to study the influence of flow blockage on the velocity, turbulent kinetic energy and sediment deposition in the wake. They found that flow blockage is an effective predictor of the extent of the diminished velocity region and streamwise exit velocity directly behind the cluster. However, we find that this metric is not the sole predictor of the diminished velocity region, since individuals B and D are similar to each other with regard to flow blockage, but they give rise to diminished velocity regions of different lengths as is evident in Figures~\ref{fig:midline}, \ref{fig:midline_sim}, and~\ref{fig:linecut}. Another experimental study that considered the drag acting on individual cylinders in a group was conducted by \cite{Shan2019}, where they placed randomly- as well as uniformly-distributed tree models in a uniform inflow. They demonstrated that in uniform arrangements, the trees on the frontline experienced the highest drag, while the rest were increasingly sheltered, i.e., they did not contribute significantly to the total drag. However, in the case of random distributions, they found significant variation in the locations of cylinders experiencing the largest drag. A similar conclusion may be drawn from the sensitivity analysis presented in Figure \ref{fig:sense-highlight}, where the cylinders that most influence the drag are not always the ones in the front of the array. A prior study by \citet{Cheer1987} also draws a correlation between drag and spacing in cylinder arrays, albeit at very low Reynolds numbers. They found that bristled appendages of small organisms, which were treated as porous cylinder arrays, functioned to provide locomotion in certain conditions and were used to catch particulate food in other conditions. They concluded that it was a combination of `leakiness' and frontal area that determined the hydrodynamic forces acting on the appendages (i.e., the simplified cylinder arrays), which is similar to observations presented here. Overall, the results presented here demonstrate that even minor changes to internal flow patterns within a porous array can lead to substantial changes in the wake enstrophy and drag. Thus, studying the influence of the relative geometrical arrangement of cylinders can lead to an improved understanding, and consequently to better models of flow in porous arrays.

We note that there is no strict reason for specifically selecting nine cylinders for the arrays in the present study, except as a balance between sufficient generality and too many degrees of freedom. As discussed earlier, mangrove trees usually consist of a central trunk surrounded by a number of auxiliary roots, and the arrays used here are meant to be simplified model representations of this structure. Choosing more cylinders for the arrays would lead to an increased degree of freedom in the array configurations, in turn allowing more diverse configurations to arise. However, the increased dimensionality would pose difficulties with regard to optimization and computational cost. Thus, nine cylinders were selected as a reasonable compromise. Furthermore, we note that the stopping criterion for the NSGA II optimization algorithm is generally user-defined. In the present work, there were no significant changes to the Pareto front for several generations leading up to generation 33, which led to the decision to stop further iterations. The convergence of the optimization is further supported by the fact that none of the manually-designed arrangements presented in \S\ref{subsec:manual} perform significantly better than the Pareto-optimal individuals.

In the sensitivity analysis presented in \S\ref{subsec:sense}, we observed that enstrophy shows a stronger dependence on $\theta_i$, whereas drag shows a stronger dependence on $r_i$. This can be explained by the fact that changing $\theta_i$ does not impact the projected frontal area of the array significantly, whereas perturbing $r_i$ can result in notable changes to the area depending on a cylinder’s location. This makes drag more sensitive to radial perturbations. Enstrophy on the other hand was observed to be sensitive to disruption in boundary layer attachment among neighbouring cylinders. Such disruptions occurred more readily upon perturbing the cylinders in the azimuthal direction than in the radial direction, which explains the heightened sensitivity of enstrophy to $\theta_i$.

One aspect that differs between the present study and isolated mangrove trees is the tendency of natural peripheral roots to be distributed in concentric arrangements around the central trunk. This is likely related to the function of peripheral roots as mechanical support for the main trunk against wind and waves, which is not considered as an objective in the present optimization. We also note that the results and observations presented in the current work were obtained at a relatively modest Reynolds number, and may need to be revisited when considering significantly higher Reynolds numbers. Nonetheless, the optimization procedure and analyses presented here, which leverage a complementary combination of numerical and experimental approaches, can be extended to other related scenarios in an effort to minimize the labour-intensive and time-consuming nature of discovering optimal configurations.

\section{Conclusion}
\label{sec:conclusion}

In the present work, two-dimensional Direct Numerical Simulations of flow around porous cylinder arrays have been coupled with a multi-objective optimization algorithm to study the dependence of drag and wake enstrophy on the relative positioning of individual cylinders. In a group of nine cylinders placed in uniform flow, the radial and azimuthal positions of eight surrounding cylinders with respect to a central cylinder are modified by the optimization algorithm in an attempt to simultaneously minimize enstrophy and maximize drag. These characteristics are desirable for coastal-defense structures, since low enstrophy is more conducive to promoting sediment deposition and suppressing resuspension, whereas high drag increases velocity deficit in the wake, mitigating the impact of the incoming flow. Drag and enstrophy show a general monotonic relationship, which makes multi-objective optimization a suitable approach for attaining the desired conflicting metrics simultaneously. The optimization process converges to a set of Pareto-optimal individuals after iterating through several generations. Four such individuals were selected for further analysis: a low-enstrophy low-drag individual (A), a high-enstrophy high-drag individual (D), and two individuals with characteristics in between the two extremes (individuals B and C).

Using Particle Image Velocimetry, the streamwise velocity along streamwise and cross-stream line cuts was obtained for the four selected arrays. Individual A was observed to display the least amount of velocity deficit, whereas individual D displayed the largest velocity deficit in the wake. The differences in drag and enstrophy among the four individuals are related to how the flow behaves within and around the carrays, which was analyzed further using data from the numerical simulations. Individual A was observed to have noticeable gaps within the array that created a path for the upstream flow to pass through the interior, which had the dual impact of reducing both drag and wake enstrophy. The lower enstrophy was determined to result from vorticity cancellation among boundary layers of oppositely signed vorticity, which formed due to flow being directed through closely spaced neighbouring cylinders in the array's interior. The high-drag configurations were observed to allow minimal flux through the porous interior, with the majority of the freestream being redirected around the arrays instead of passing through them. This resulted in shedding patterns that closely resembled those of rigid impervious objects. The high-drag arrays were also observed to have larger projected frontal areas.

A sensitivity analysis was conducted to determine the influence of an individual cylinder's position on the resultant drag and enstrophy. None of the perturbed individuals were better than the original Pareto-optimal individuals in both fitness metrics, which indicates that the Pareto-optimal solutions obtained using the multi-objective optimization are robust. The perturbations tended to have a greater influence on wake enstrophy than on drag, which indicates that enstrophy is more sensitive than drag to specific cylinder placement within porous arrays. The most sensitive cylinders were found to be those which could significantly alter the internal flow through the array when perturbed. The least sensitive cylinders tended to be isolated from other cylinders in the array.

To better understand the impact on particle sedimentation and erosion, tracer particles and Lagrangian coherent structures were used to examine particle transport through and around the porous arrays. The arrangement with lowest enstrophy (individual A) was found to retain the largest number of particles in the steady state shedding state, while the arrangement with a large recirculation region and a delayed start for vortex shedding (individual B) was found to retain a comparable number of particles in the initial transient state. High array-drag was observed to reduce particle retention in the quasi-steady shedding state, and these observations were supported by the behaviour of the Lagrangian coherent structures, where the ridges did not vary with time for individual A but displayed significant time-variation for individuals B, C, and D. Thus, although increased drag leads to lower streamwise wake velocity along the midline, the flow speed experienced by particles in the near-wake region can be relatively high on account of strong vorticity (and hence, high enstrophy) induced by the primary shedding vortices, making it more difficult for particles to sediment.

Based on the characteristics observed for the four selected Pareto-optimal arrays, several manually-designed array configurations were examined and their performance was compared to the designs obtained via optimization. The relationship between wake enstrophy and boundary layer interactions among neighbouring cylinders, and its dependence on the relative positioning of cylinders within the arrays was confirmed using simple designs resembling square array configurations, and even using an 18-cylinder configuration that was shown to generate low wake enstrophy. One of the 9-cylinder designs was developed with the aim of achieving both high drag and low wake enstrophy simultaneously, and relied on a large projected frontal area to maximize drag, and a large posterior cavity where the primary vortices being shed from the array generated strong secondary vortices resulting in reduced wake enstrophy. However, none of the manually-designed configurations were able to surpass the Pareto front, indicating that it is difficult to exceed the performance of the designs obtained via optimization.

We note that the selection of the Reynolds number ($Re_d = 500$) and porosity ($\phi = 0.316$) in the current work is not a limitation of the optimization method itself, but rather of using 2D simulations and for keeping computational cost manageable. This is because the overall flow patterns are influenced not by the individual cylinder diameter $d$, but by the array size which can include up to 5 cylinders lined up in the cross-stream direction, which corresponds to $Re=2500$. Results from 2D simulations may not be a valid representation at much higher $Re$ values due to the absence of the vortex-stretching term. The porosity value, on the other hand, is determined by the use of 9 cylinders within the given reference area, a number which was selected as a balance between sufficient generality and too many degrees of freedom in the array configurations.

In conclusion, the observations presented here indicate that the internal geometric arrangement of cylinders in porous arrays plays a critical role in regulating the internal flow patterns, which in turn influence the overall wake and drag characteristics. These aspects may not always be accounted for when using randomized array configurations or using pre-specified uniform arrangements, and relative cylinder positions should be factored in alongside other common parameters such as cylinder spacing, number, and diameter when studying such porous arrays. The findings from this study can prove to be useful for designing improved coastal protection infrastructure which can dissipate wave energy effectively, promote sediment deposition, and reduce erosion.

\section*{Declaration of Interests.} 
The authors report no conflict of interest.

\section*{Acknowledgements}
We gratefully acknowledge support from the National Science Foundation through Grant No. CBET 2103536 (Program Officer: Ronald D. Joslin). We thank Prof. Petros Koumoutsakos for providing access to computational resources via the Swiss National Supercomputing Center (CSCS) project ID `S929'.

\appendix

\section{Validating the Numerical Simulations}
\label{app:validation}

To validate the Navier-Stokes simulations used in this work, the wakes for staggered configurations of two cylinders in uniform inflow are compared with experimental results from \citet{Sumner2000}. They identified nine different flow patterns by varying the normalized distance $P/d$ between the cylinders and the angle $\alpha$ between them, where $P$ is the distance between the centres of the cylinders and $d$ is their diameter. The Reynolds number in these experiments is based on the individual cylinder diameter $d$, and ranges from $Re_d = U_\infty d/\nu= 800$ to $1900$. We provide a qualitative comparison of the wakes obtained for various flow patterns in Figure~\ref{fig:qualitative}, which shows that the flow patterns from the numerical simulations match well with those observed in the experiments.
\begin{sidewaysfigure}
\captionsetup{justification=centering}
\centering
\begin{subfigure}{0.19\textwidth}
  \centering
  \includegraphics[scale=0.12]{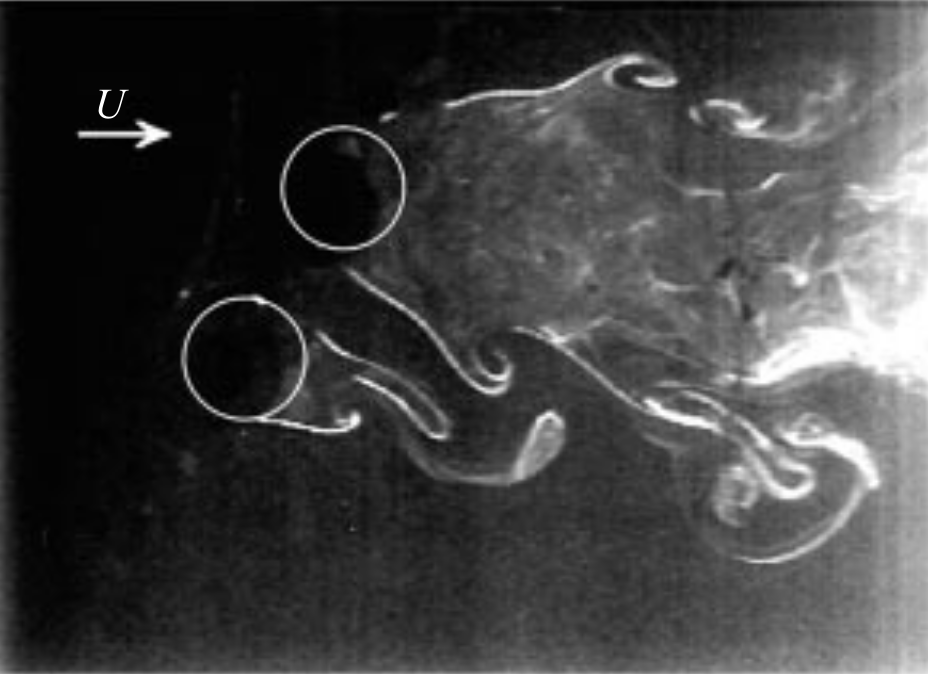}
\end{subfigure}
\begin{subfigure}{0.19\textwidth}
  \centering
  \includegraphics[scale=0.12]{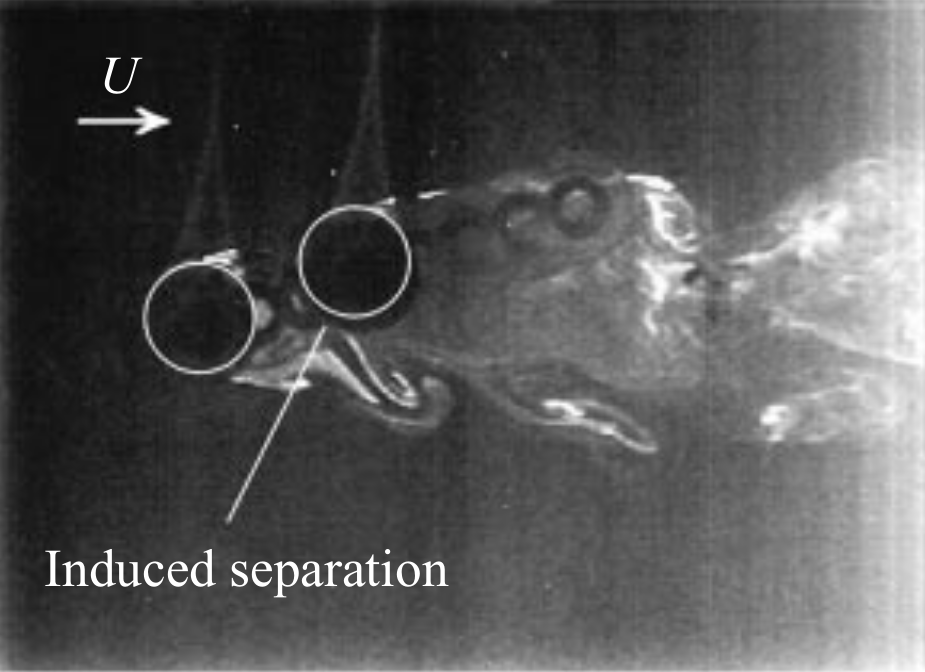}
\end{subfigure}
\begin{subfigure}{0.19\textwidth}
  \centering
  \includegraphics[scale=0.24]{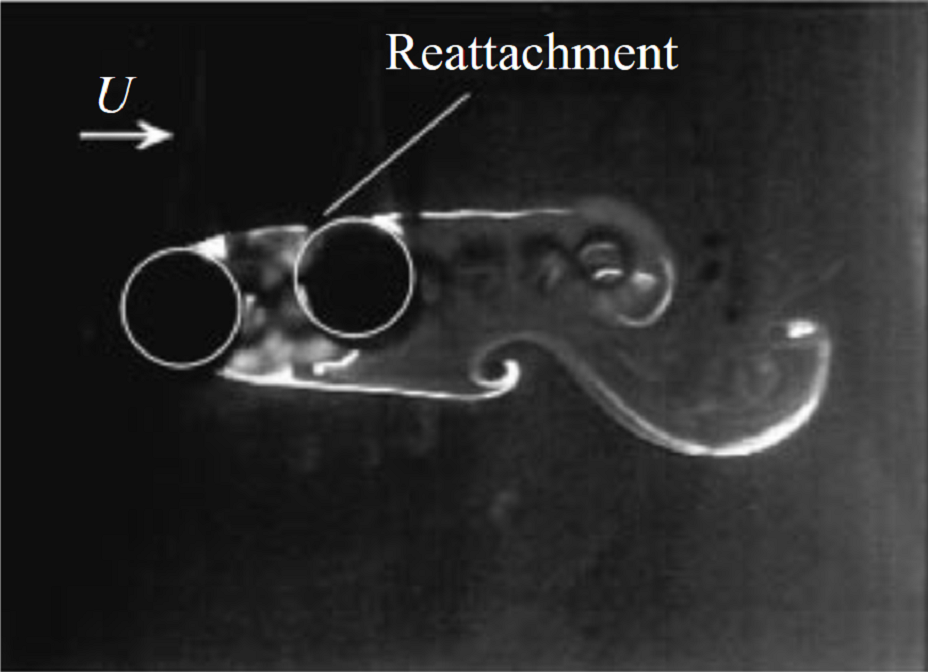}
\end{subfigure}
\begin{subfigure}{0.19\textwidth}
  \centering
  \includegraphics[scale=0.24]{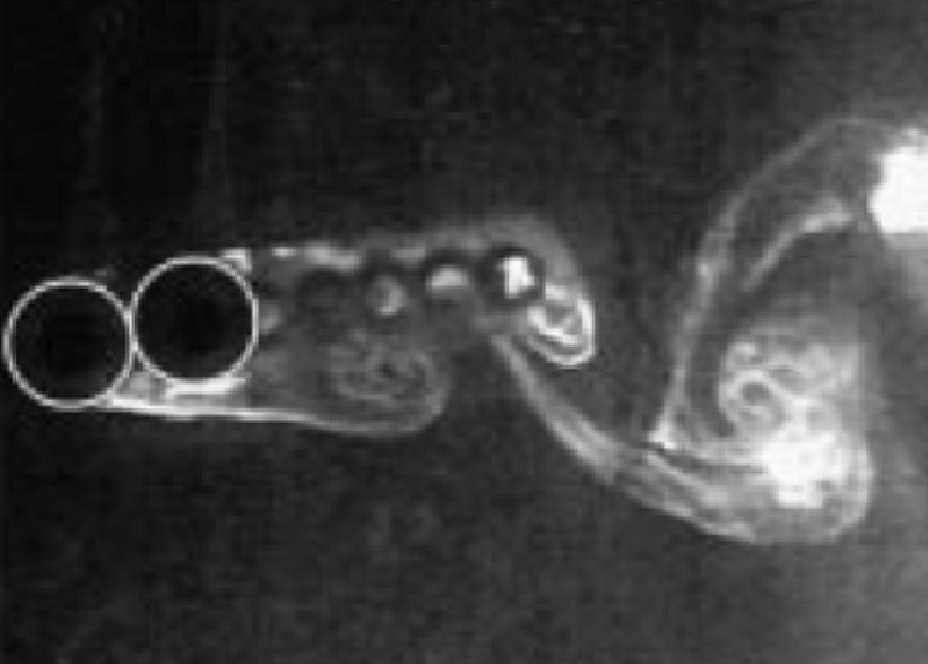}
\end{subfigure}
\begin{subfigure}{0.19\textwidth}
  \centering
  \includegraphics[scale=0.24]{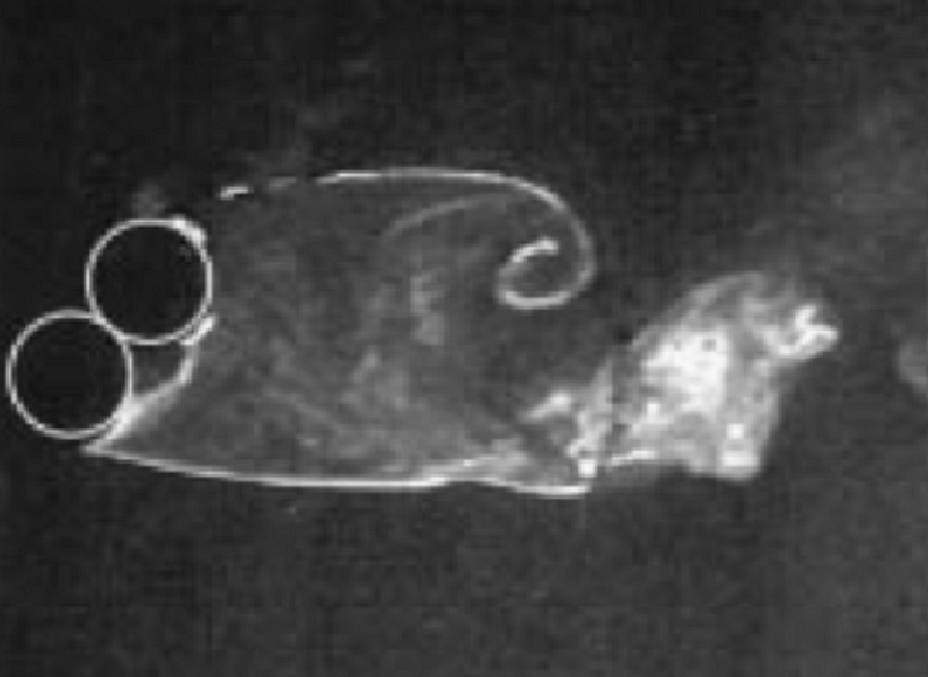}
\end{subfigure}
\\
\begin{subfigure}{0.19\textwidth}
  \centering
  \includegraphics[scale=0.16]{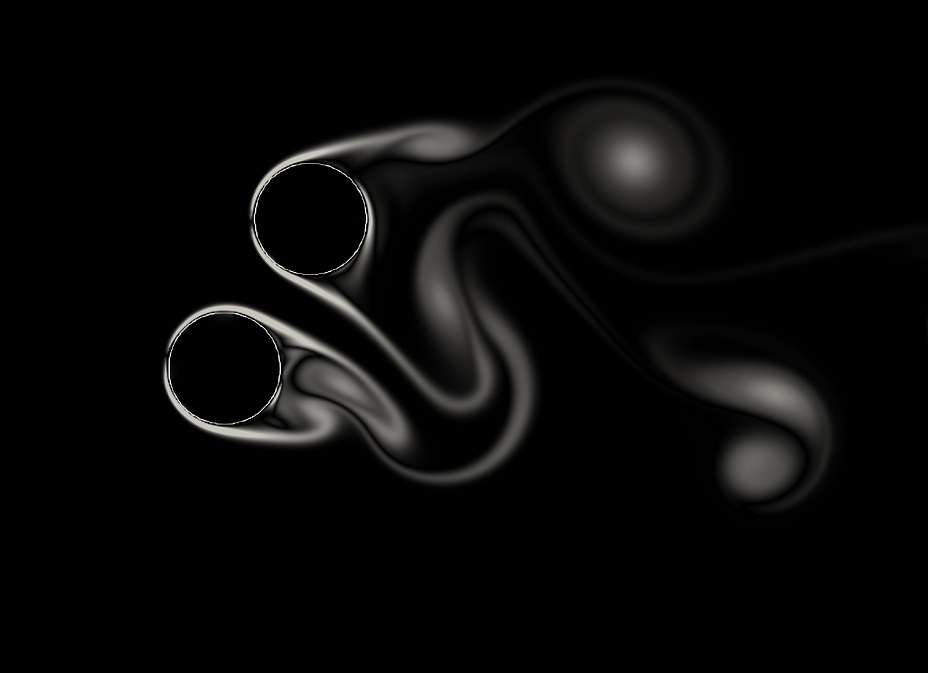}
  \subcaption{}
  \label{fig:a-exp}
\end{subfigure}
\begin{subfigure}{0.19\textwidth}
  \centering
  \includegraphics[scale=0.16]{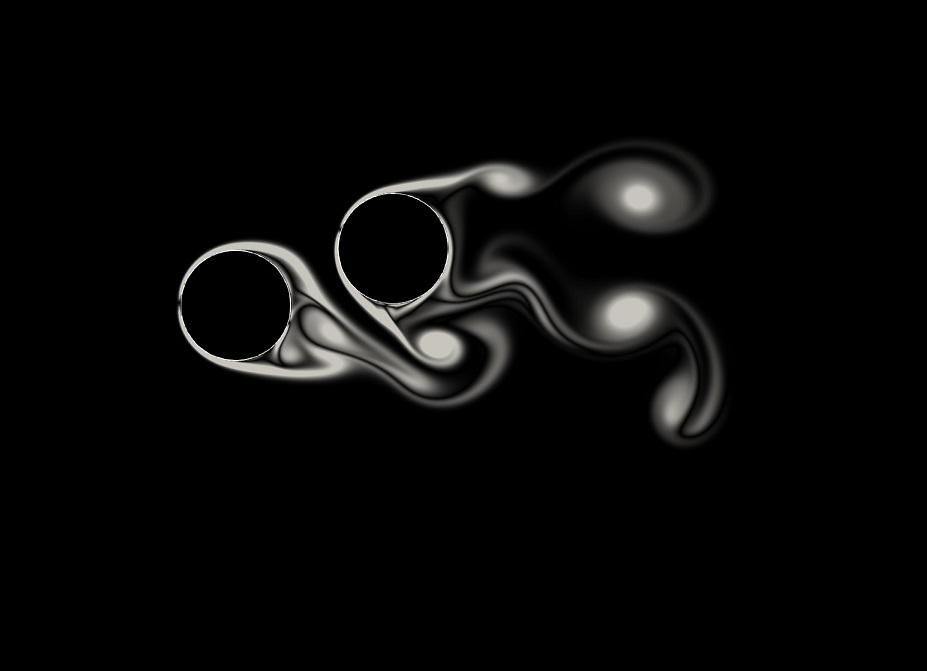}
  \subcaption{}
  \label{fig:b-exp}
\end{subfigure}
\begin{subfigure}{0.19\textwidth}
  \centering
  \includegraphics[scale=0.16]{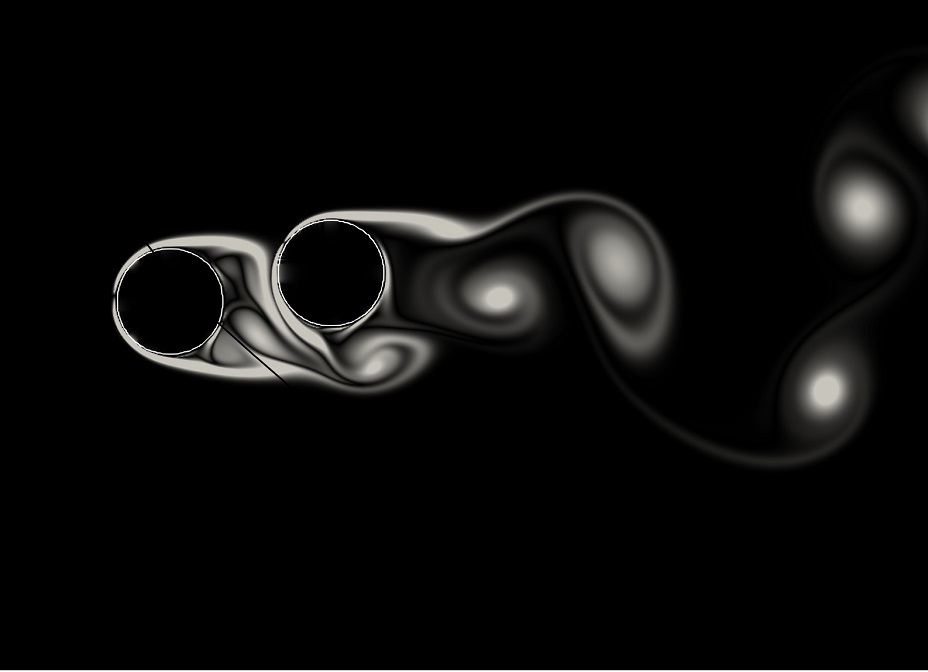}
    \subcaption{}
    \label{fig:c-exp}
\end{subfigure}
\begin{subfigure}{0.19\textwidth}
  \centering
  \includegraphics[scale=0.16]{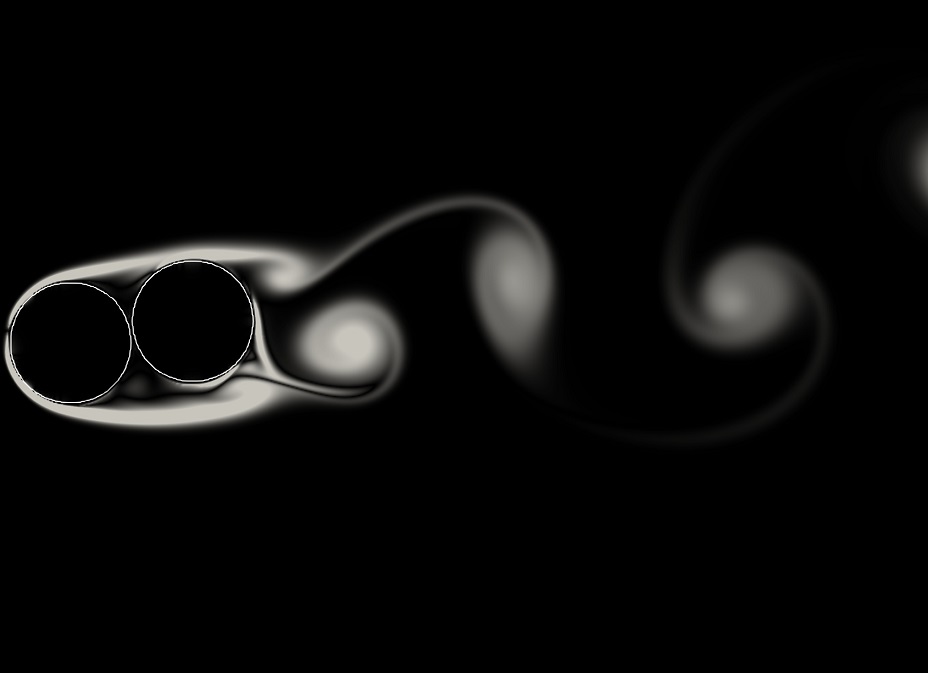}
    \subcaption{}
    \label{fig:d-exp}
\end{subfigure}
\begin{subfigure}{0.19\textwidth}
  \centering
  \includegraphics[scale=0.16]{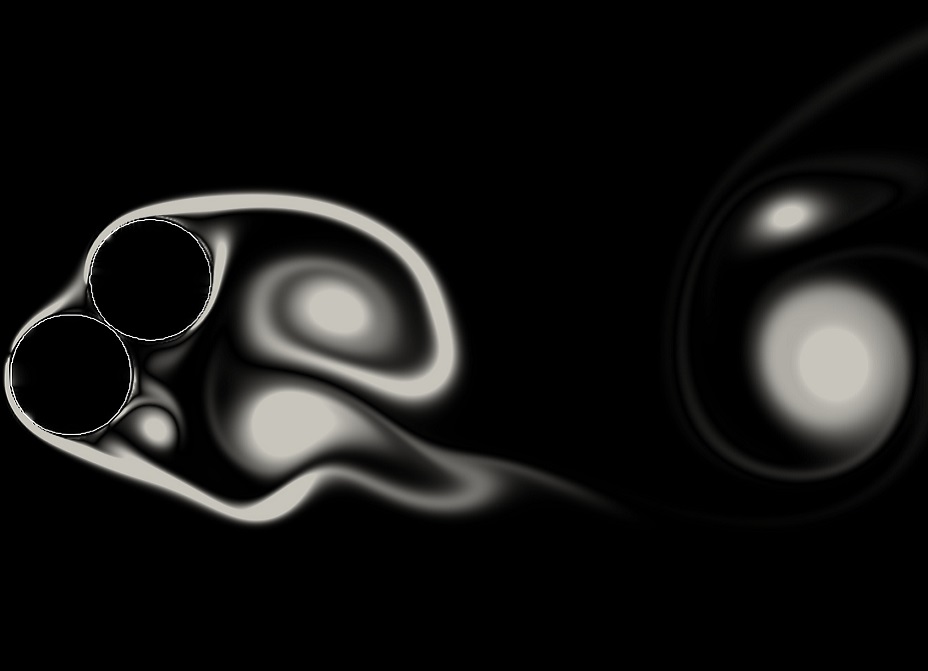}
    \subcaption{}
    \label{fig:e-exp}
\end{subfigure}
	\caption {Qualitative comparison of vortex shedding patterns (top row - experiments, bottom row - simulations). The experiments from \citet{Sumner2000} use dye visualization, whereas the simulation results are visualized using the vorticity magnitude. As characterized by \citet{Sumner2000}, the individual panels depict: \subref{fig:a-exp} vortex pairing, splitting, and enveloping; \subref{fig:b-exp} induced separation; \subref{fig:c-exp} shear layer reattachment; \subref{fig:d-exp} single bluff-body pattern 1; and \subref{fig:e-exp} single bluff-body pattern 2.}
\label{fig:qualitative}
\end{sidewaysfigure}
A quantitative comparison for a different set of experiments and simulations is presented in Figure~\ref{fig:Stvalpd}, which compares the Strouhal number for vortex shedding ($St = fd/U_\infty$, where $f$ is the frequency of vortex shedding, and $U_\infty$ is the inflow velocity). The results are shown for various $P/d$ values, with the angle between the cylinders kept constant at $\alpha$ = $60\degree$. The specific angle was selected to match the configurations for which quantitative Strouhal number data is provided by \citet{Sumner2000}. We observe good agreement between the data from the experiments and the simulations.
\begin{figure}
\centering
\includegraphics[scale=1.5]{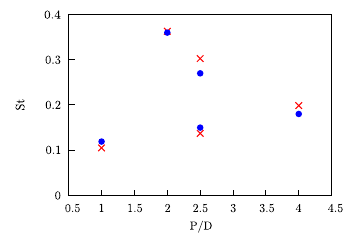}
	\caption{Comparing Strouhal numbers for vortex shedding in various configurations from the present simulations ($\bullet$) and experiments ($\times$) from \citet{Sumner2000}.} 
\label{fig:Stvalpd}
\end{figure}

We note that the simulations used a coarse grid ($1024^2$) during the optimization runs, in order to manage the high computational cost associated with the optimization procedure. However, the grid resolution was increased to at least double this value for the analyses presented in the Results section. To ensure that this grid resolution ($2048^2$ grid cells) was sufficient for cases with small separation between neighbouring cylinders' edges, additional simulations were conducted for Individual A with $4096^2$ grid cells. The time-evolution of drag at the two resolutions was compared, and no notable difference was observed between the results. Moreover, Figures~\ref{fig:d-exp} and~\ref{fig:e-exp}, indicate that the simulation results are consistent with experiments even in cases where the cylinders are in contact, i.e., for $P/d = 0$.

\begin{figure}
\centering
\includegraphics[width=0.8\textwidth]{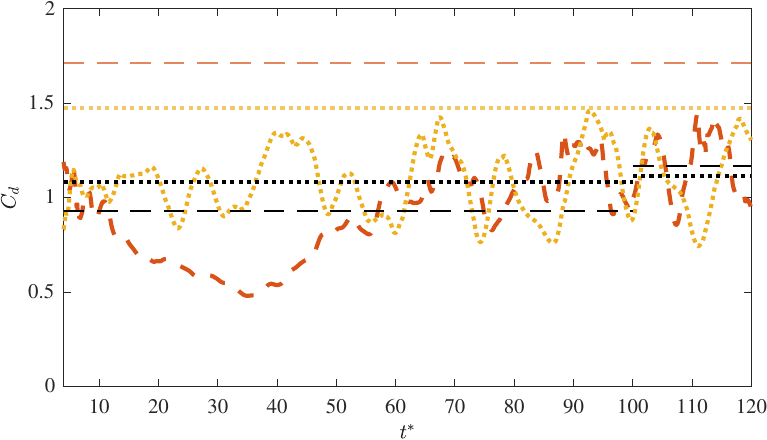}
	\caption{The time-varying drag coefficient for arrays B and C. The line-types for the curves correspond to those shown in Figure~\ref{fig:midline}. The coloured horizontal lines represent $C_d$ computed from the experiments for array B (dashed orange line) and for array C (dotted yellow line). The horizontal black lines indicate average $C_d$ computed from DNS data for arrays B (dashed line) and C (dotted line). The average $C_d$ which includes the initial transient startup is shown from $t^*=4$ up to $t^*=100$, and the average $C_d$ in the steady-shedding state is shown from $t^*=100$ to $t^*=120$, with the drag for array B being higher than that for array C.} 
\label{fig:timeVaryingCd}
\end{figure}

\section{Particle Tracking and Lagrangian Coherent Structures}

Snapshots from particle tracking in the initial transient state, and the quasi-steady vortex shedding state are shown in Figures~\ref{fig:particles_tr} and~\ref{fig:particles_ss}, respectively. The forward Finite Time Lyapunov Exponent (FTLE) is a time-dependent scalar field that acts as a measure of the separation of flow trajectories, and the Lagrangian Coherent Structures (LCSs) are defined as the locally maximizing surfaces or ridges in the FTLE scalar field. These lines represent the greatest separation of particle trajectories within the integration time period, which results in minimal flux across the Lagrangian Coherent Structures \citep{Shadden2005}.

\begin{figure}
\centering
\begin{subfigure}{0.48\textwidth}
  \centering
  \includegraphics[width=0.49\textwidth]{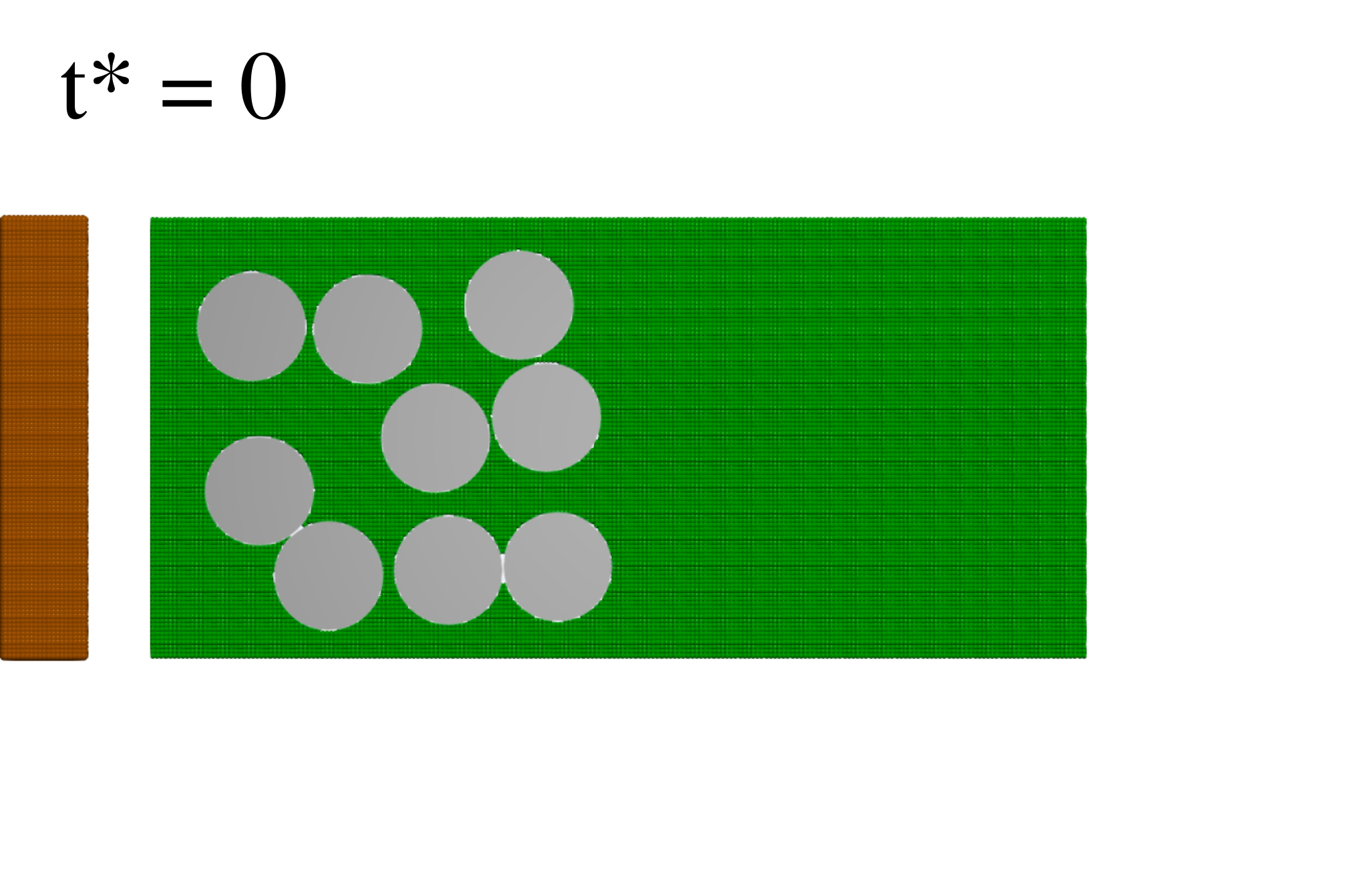}
  \includegraphics[width=0.49\textwidth]{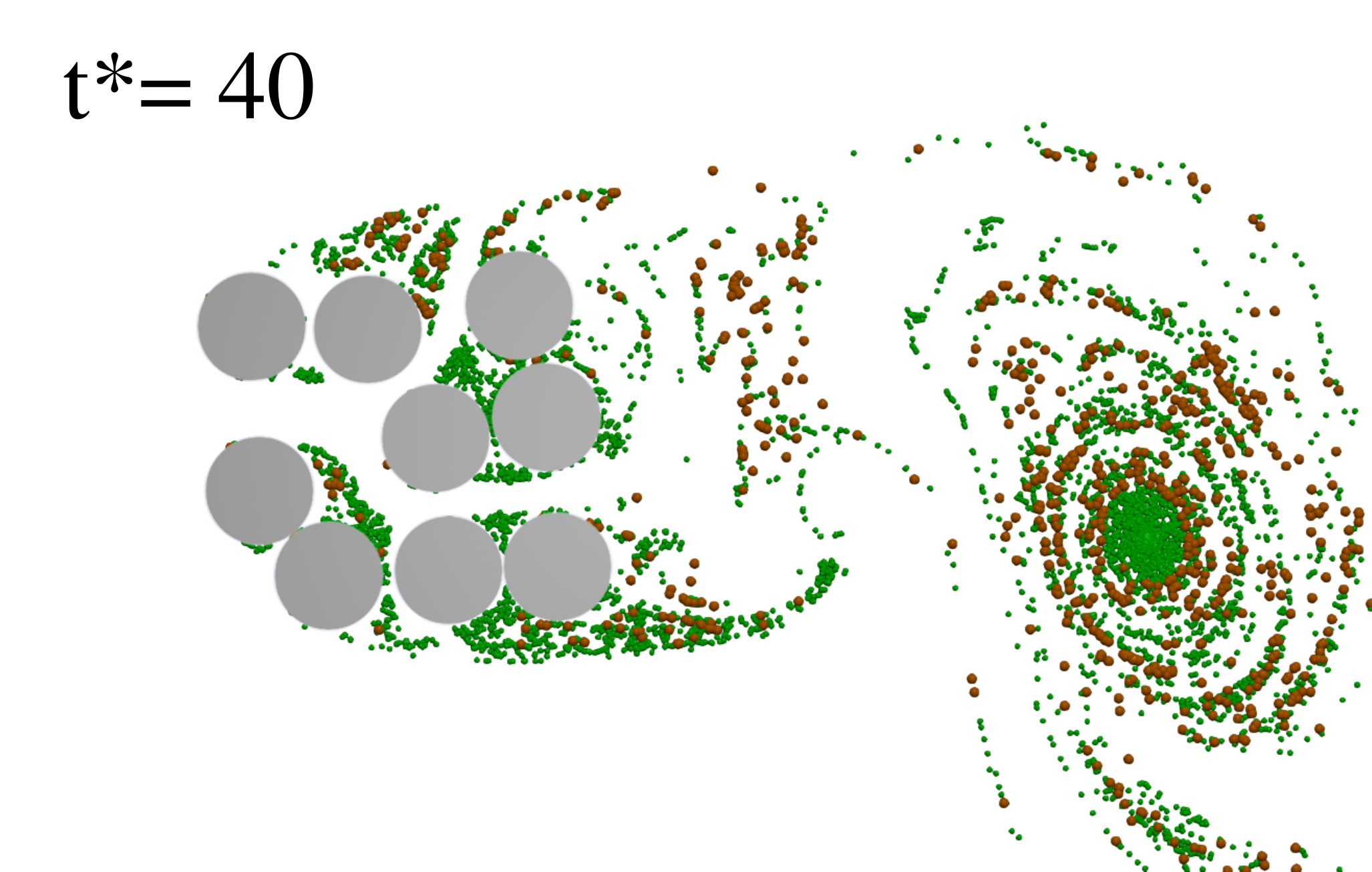}
  \subcaption{}
  \label{fig:part_tr1}
\end{subfigure}
\hfill
\begin{subfigure}{0.48\textwidth}
  \centering
  \includegraphics[width=0.49\textwidth]{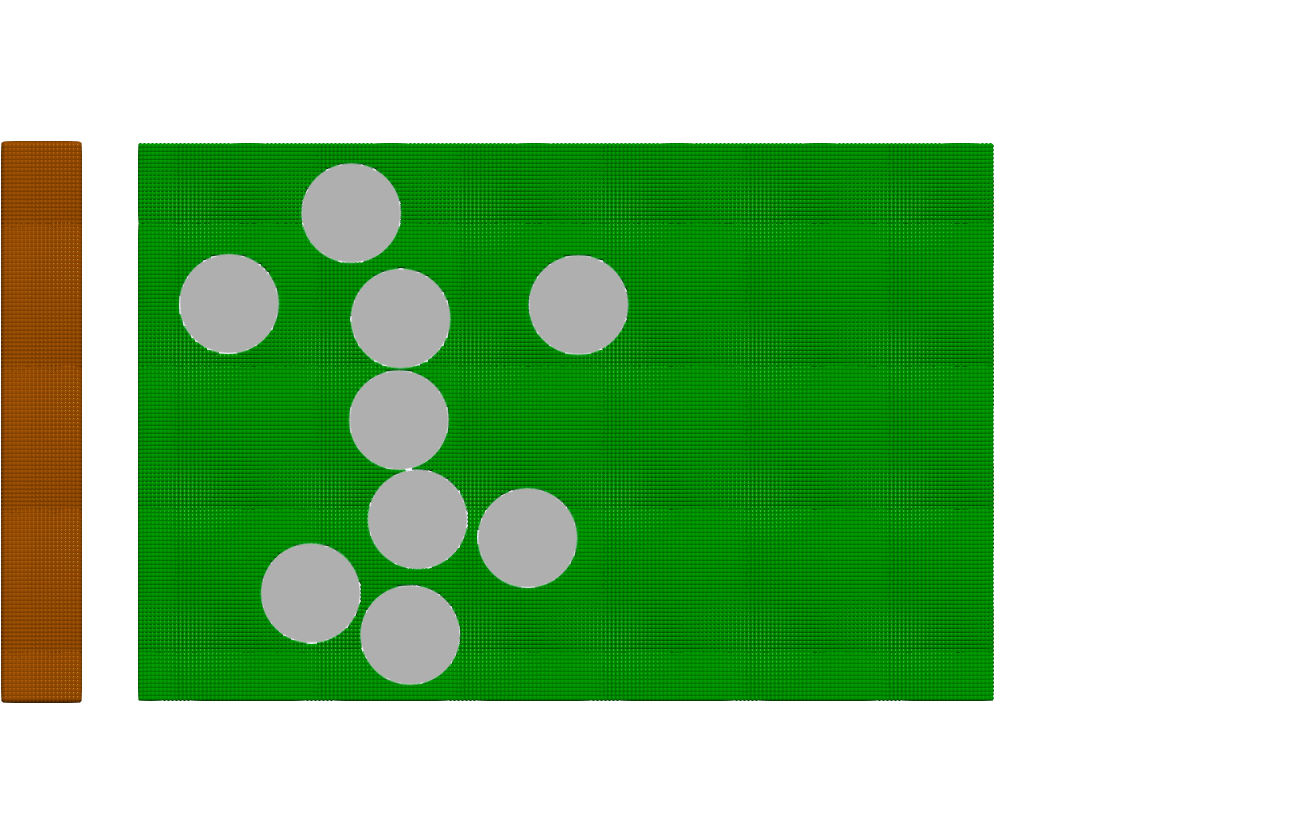}
  \includegraphics[width=0.49\textwidth]{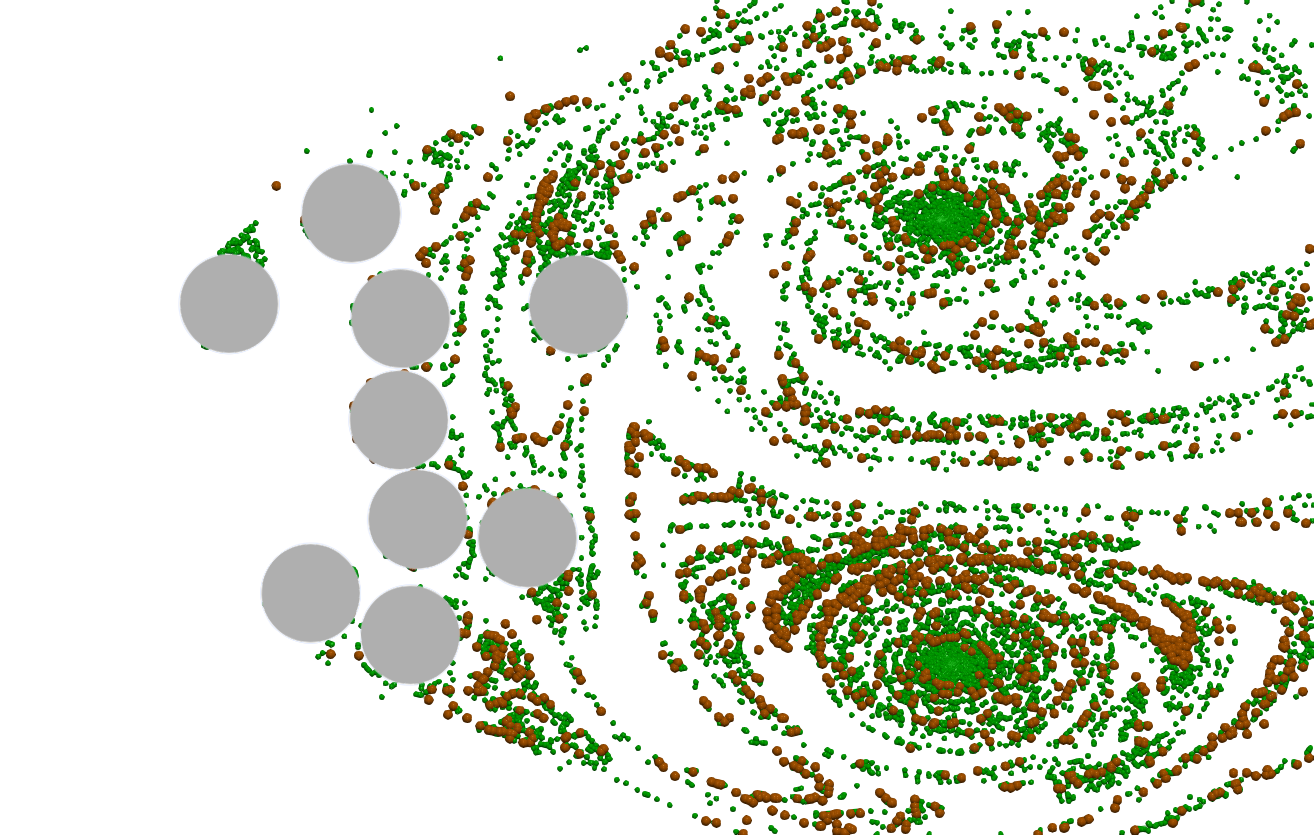}
  \subcaption{}
  \label{fig:part_tr2}
\end{subfigure}

\begin{subfigure}{0.48\textwidth}
  \centering
  \includegraphics[width=0.49\textwidth]{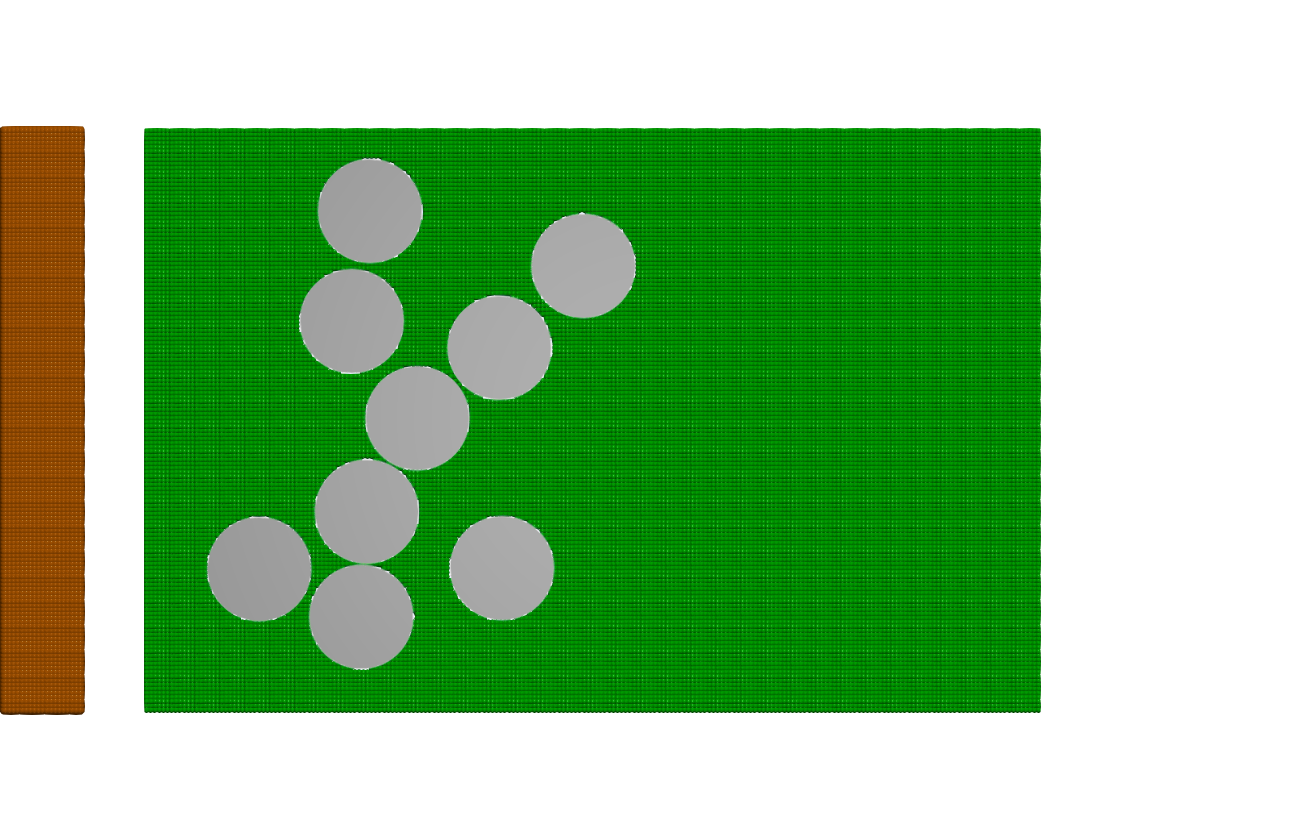}
  \includegraphics[width=0.49\textwidth]{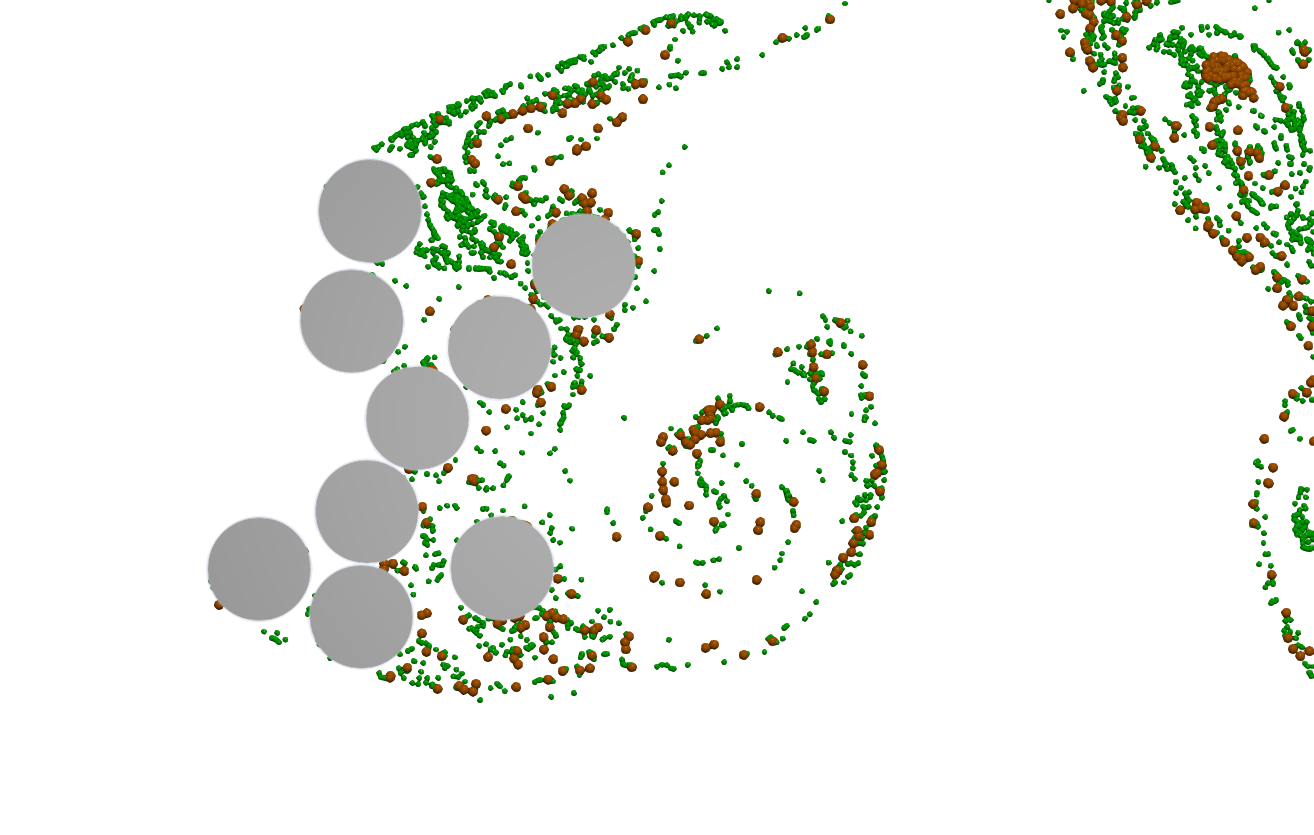}
  \subcaption{}
  \label{fig:part_tr3}
\end{subfigure}
\hfill
\begin{subfigure}{0.48\textwidth}
  \centering
  \includegraphics[width=0.49\textwidth]{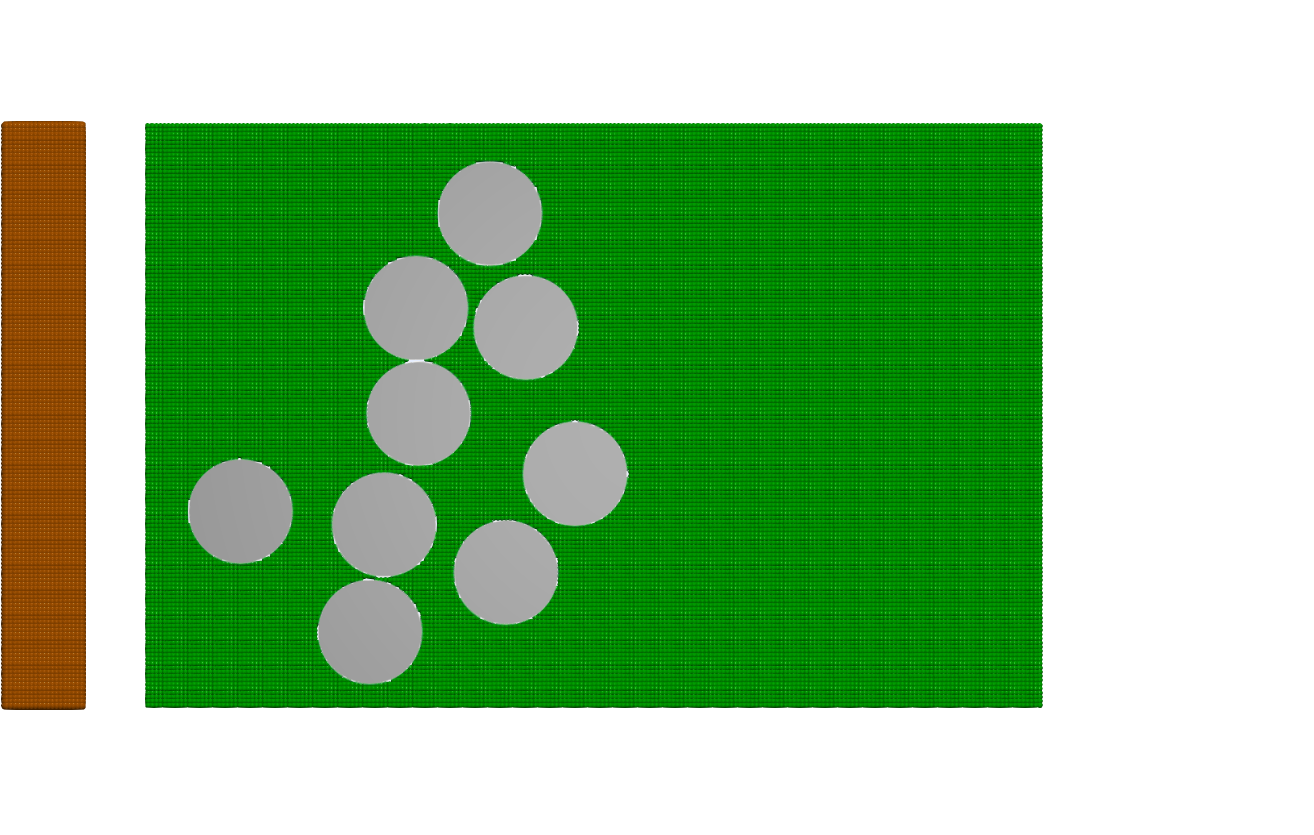}
  \includegraphics[width=0.49\textwidth]{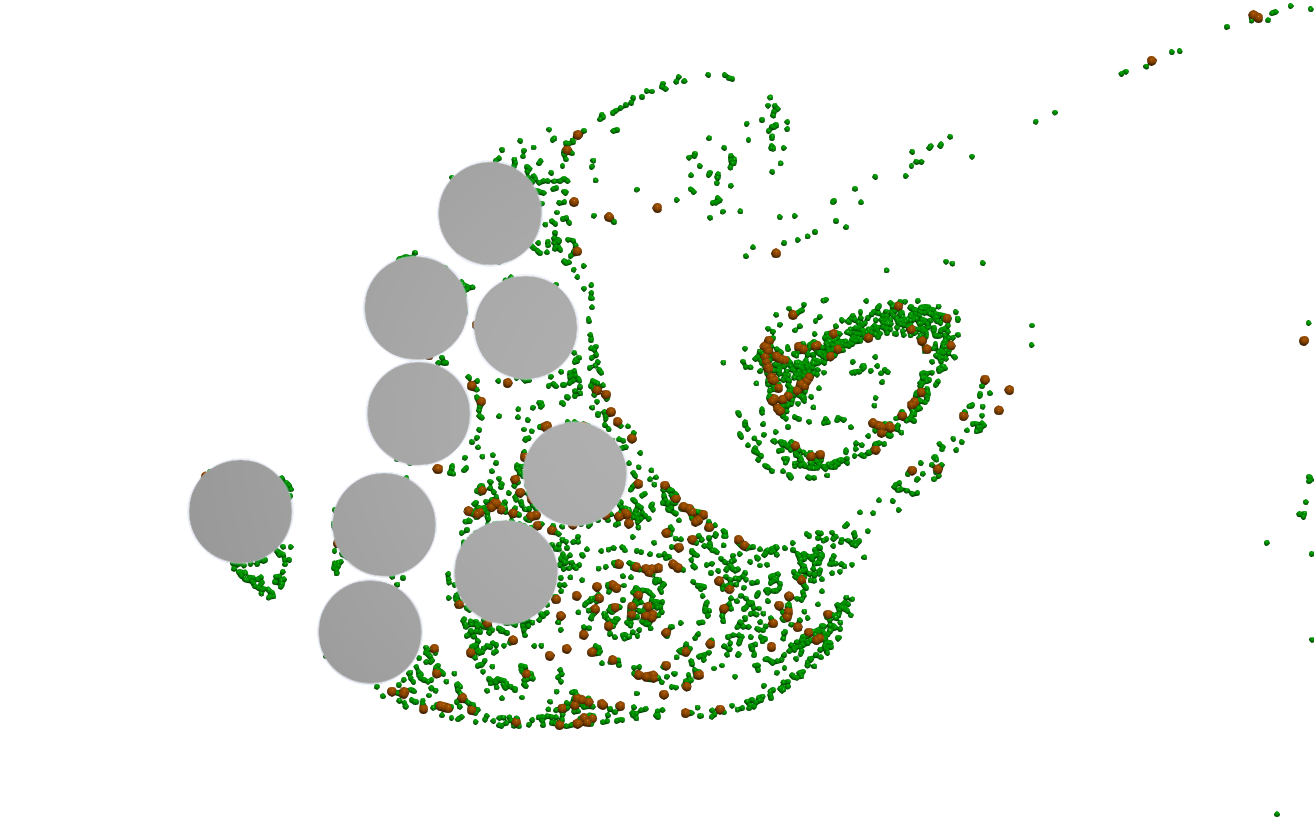}
  \subcaption{}
  \label{fig:part_tr4}
\end{subfigure}
	\caption{Particles initialized in DNS at $t^{*}=0$ (i.e., at the start of the transient state), and their position at $t^{*}=40$ for each of the four selected Pareto-optimal arrays A though D, shown respectively in panels \subref{fig:part_tr1} to \subref{fig:part_tr4}. $t^{*}$ is the non-dimensional time defined as $t^{*} = tU_{\infty}/d$, where $d$ is the diameter of an individual cylinder.  The brown particles were seeded upstream of the arrays, and the green particles were seeded in the vicinity and near-wake regions of the cylinders. The corresponding animations are provided in Supplementary Movie 2.}
\label{fig:particles_tr}
\end{figure}

\begin{figure}
\centering
\begin{subfigure}{0.48\textwidth}
\fbox{
  \begin{minipage}{\textwidth}
  \centering
  \includegraphics[width=0.49\textwidth]{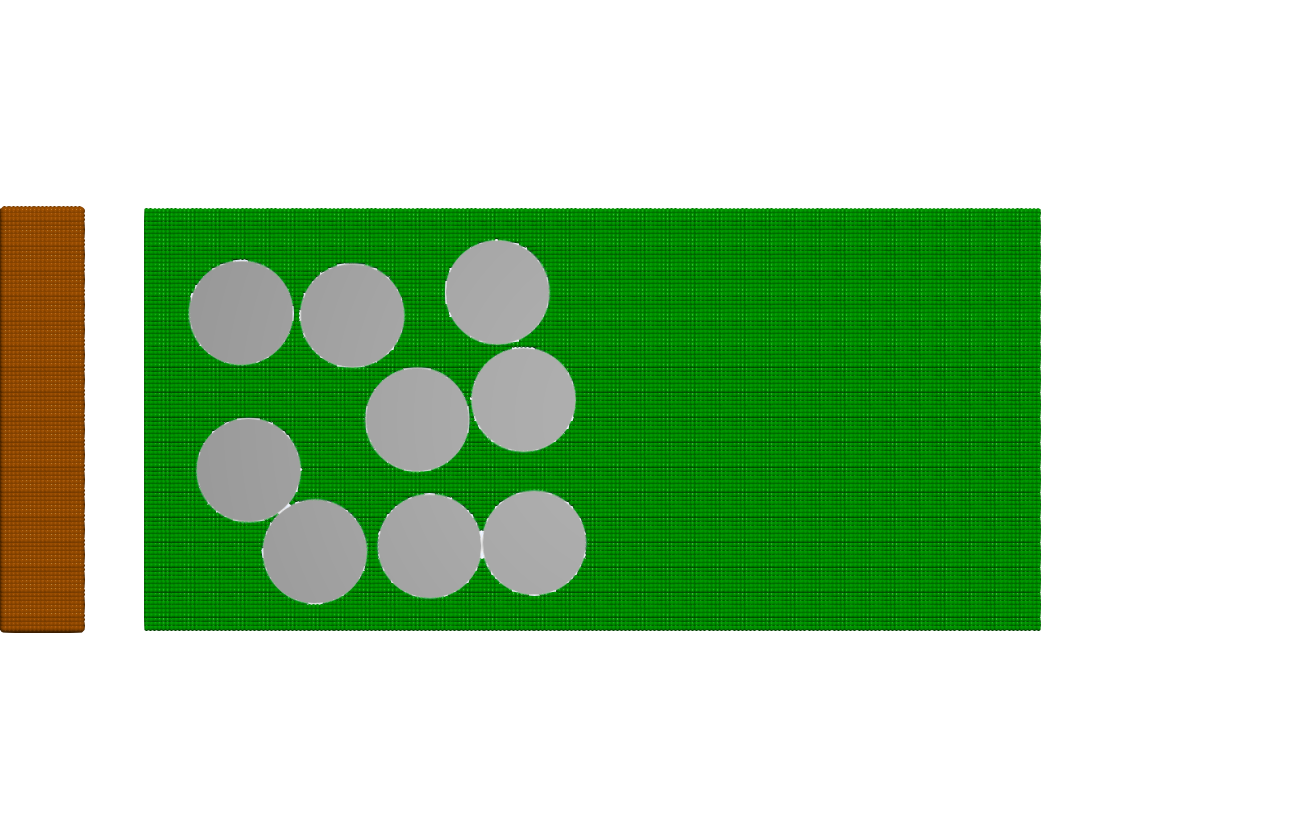}
  \includegraphics[width=0.49\textwidth]{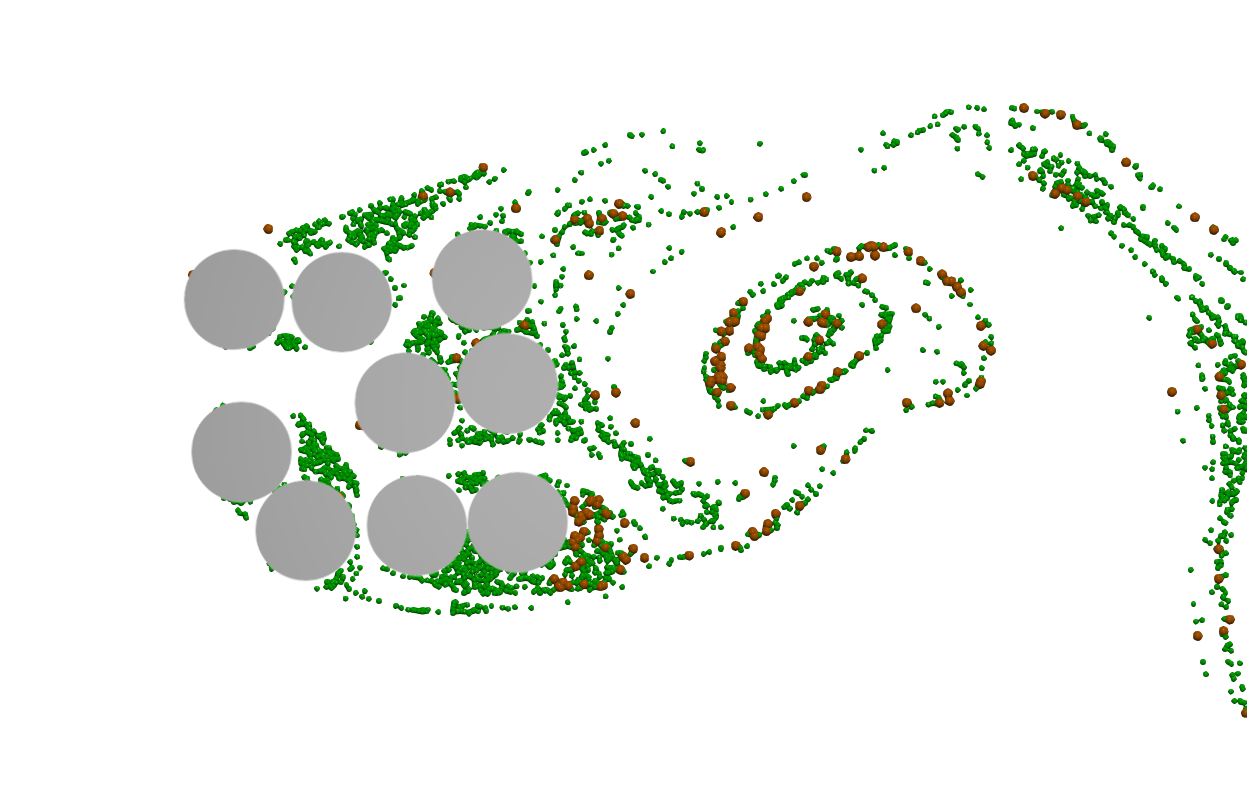}
  \includegraphics[width=0.49\textwidth]{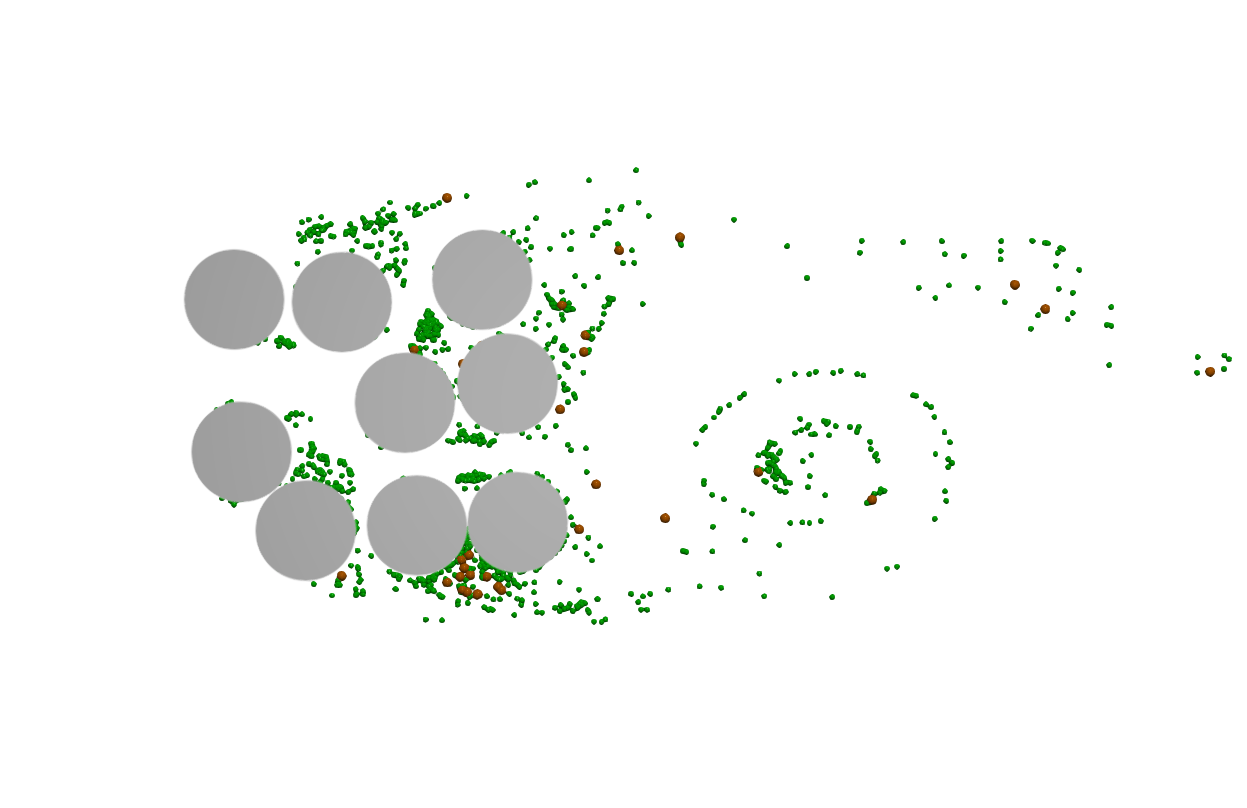}
  \includegraphics[width=0.49\textwidth]{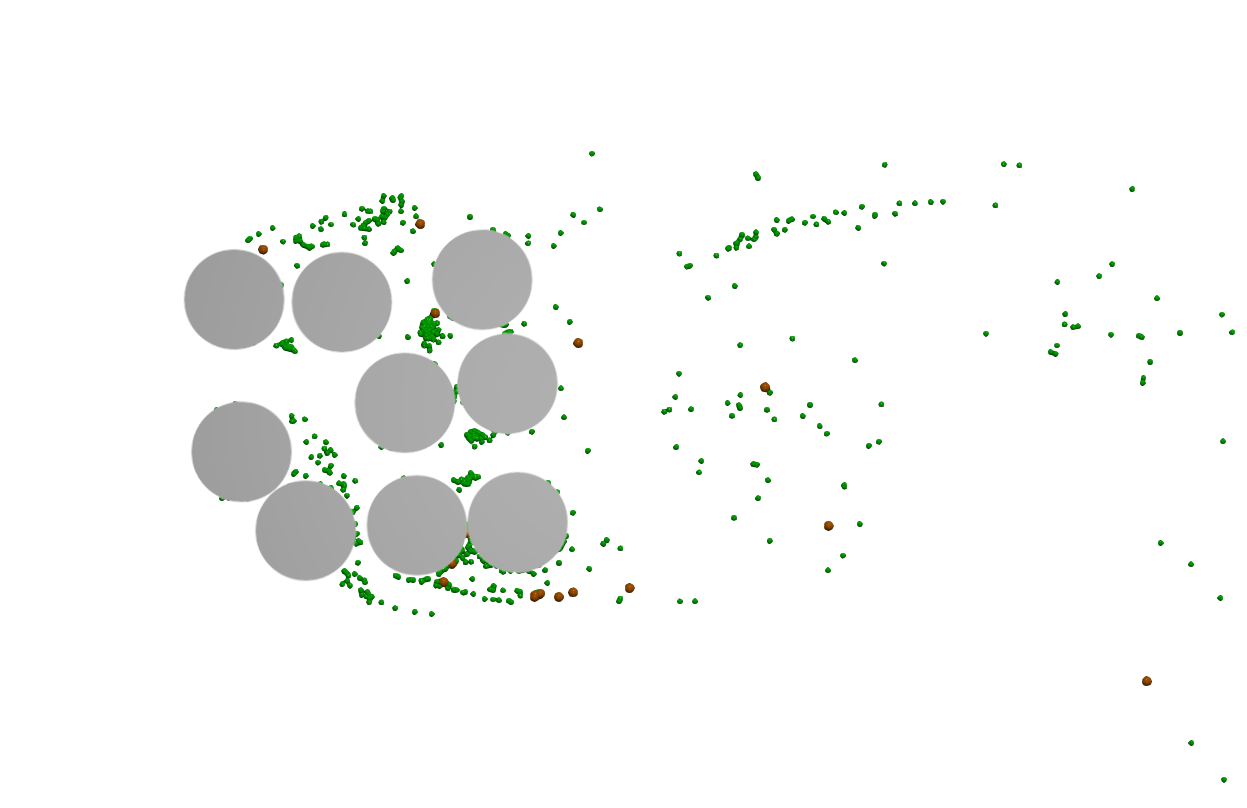}
  \end{minipage}}
  \subcaption{}
  \label{fig:part_ss1}
\end{subfigure}
\hfill
\begin{subfigure}{0.48\textwidth}
\fbox{
  \begin{minipage}{\textwidth}
  \centering
  \includegraphics[width=0.49\textwidth]{p2_t=0.png}
  \includegraphics[width=0.49\textwidth]{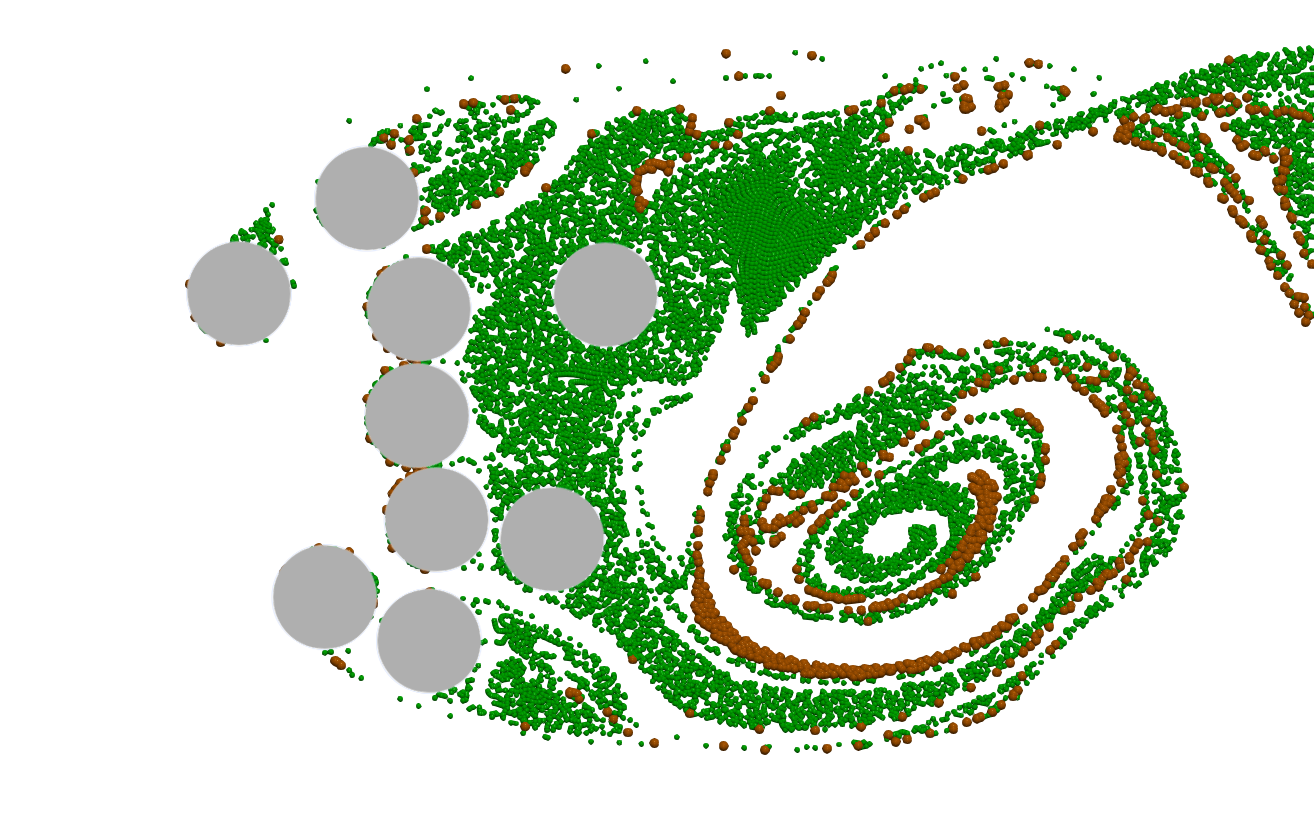}
  \includegraphics[width=0.49\textwidth]{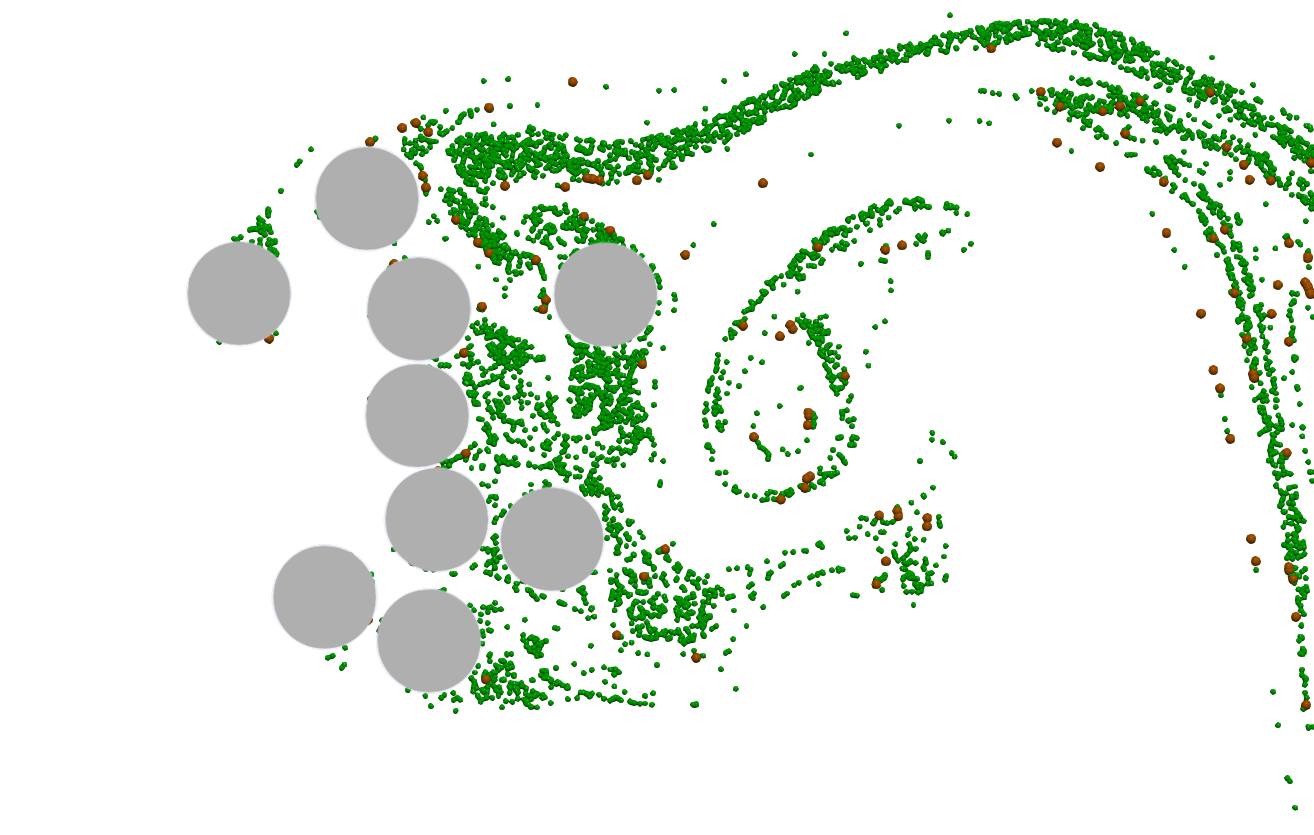}
  \includegraphics[width=0.49\textwidth]{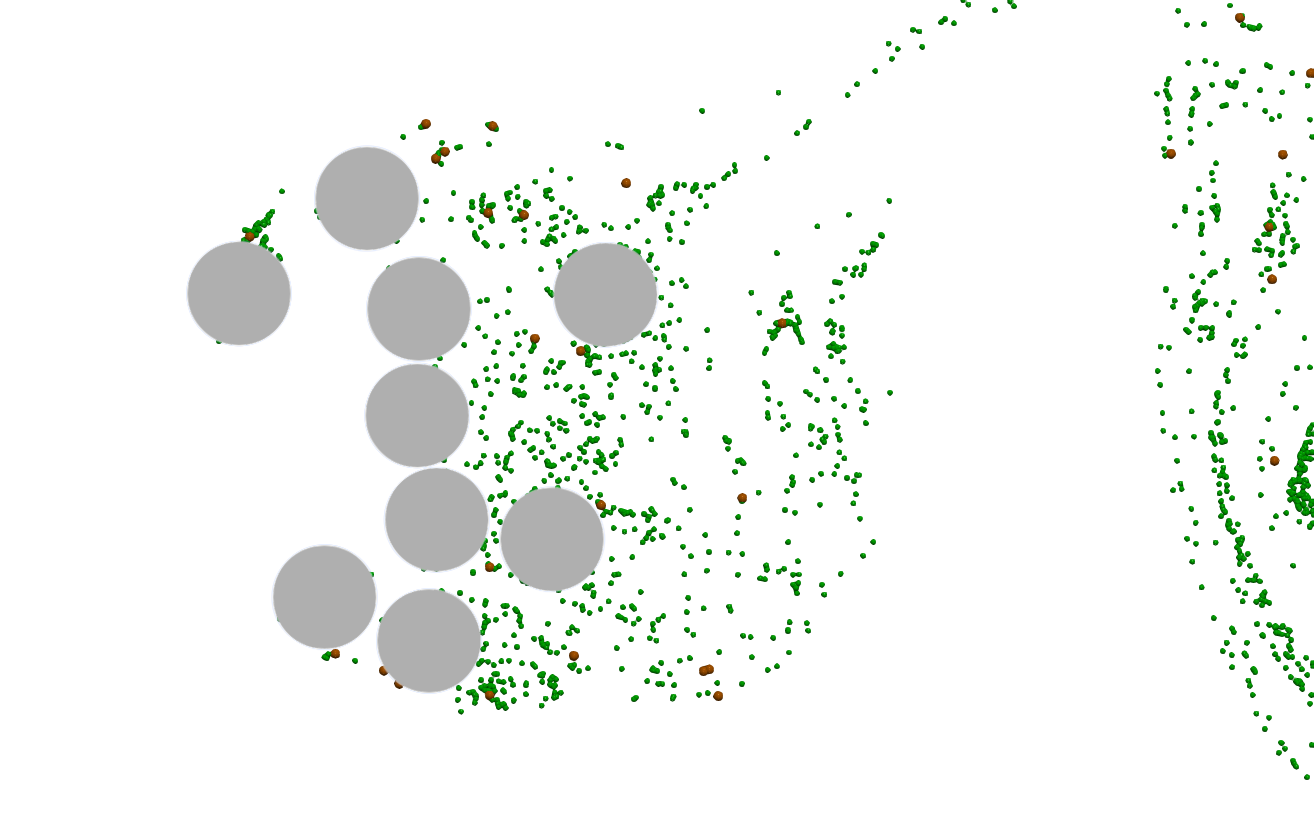}
    \end{minipage}}
  \subcaption{}
  \label{fig:part_ss2}
\end{subfigure}

\begin{subfigure}{0.48\textwidth}
\fbox{
  \begin{minipage}{\textwidth}
  \centering
  \includegraphics[width=0.49\textwidth]{p3_t=0.png}
  \includegraphics[width=0.49\textwidth]{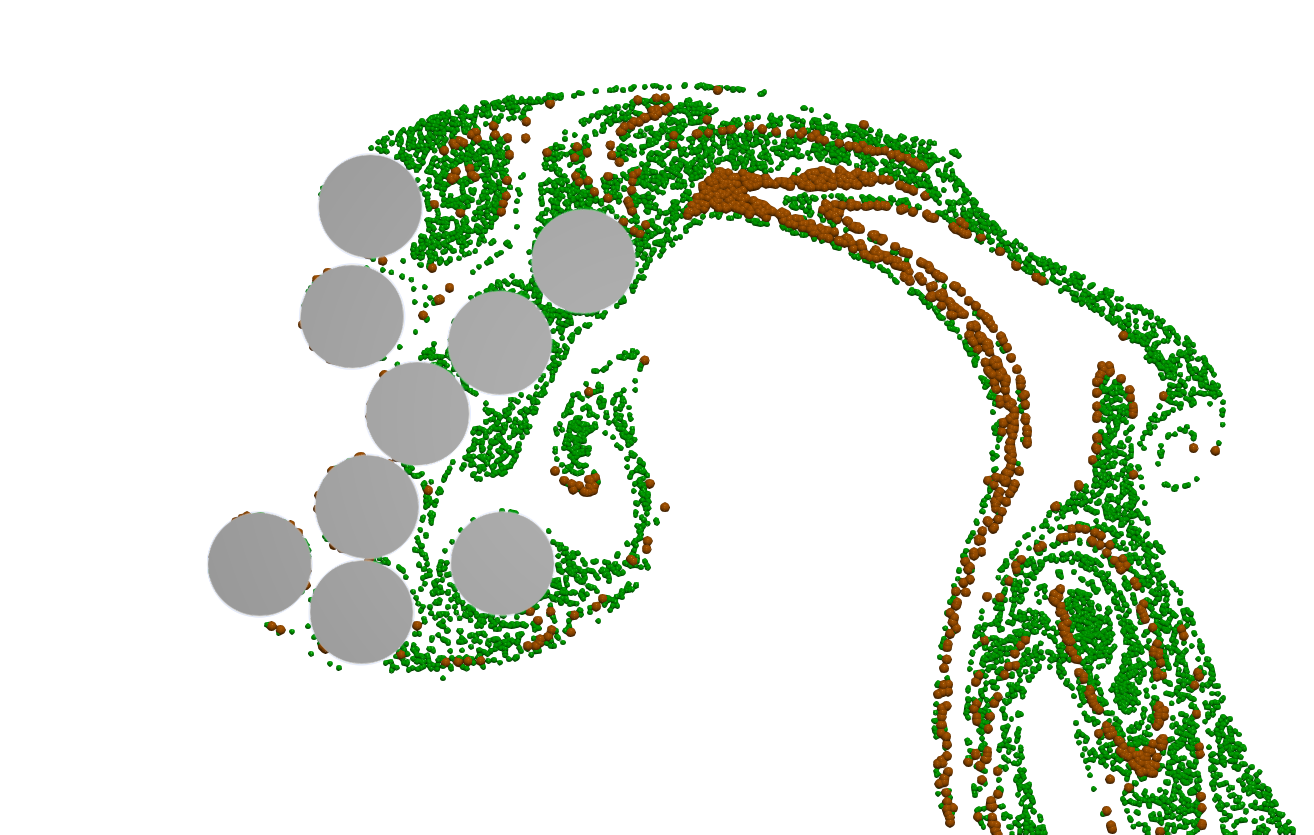}
  \includegraphics[width=0.49\textwidth]{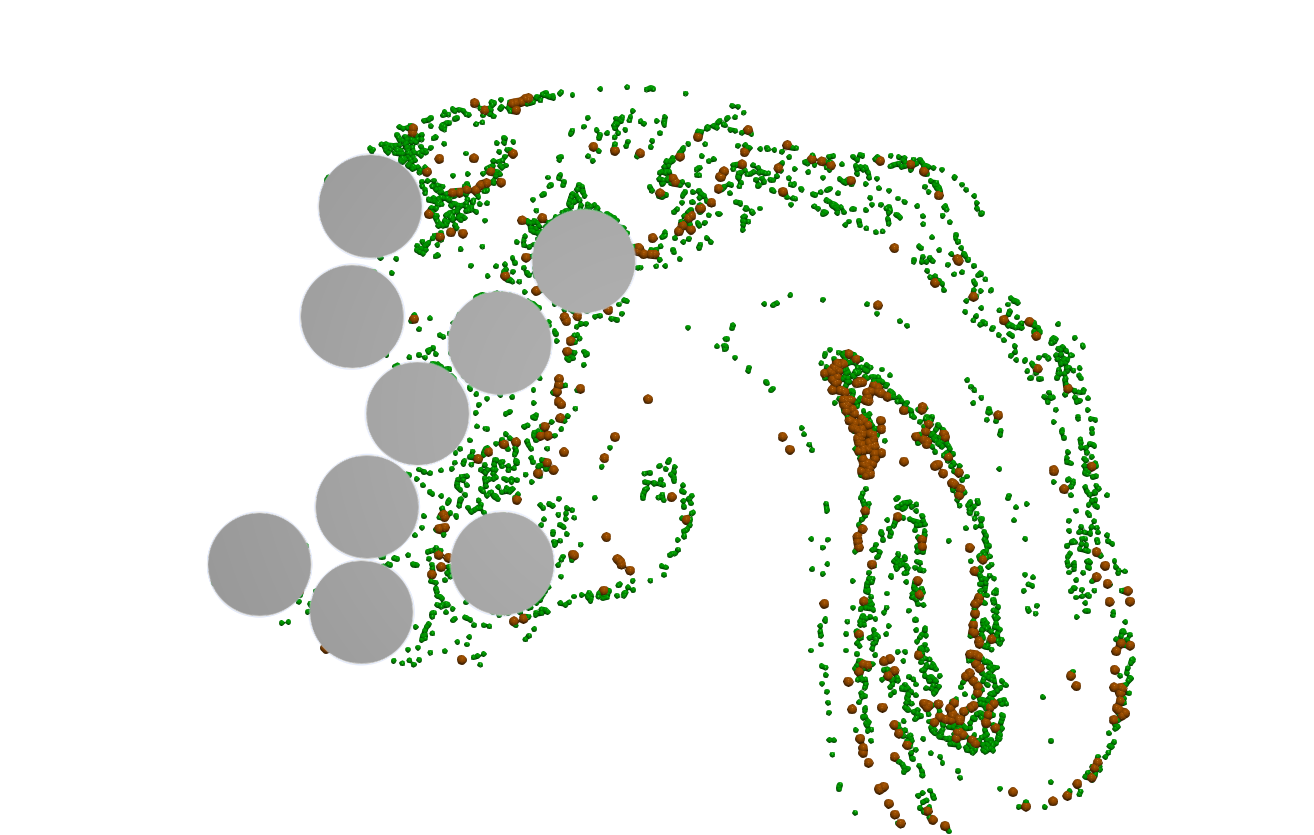}
  \includegraphics[width=0.49\textwidth]{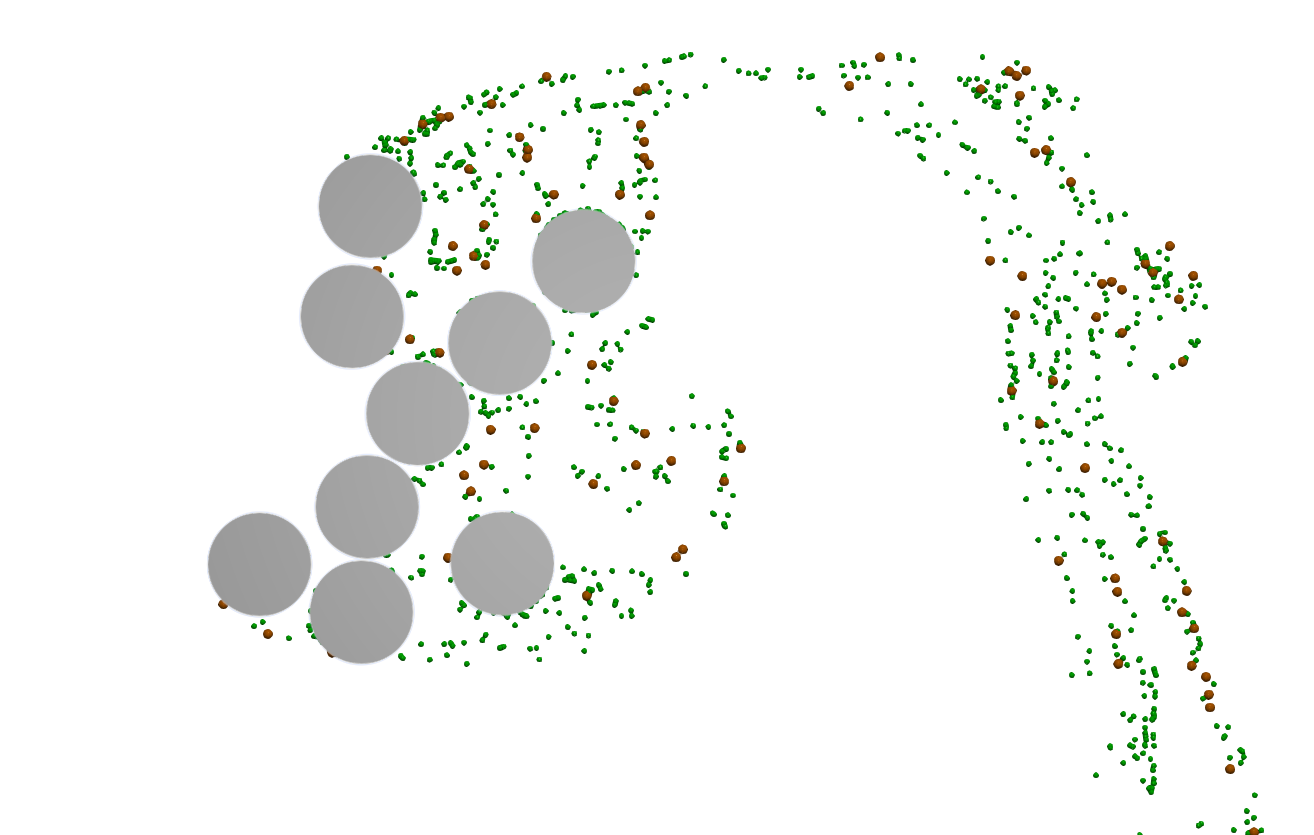}
  \end{minipage}}
  \subcaption{}
  \label{fig:part_ss3}
\end{subfigure}
\hfill
\begin{subfigure}{0.48\textwidth}
\fbox{
  \begin{minipage}{\textwidth}
  \centering
  \includegraphics[width=0.49\textwidth]{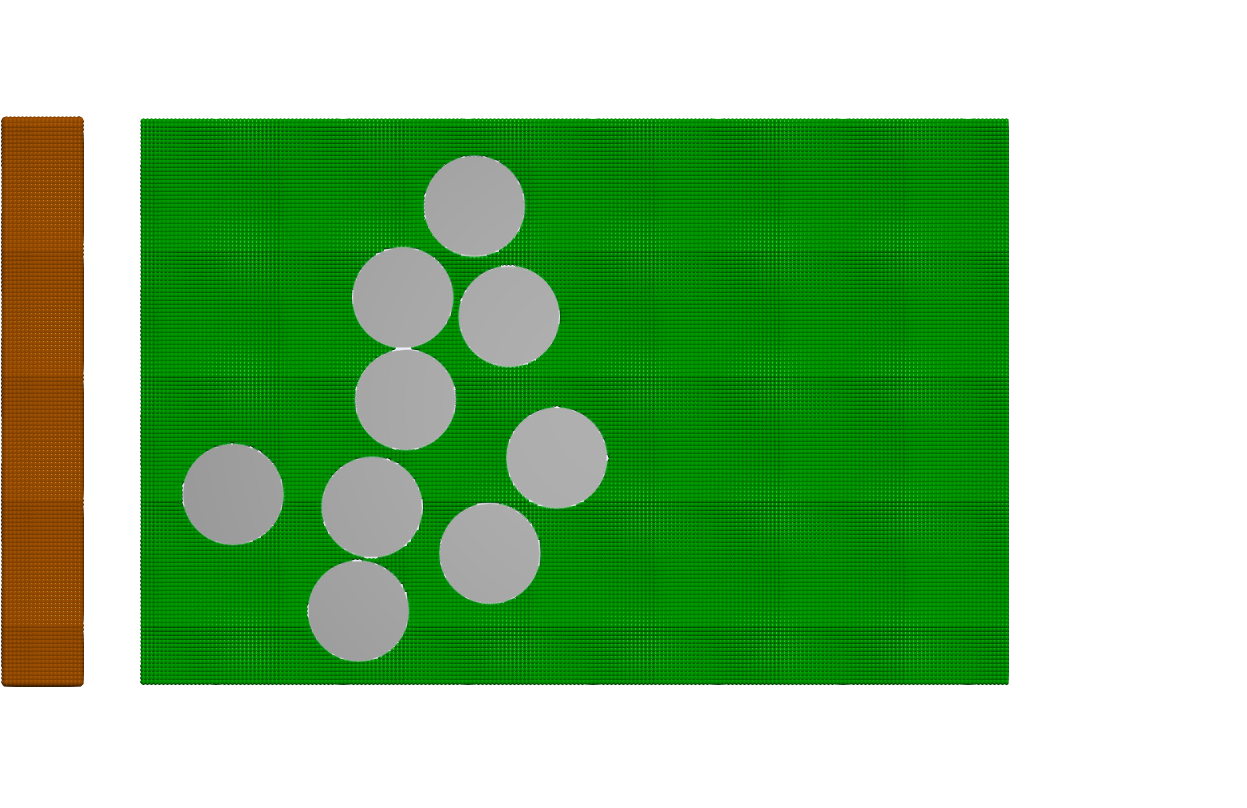}
  \includegraphics[width=0.49\textwidth]{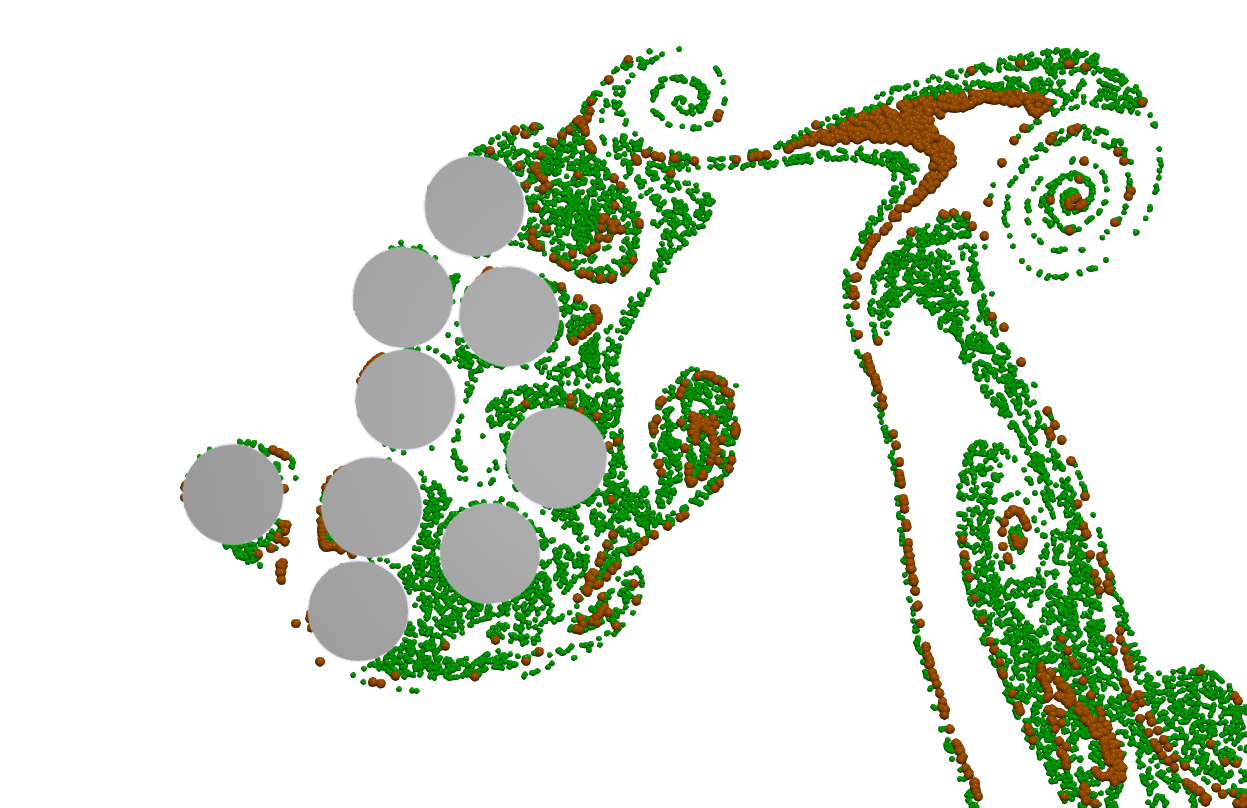}
  \includegraphics[width=0.49\textwidth]{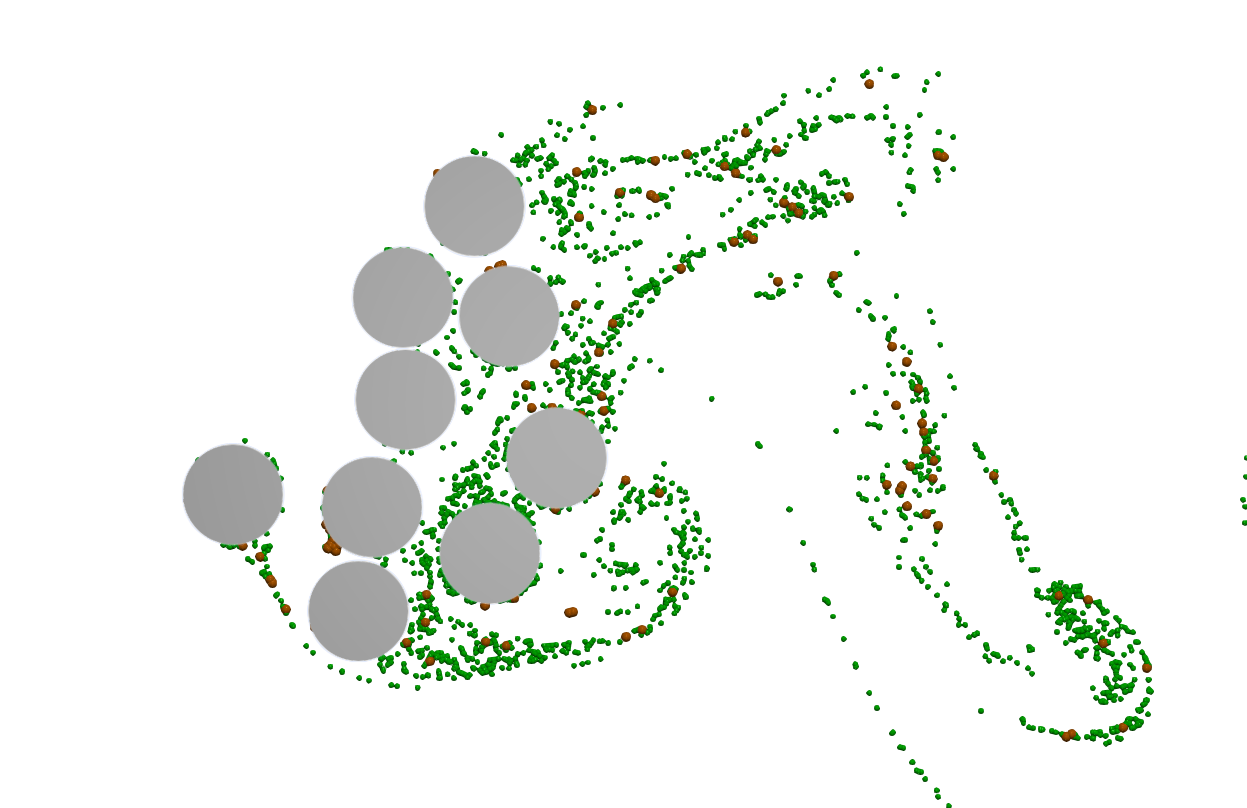}
  \includegraphics[width=0.49\textwidth]{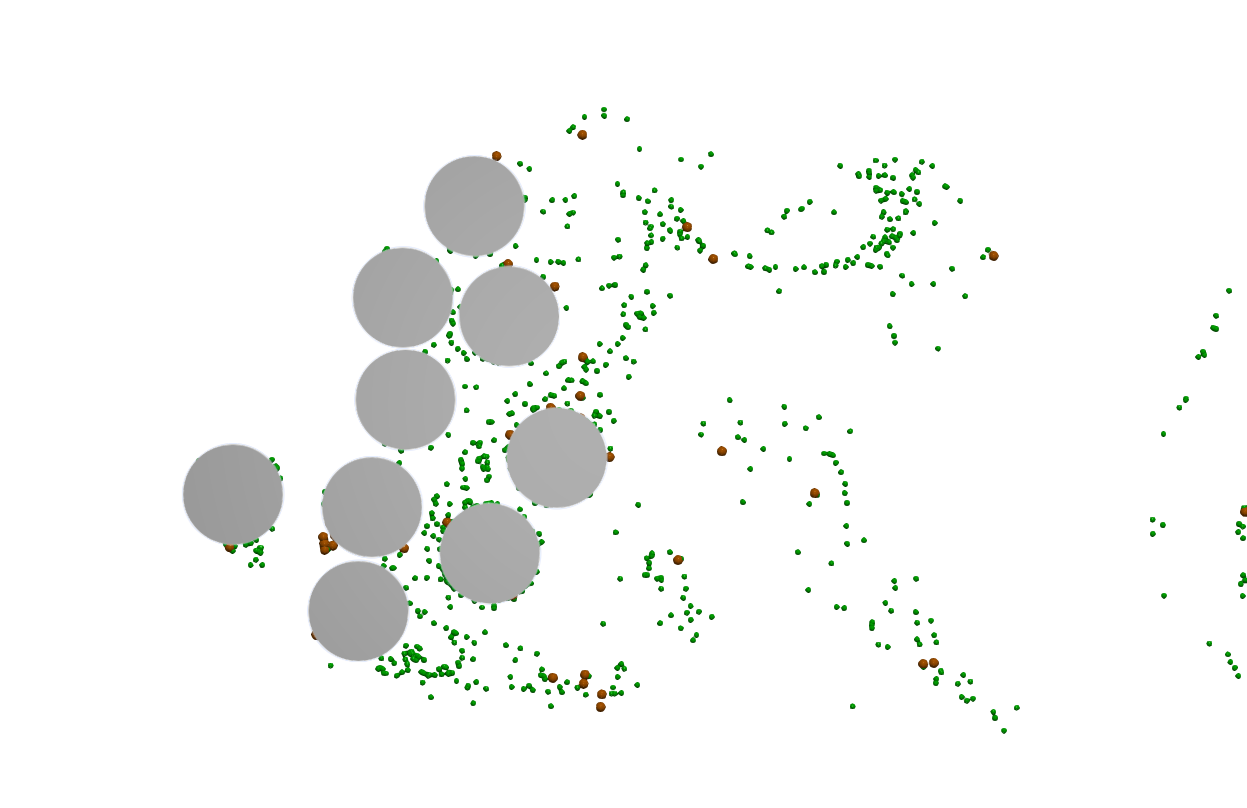}
  \end{minipage}}
  \subcaption{}
  \label{fig:part_ss4}
\end{subfigure}
	\caption{Particles seeded in DNS in the steady shedding state at $t=t_{o}$, $t_{o}+T$, $t_{o}+2T$ and $t_{o}+3T$, where $T$ is the respective shedding time period for each of the arrangements in panels (a) through (d). The corresponding animations are provided in Supplementary Movie 3.}
\label{fig:particles_ss}
\end{figure}

\bibliographystyle{jfm}
\bibliography{mangrove}

\end{document}